\documentclass[11pt,a4paper]{article}
\usepackage[utf8]{inputenc} 
\usepackage[english]{babel}
\usepackage[T1]{fontenc}
\usepackage{graphicx,amssymb,amsmath,wasysym}
\usepackage[bookmarks=false]{hyperref}

\begin{document}
\baselineskip=15.5pt

\thispagestyle{empty}

\begin{flushright}
CFTP/09-040\\
LPT-09-110
\end{flushright}

\vspace{.5cm}

\begin{center}

{\Large\sc{\bf Dark matter detection in the BMSSM}}

\vspace*{9mm}
\setcounter{footnote}{0}
\setcounter{page}{0}
\renewcommand{\thefootnote}{\arabic{footnote}}

\mbox{ {\sc Nicolás BERNAL$^{1}$, Andreas GOUDELIS$^{2, 3}$}}

\vspace*{0.9cm}

{\it $^1$Centro de Física Teórica de Partículas (CFTP),\\
Instituto Superior Técnico, Avenida Rovisco Pais, 1049-001 Lisboa, Portugal}\\
\vskip 0.3cm
{\it $^2$Laboratoire de Physique Théorique d'Orsay, UMR8627--CNRS,\\
Université Paris--Sud, Bât. 210, F--91405 Orsay Cedex, France}\\
\vskip 0.3cm
{\it $^3$Istituto Nazionale di Fisica Nucleare, 
Sezione di Padova, I-35131, Padova, Italy}\\
\vskip5mm
E-mails: {\tt nicolas.bernal@cftp.ist.utl.pt, andreas.goudelis@th.u-psud.fr}
\vskip5mm

\end{center}

\vspace{1cm}

\begin{abstract}
The addition of non-renormalizable terms involving the Higgs fields
to the MSSM (BMSSM) ameliorates the little hierarchy problem of the
MSSM. For neutralino dark matter, new regions for which the relic abundance
of the LSP is consistent with WMAP (as the bulk region and the stop coannihilation region)
are now permitted. In this framework, we analyze in detail the direct dark matter detection
prospects in a XENON-like experiment. On the other hand, we study the
capability of detecting gamma-rays, antiprotons and positrons produced
in the annihilation of neutralino LSPs in the Fermi and oncoming AMS-02 experiments.
\end{abstract}

\newpage

\tableofcontents

\newpage

\section{Introduction}
\label{sec:intro}
The smallness of the quartic Higgs coupling in the framework of the
minimal supersymmetric standard model (MSSM) poses a problem. The tree
level bound on the Higgs mass is violated, and
large enough loop corrections to satisfy the lower bound on the Higgs
mass suggest that the stop sector has rather peculiar features: at
least one of the stop mass eigenstates should be rather heavy and/or
left-right-stop mixing should be substantial \cite{Djouadi:2005gj}.

The situation is different if the quartic Higgs couplings are affected
by new physics. If the new physics appears at an energy scale that is
somewhat higher than the electroweak breaking scale, then its effects
can be parametrized by non-renormalizable terms. The leading
non-renormalizable terms that modify the quartic couplings are
\cite{Strumia:1999jm,Brignole:2003cm,Casas:2003jx,Pospelov:2005ks,
Pospelov:2006jm,Dine:2007xi,Batra:2008rc,Antoniadis:2008es,Antoniadis:2009rn}:
\begin{equation}\label{eq:wdst}
W_{\rm BMSSM}=\frac{\lambda_1}{M}(H_uH_d)^2+\frac{\lambda_2}{M}{\cal
  Z}(H_uH_d)^2,
\end{equation}
where ${\cal Z}$ is a SUSY-breaking spurion:
\begin{equation}
{\cal Z}=\theta^2m_{\rm susy}.
\end{equation}
The first term in equation~\eqref{eq:wdst} is supersymmetric, while the
second breaks supersymmetry (SUSY). In the scalar potential, the
following quartic terms are generated:
\begin{equation}\label{eq:dstsp}
2\epsilon_1 H_uH_d(H_u^\dagger H_u+H_d^\dagger H_d)
+\epsilon_2(H_uH_d)^2,
\end{equation}
where
\begin{equation}
\epsilon_1\equiv\frac{\mu^*\lambda_1}{M},\ \ \
\ \ \ \epsilon_2\equiv-\frac{m_{\rm susy}\lambda_2}{M}.
\end{equation}

However, let us note that these operators Beyond the MSSM (BMSSM) may destabilize the scalar potential.
If $4|\epsilon_1|>\epsilon_2$, the effective quartic coupling along one
of the D-flat directions is negative, causing a remote vacuum to form
in the presence of which the electroweak vacuum could become
metastable \cite{Blum:2009na}. 

The interplay between the Higgs sector, the stop sector, and the
non-renormalizable (NR) operators has interesting consequences for the
MSSM baryogenesis \cite{Blum:2008ym,Bernal:2009hd}. The window for MSSM baryogenesis
is extended and, more importantly, can be made significantly more
natural. In addition, these operators have implications for yet another
cosmological issue, and that is Dark Matter (DM) \cite{Cheung:2009qk,
Berg:2009mq,Bernal:2009hd}. The phenomenology of the BMSSM
at colliders has been studied in reference \cite{Carena:2009gx};
implications for fine-tuning have been analyzed in reference \cite{Cassel:2009ps}.

One of the attractive features of the MSSM is the fact that the
lightest R-parity-odd particle (LSP)
is a natural candidate for being the dark matter particle.
Progress in experimentally constraining the MSSM parameter space
restricts, however, the regions where the dark matter is
quantitatively accounted for to rather special regions of the MSSM:
the focus point region, with surprisingly heavy sfermions; the funnel
region, where the mass of the CP-odd neutral Higgs scalar is very
close to twice the mass of the LSP; the co-annihilation region, where
the mass of the scalar partner of the right-handed tau is very close
to the mass of the LSP; and the bulk region, where the bino-LSP and
the sleptons are light.

The effects of the NR operators are potentially important for two of
these four regions. First, these operators give rise to a new
Higgs-Higgs-higgsino-higgsino interaction Lagrangian,
\begin{equation}\label{eq:hhHH}
-\frac{\epsilon_1}{\mu^*}\left[2(H_uH_d)(\widetilde H_u\widetilde H_d)
+2(\widetilde H_u H_d)(H_u\widetilde H_d)+(H_u\widetilde H_d)
(H_u\widetilde H_d)+(\widetilde H_u H_d)(\widetilde H_u H_d)\right]
+{\rm h.c.},
\end{equation}
which contributes to the annihilation process of two higgsinos to two
Higgs particles. This effect is relevant when the dark matter particle
has a significant component of higgsinos, as is the case in the focus
point region. Second, as mentioned above, these operators modify the
relation between the light Higgs mass and the stop masses. This effect
can be important in the bulk region within models where the slepton and
stop masses are related, such as the mSUGRA models. 

The effect of these operators on the relic density was studied in detail
in references \cite{Berg:2009mq, Bernal:2009hd}. It was found that new regions
yielding the correct relic density can arise, especially in the bulk and 
co-annihilation region.
In this work, we examine how the detection prospects of the MSSM are modified 
by the introduction of terms in equation \eqref{eq:dstsp} in the superpotential,
focusing ourselves on the re-opened regions.

We shall evaluate the detection perspectives for the main four kinds of detection
usually considered in the literature, namely direct detection in a XENON-like 
experiment, gamma-ray detection from dark matter annihilations in the galactic center
for the Fermi mission \cite{FermiSite} as well as positron and antiproton detection coming again from 
dark matter annihilations in the galactic halo and for the oncoming AMS-02 \cite{AMSsite} experiment.

\section{The model}
\label{sec:model}

The BMSSM framework, if relevant to the little hierarchy problem that
arises from the lower bound on the Higgs mass, assumes a new physics
scale at a few TeV. Since the new degrees of freedom at this scale are
not specified, the effect of the new threshold on the running of parameters
from a much higher scale cannot be rigorously taken into account. It
therefore only makes sense to study the BMSSM effects in a framework
specified at low energy. In order to demonstrate some of the most
interesting consequences of the BMSSM operators for dark matter, we shall
employ the two sets of parameters explored in reference \cite{Bernal:2009hd}:
a model where all sfermion masses are
correlated, and a model where the only light sfermions are the stops.
The first model demonstrates how the so-called bulk region is re-opened,
even for correlated stop and slepton masses. The second model incorporates
the interesting process of stop co-annihilation. For both models we focus
our attention mainly on regions where the stops are light, since the main
motivation for the BMSSM operators is to avoid a heavy stop (which is the
cause of the little hierarchy problem). Previous analysis in the context
of the MSSM with a light stop was done in references \cite{Davidson:1999ii,
Balazs:2004bu}.

Within this framework, we calculate the dark matter relic density,
and the direct and indirect detection prospects
in the presence of
the new $\epsilon_i$ couplings. We used a modified version of the code
{\tt micrOMEGAs} \cite{Belanger:2006is,Belanger:2008sj}, where we implemented the BMSSM
Higgs-Higgs-higgsino-higgsino couplings of equation~\eqref{eq:hhHH}, in order to calculate
the relic density as well as the cross-sections and decay channels relevant
for dark matter detection.
The leading $\epsilon_i$-induced corrections to the
spectrum, were implemented using the code
{\tt SuSpect} \cite{Djouadi:2002ze}.

\subsection{Correlated stop-slepton masses}
\label{ssec:cssm}
The first scenario considered contains correlated stop and slepton masses,
just as the most studied MSSM scenarios, such as  the mSUGRA
\cite{Chamseddine:1982jx,Barbieri:1982eh,Hall:1983iz,Ohta:1982wn} or cMSSM
frameworks. In this case, the neutralino LSP is a bino-like state
annihilating to the standard model leptons via light slepton exchange.
However, this scenario, known as the `bulk region', is highly constrained
due to the experimental lower bound on the Higgs mass.
In general, in order to fulfill such a constraint either heavy
or mixed stops are required \cite{Belyaev:2007xa}. In the BMSSM, nevertheless, it is 
possible to re-open the bulk region regardless of the structure of the stop sector.

In order to allow for a simple comparison with mSUGRA-models, we focus the attention to
the parameters
\begin{equation}\label{eq:msugra}
\tan\beta,\ m_{1/2},\ m_0,\ A_0,\ {\rm sign}(\mu).
\end{equation}
Let us emphasize again that one should {\it not}
think about this set of parameters as coming from an extended mSUGRA
model, since the effects of the BMSSM physics at the few TeV scale on
the running are not (and cannot) be taken into account. In addition,
we have two extra BMSSM parameters: $\epsilon_1$ and $\epsilon_2$.

In practice, we make discrete choices of $\tan\beta$, $A_0$,
sign$(\mu)$, $\epsilon_1$ and $\epsilon_2$, and scan over $m_0$ and $m_{1/2}$. We
focus our attention on moderate values of $m_{1/2}$ and $m_0$ because
we are mainly interested in light sfermions and the bulk region.
We also use $A_0=0$ GeV and $\mu>0$ in the whole analysis.
 
As pointed out in reference \cite{Bernal:2009hd}, we would like to
emphasize several points regarding the present scenario:
\begin{enumerate}
\item A generic point in the former parameter space usually gives rise to a
  too low annihilation cross-section of neutralino LSP and hence to a
  too large relic density, in conflict with the WMAP measurements \cite{Dunkley:2008ie}.
\item However, for moderate $m_{1/2}$ and low $m_0$ values, there is a
  region where the LSP is almost degenerate in mass with the lightest stau
  ($\tilde\tau_1$), enhancing the co-annihilation cross-section
  $\chi_1^0-\tilde\tau$.
\item Another region giving rise to relic density in agreement with the
  WMAP measurements appears for $m_{1/2}\sim 120$ GeV.  
  This is the `$h$-pole' and the '$Z$-pole' region in which
  $m_h\sim m_Z\sim 2\,m_{\chi_1^0}$, and the $s$-channel Higgs and $Z$
  boson exchange is nearly resonant, allowing the
  neutralinos to annihilate efficiently \cite{Djouadi:2005dz}.
  Let us note that for the ordinary mSUGRA case, this region is already excluded
  by LEP measurements.  
\item For negative enough $\epsilon_1$ values, the uplift of the Higgs mass
  generates a splitting among the `$h$-pole' and the `$Z$-pole' regions, with the 
  former now evading LEP constraints.
\end{enumerate}

The latter point concerning the Higgs boson mass is the most significant
effect of the BMSSM operators. Within the MSSM with mSUGRA-like
correlations, the bound on the Higgs mass strongly constrains
$m_{1/2}$. In contrast, in the presence of $\epsilon_1={\cal O}(-0.1)$,
the full region for which the correct value of the
relic abundance is obtained is allowed.
Let us emphasize that in the plain MSSM scenario at stake the bulk
region is already ruled out because it gives rise to a too light Higgs boson,
in contradiction to LEP2 data \cite{Amsler:2008zzb}.
However, the introduction of the NR operators in equation \eqref{eq:wdst}
uplifts the latter mass up to $m_h\gtrsim 130$ GeV or $m_h\gtrsim 155$ GeV
for $\tan\beta=10$ or $3$ respectively.

In the $m_0$ region that we are considering here, the impact of
the BMSSM operators on the mass of the neutralino LSP is rather limited.
The reason
is that in the bulk region the LSP is mostly bino-like, while the
BMSSM operators affect the higgsino parameters.

Concerning precision electroweak data and low energy processes, it is
important to realize that the new physics that generates the
non-renormalizable operators can directly modify the constraints that
come from these measurements. Ignoring this point, it is still
possible to identify regions in the parameter space favored by the
WMAP data which satisfy all such low energy constraints. The relevance
of the BMSSM lies in the fact that constraints involving the Higgs are
decoupled from constraints involving the stop sector.

\subsection{Light stops, heavy sleptons}
\label{ssec:lshs}
In order to continue with the analysis of scenarios with light unmixed stops,
we focus on a set of low energy parameters very different from the
previous subsection. Explicitly, in addition to the BMSSM $\epsilon_i$
parameters, we consider the following set of parameters:
\begin{equation}
M_1,\ \mu,\ \tan\beta,\ X_t,\ m_U,\ m_Q, \ m_{\tilde f},\ m_A\,,
\end{equation}
where
$m_{\tilde f}$ is a common mass for the sleptons, the first and second
generation squarks, and $\tilde b_R$. We further use
$M_1=\frac53\,\tan^2\theta_W\,M_2\sim\frac12\,M_2$.
To demonstrate our main points, we fix the values of all but two
parameters as follows: $\epsilon_1=0$ or $-0.1$, $\epsilon_2=0$ or
$+0.05$, $\tan\beta=3$ or $10$, $X_t=0$, $m_U=210$ GeV, $m_Q=400$ GeV,
$m_{\tilde f}=m_A=500$ GeV. This scenario gives rise to relatively
light stops:
\begin{equation}
m_{\tilde t_1}\lesssim 150\ {\rm GeV},\ \ \ \ 370\
{\rm GeV}\lesssim m_{\tilde t_2}\lesssim 400\ {\rm GeV}.
\end{equation}
We scan over the remaining two parameters, $M_1$ and $\mu$.

Again, as pointed out in reference \cite{Bernal:2009hd}, in the
prescribed framework one can identify four regions in which
the WMAP constraint is fulfilled:
\begin{enumerate}
\item The `$Z$-pole' region in which the LSP is very light,
  $m_{\chi_1^0}\sim\frac12 M_Z\sim 45$ GeV, and the $s$-channel $Z$
  exchange is nearly resonant. This region is not ruled out only in
  scenarios where the mass splitting between $M_1$ and $M_2$ at the
  electroweak scale is very large.
\item The `$h$-pole' region in which the LSP is rather light,
  $m_{\chi_1^0}\sim\frac12 M_h$, and the $s$-channel $h$ exchange is
  nearly resonant, allowing the neutralinos to annihilate efficiently
  \cite{Djouadi:2005dz}.
\item The `mixed region' in which the LSP is a higgsino--bino
  mixture \cite{ArkaniHamed:2006mb}, $M_1 \sim\mu$, which enhances
  (but not too much) its annihilation cross-sections into final states
  containing gauge and/or Higgs bosons: $\chi_1^0 \chi_1^0 \to W^+
  W^-$, $ZZ$, $Zh$ and $hh$.
\item The `stop co-annihilation' region, in which the LSP is
  almost degenerate in mass with the lightest stop ($\tilde t_1$).
  Such a scenario leads to an enhanced annihilation of sparticles
  since the $\chi_1^0-\tilde t_1$ co-annihilation cross-section
  \cite{Griest:1990kh,Boehm:1999bj,Ellis:2001nx} is much larger than that of the
  LSP.
\end{enumerate}

Let us first consider the case where $\epsilon_1=\epsilon_2=0$.
The region at $M_1\sim m_Z/2\sim m_h/2$ corresponds to the
$s$-channel exchange of an almost on-shell Higgs or $Z$ boson. Note
that when $2\,m_{\chi_1^0}$ is too close to the Higgs or $Z$ mass
pole, the LSP annihilation is too efficient and leads to a much too
small $\Omega_{\rm DM}\,h^2$. In any case, for the Higgs mass values
obtained here, $m_h \sim 85(98)$ GeV for $\tan\beta=3(10)$, this
region is already excluded by the negative searches for chargino pairs
at LEP2 \cite{Amsler:2008zzb}.

The region close to $M_2\sim\mu\sim 200$ GeV corresponds to the LSP
being a bino--higgsino mixture with sizeable couplings to $W$, $Z$ and
Higgs bosons, allowing for reasonably large rates for neutralino
annihilation into $\chi_1^0 \chi_1^0 \to W^+W^-$, $ZZ$, $hZ$ and $hh$
final states. Above and below the band, the LSP couplings to the
various final states are either too strong or too weak to generate the
relevant relic density.  

Finally, for larger $\mu$ values, the mass of the lightest neutralino approaches the mass of the
lightest stop leading to an enhanced co-annihilation cross-section:
$\chi_1^0\,\tilde t_1\to W^+\,b$, $g\,t$ ($\sim 90\%$). Also, to a
lesser extent ($\sim 5\%$), the annihilation cross-section of the stop
NLSP contributes to the total cross-section by the process $\tilde
t_1\,\tilde t_1\to g\,g$.

Next we consider the $\epsilon_1=-0.1$ case.
The features of the DM allowed regions are similar to the previous case.  
The main difference comes from the important enhancement of the Higgs mass
due to the presence of the BMSSM operators. In this case it is possible to
disentangle the $Z$ and the $h$ peaks, since the Higgs-related peak moves to
higher $M_2$ values, due to the increase of the Higgs mass: $m_h=122(150)$
GeV for $\tan\beta=10(3)$. Furthermore, the latter peak is no longer excluded
by chargino searches.

\section{`Detectability' definition}\label{detectability}
Let us now define what we shall be meaning by saying that a certain parameter space point
is detectable.
We employ a method based on the $\chi^2$ quantity.
Consider whichever mode of detection: direct or indirect in any of the three channels
($\gamma$-rays, $e^+, \bar{p}$) we shall be considering. In all three modes, what 
is finally measured is a number of events per energy bin.
Let us call $N^{sig}_i$ the number of signal (dark matter - induced) events in
the $i$-th bin, the nature of which depends on the specific experiment, 
$N^{bkg}_i$ the corresponding
background events in the same bin, and $N^{tot}_i$ the sum of the two.
The variance $\chi^2_i$ in every bin is defined as:
\begin{equation}
\chi^2_i =
\frac{(N_i^{tot}-N_i^{bkg})^2}{N_i^{tot}} \ .
\end{equation}
Then, the condition that we impose to characterize a point as detectable, is that at least 
in one energy bin 
$\chi^2_i\gtrsim 5.8$. In Gaussian error terms, this corresponds to a $95\%$ CL.

\section{Direct detection}\label{dd}
\subsection{Differential event rate}
\label{Xenon}

In spite of the experimental challenges, a number of efforts worldwide
are actively pursuing to directly detect WIMPs with a variety of targets
and approaches (for previous works see, e.g. references \cite{Cerdeno:2001up,
Baek:2005wi, Ellis:2005mb}). Many direct dark matter detection experiments are
now either operating or in preparation.
All these experiments measure the number $N$ of elastic
collisions between WIMPs and target nuclei in a detector,
per unit detector mass and per unit of time, as a function of the
nuclear recoil energy $E_r$.
The detection rate in a detector depends on the density
$\rho_0\simeq0.385$ GeV cm$^{-3}$ \cite{Catena:2009mf} and velocity distribution $f(v_\chi)$ of WIMPs near the Earth.
We assume a Maxwellian halo for WIMP's velocity in the rest frame of our galaxy
(for a recent treatment of non-Maxwellian case, see e.g. reference \cite{Kuhlen:2009vh}),
taking into account the orbital motion of the solar system around the galaxy,
and neglecting the motion of the Earth around the Sun \cite{Jungman:1995df}:
\begin{equation}
f(v_\chi)=\frac{1}{\sqrt{\pi}}\frac{v_\chi}{1.05\,v_0^2}\left[e^{-(v_\chi-1.05\,v_0)^2/v_0^2}-e^{-(v_\chi+1.05\,v_0)^2/v_0^2}\right],
\end{equation}
where $v_0\simeq 220$ km/s is the orbital speed of the Sun around the galactic center.
In general, the differential event rate per unit detector mass and
per unit of time can be written as:
\begin{equation}
\frac{dN}{dE_r}=\frac{\sigma_0\,\rho_0}{2\,m_r^2\,m_\chi}\,
F(E_r)^2\int_{v_\text{min}(E_r)}^{\infty}\frac{f(v_\chi)}{v_\chi}dv_\chi\,,
\label{Recoil}
\end{equation}
where $\sigma_0$ is related to the
WIMP-nucleon cross-section, $\sigma_{\chi-p}$, by $\sigma_0=\sigma_{\chi-p}\cdot (A\,m_r/M_r)^2$, with $M_r=\frac{m_\chi\,m_p}{m_\chi+m_p}$ the WIMP-nucleon reduced mass, $m_r=\frac{m_\chi\,m_N}{m_\chi+m_N}$ the
WIMP-nucleus reduced mass, $m_\chi$ the WIMP mass, $m_N$ the nucleus mass,
and $A$ the atomic weight. $F$ is the nuclear form factor; in the following analysis the Woods-Saxon form factor \cite{Engel:1991wq} will be used
(a more complete discussion can be found in reference \cite{Ellis:2008hf}).
Let us note that we are assuming identical WIMP-proton and WIMP-neutron cross-sections, and that we are ignoring the spin-dependent interactions.
The integration over velocities is limited to those which
can give place to a recoil energy $E_r$, thus there is a minimal velocity
given by $v_\text{min}(E_r)=\sqrt{\frac{m_N\,E_r}{2\,m_r^2}}$.

A word of caution is also needed here to clarify the results we shall present. 
As it is obvious from equation \eqref{Recoil}, the sensitivity
of a direct detection experiment depends strongly on the local DM density. 
The value $\rho_0\simeq0.385$ GeV cm$^{-3}$ stated previously, refers to the 
\textit{overall} density. If DM consists of multiple components, the differential event rate
will depend on the \textit{partial} density of the $i$-th component (along with its 
velocity distribution).

We already saw that the WMAP constraints can be satisfied
in rather restricted regions of the parameter space. 
So, in most of the parameter space the model's relic
density is larger than the $\Lambda$CDM one as inferred from WMAP, whereas in some cases it can
also become smaller.
In the case $\Omega_{\rm \chi_1^0}\,h^2 > \Omega_{\rm DM}\,h^2$, we can speak of a 
`hard exclusion' by experimental data, since the predicted relic density cannot be larger
than the measured one. But in the case $\Omega_{\rm \chi_1^0}\,h^2 < \Omega_{\rm DM}\,h^2$,
things are more complicated. If neutralinos are not enough to explain the total DM relic density, 
nothing excludes them being only one of the components of the total DM density. 
In this case, the local density of neutralinos should be renormalised so as to account
for this feature. In this sense, when computing whether a parameter space point is 
detectable or not, we should use the correct local density value (one could assume, for example, that 
the local density fraction scales as the relic density one). 

In the following, we shall be ignoring this point. 
We shall be computing sensitivity lines
considering the local density as being constant over the parameter space. 
In this spirit, the sensitivity lines should be read with a little caution. 
They can be read safely with  respect to the regions where both WMAP bounds are fulfilled
(the upper and the lower), whereas the reader should keep the previous remarks in mind for the
bulk of the parameter space. When we say that a region of the parameter space is detectable, 
this corresponds literally to the WMAP fully compatible regions.

\subsection{A XENON-like experiment}
The XENON experiment aims
at the direct detection of dark matter via its elastic scattering off
xenon nuclei.
It allows the simultaneous measurement
of direct scintillation in the liquid and of ionization,
via proportional scintillation in the gas. In this way, XENON discriminates
signal from background for a nuclear recoil energy as small as $4.5$ keV.
Currently, the collaboration is working with a $170$ kg detector, but the final
project is a machine containing $1$ ton of xenon.

In our study, following reference \cite{Angle:2007uj}
we will always consider $7$ energy bins between $4$ and $30$ keV. 
We could take into
account non-zero background using simulations of the recoil spectra of
neutrons in our analysis, and this would significantly degrade the sensitivity of the detector. 
However, this would involve a much more detailed
study of the detector components (shielding, etc.), and we will not carry it out. In that sense,
our results will be the most optimistic ones.
Comprehensive studies about the influence of astrophysical and background assumptions 
can be found in references \cite{Bernal:2008zk,Green:2008rd}.
Furthermore, we examine three `benchmark' experimental setups, assuming exposures
$\varepsilon=30$, $300$ and $3000$ kg$\cdot$year, which could correspond e.g. 
to a detector with $1$ ton of xenon and $11$ days, $4$ months or $3$ years of data acquisition, respectively.

Let us note that other promising direct dark matter detection 
experiments such as SuperCDMS \cite{Bruch:2010eq} and LUX 
\cite{Fiorucci:2009ak}, should give rise to sensitivities comparable to 
XENON's one.

\subsection{Results}

\subsubsection{Correlated stop-slepton masses}
\begin{figure}[ht!]
\begin{center}
\vspace{-0.2cm}\hspace{-2.5cm}
\includegraphics[width=6.3cm,angle=-90]{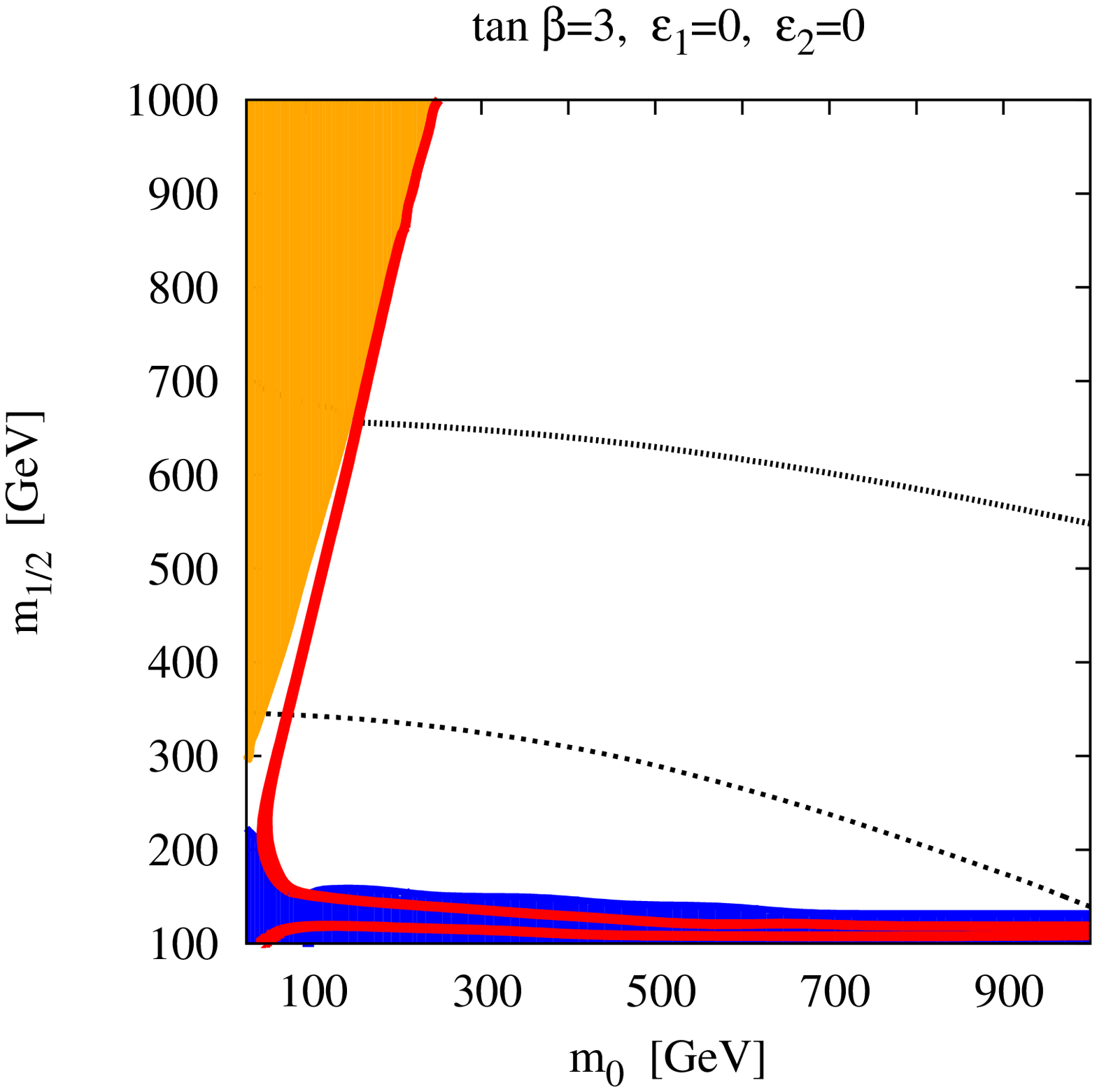}\hspace{-2.3cm}
\includegraphics[width=6.3cm,angle=-90]{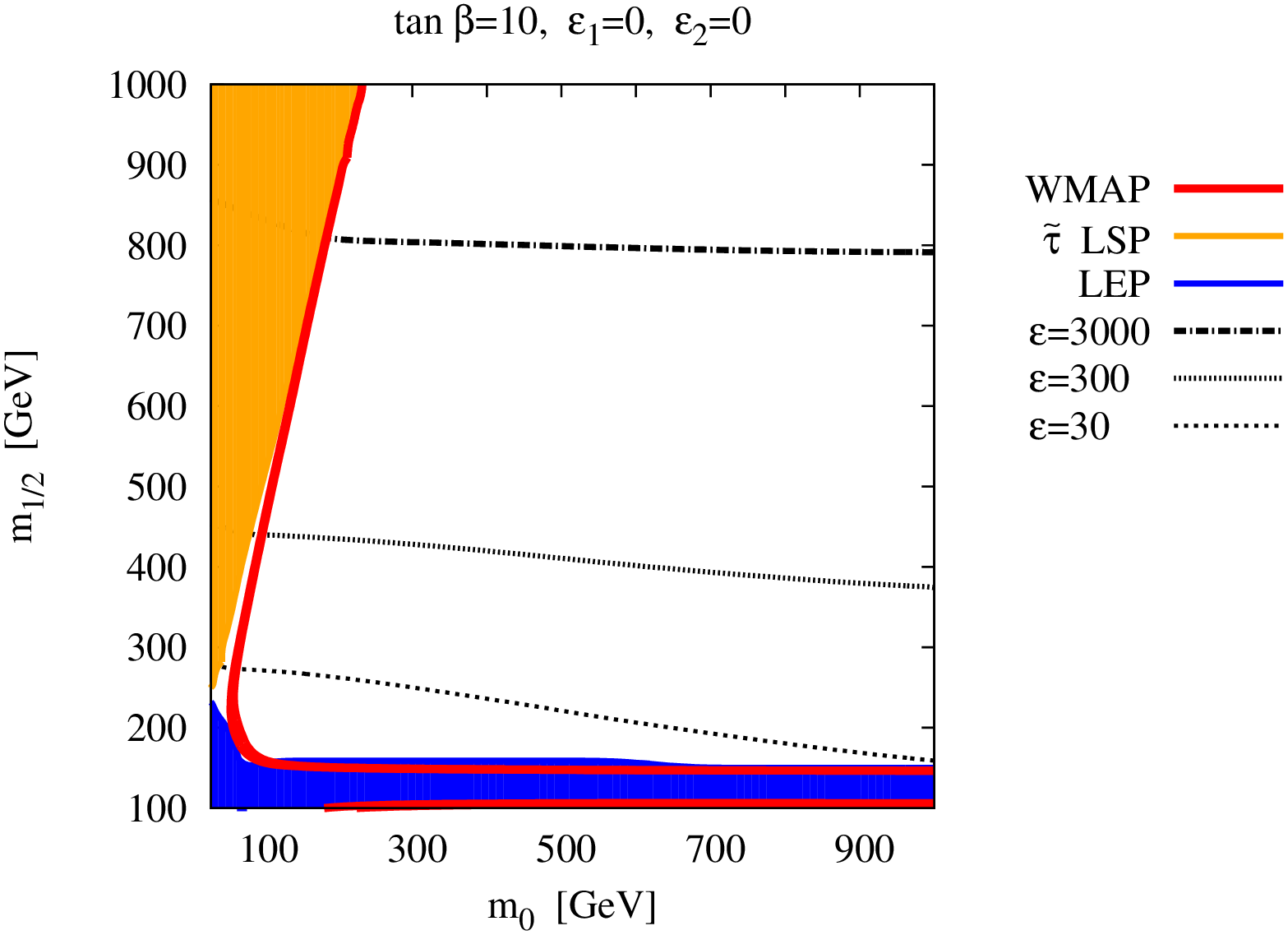}\\
\vspace{-0.4cm}\hspace{-2.5cm}
\includegraphics[width=6.3cm,angle=-90]{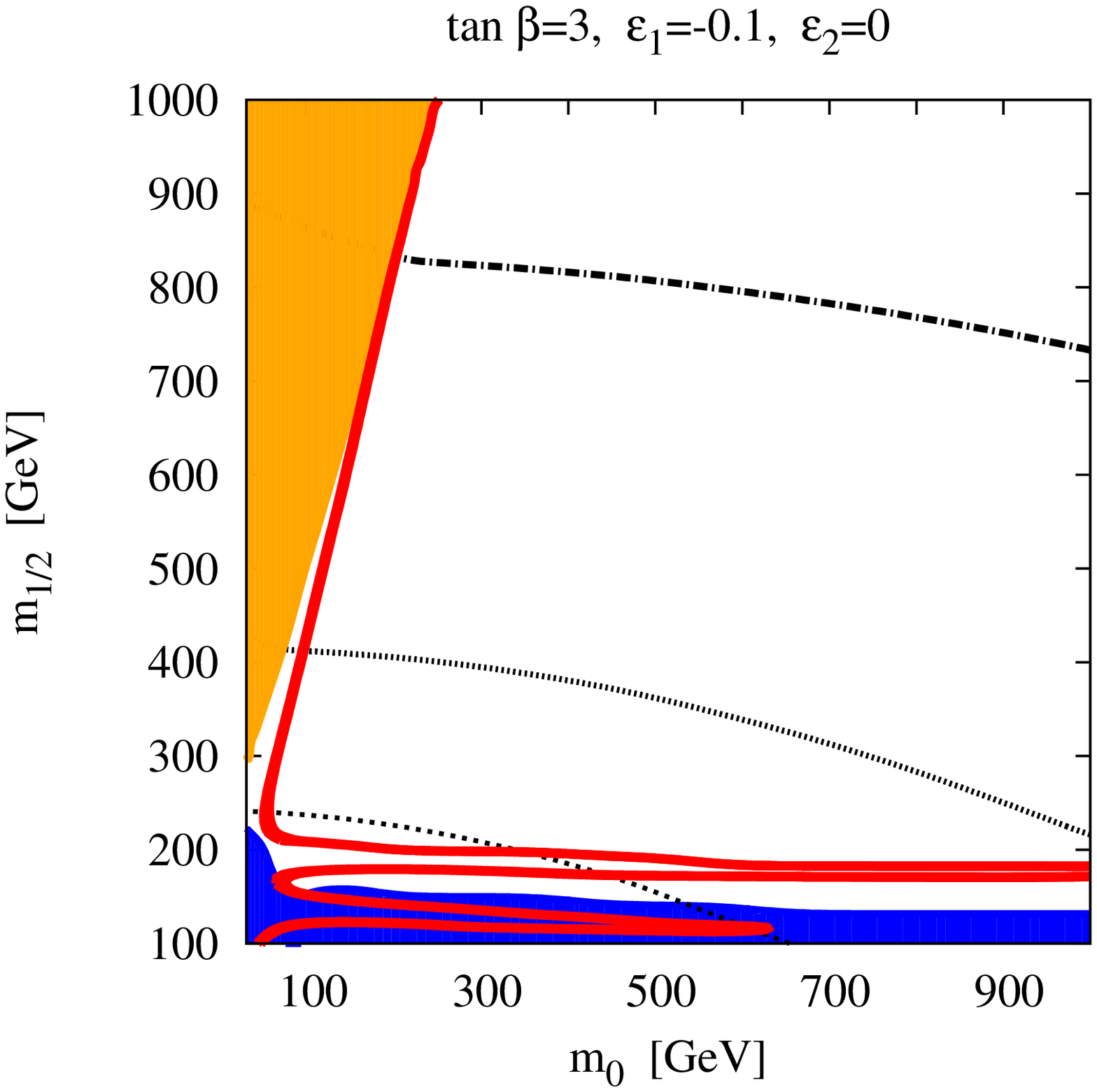}\hspace{-2.3cm}
\includegraphics[width=6.3cm,angle=-90]{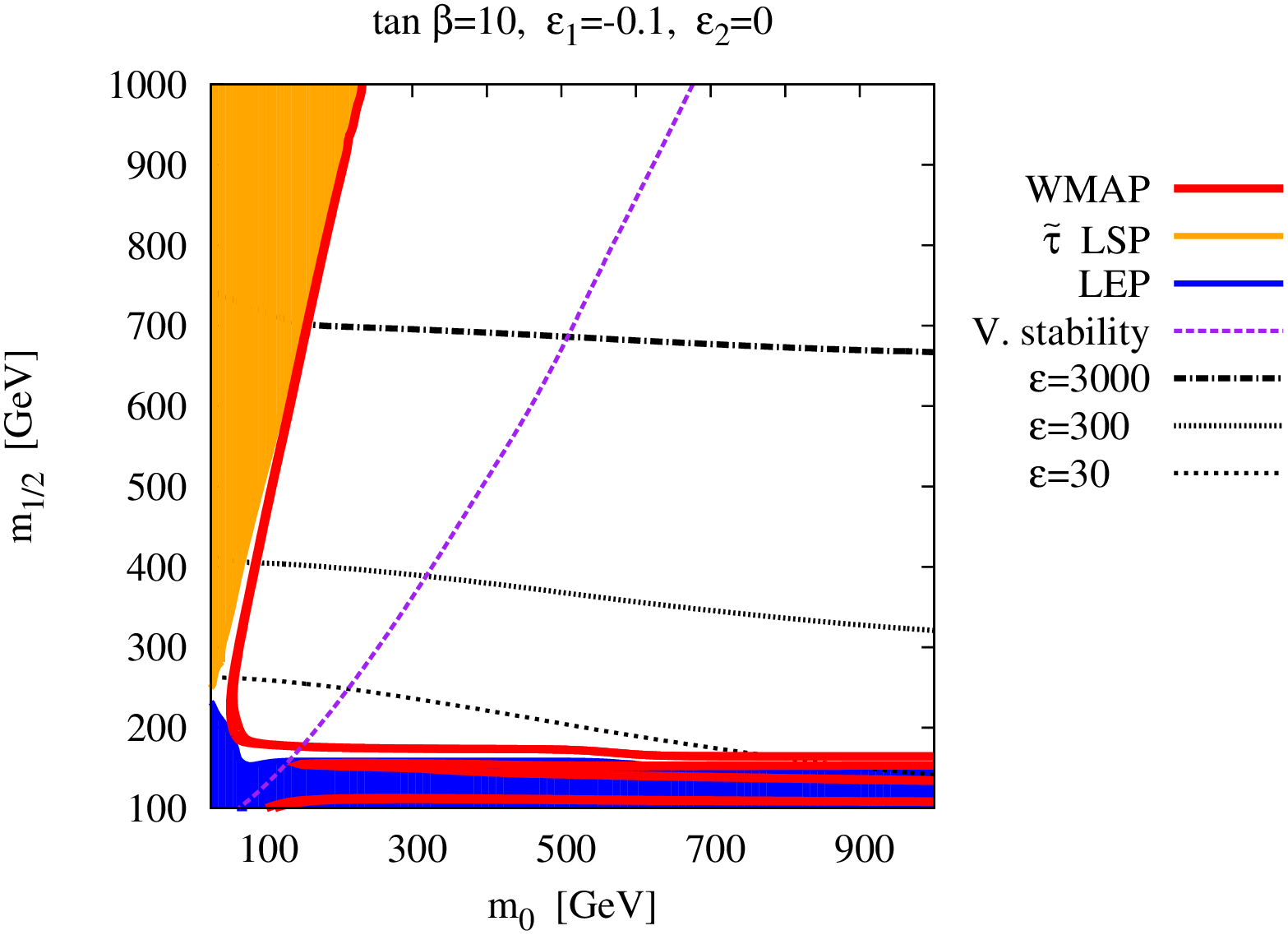}\\
\vspace{-0.4cm}\hspace{-2.5cm}
\includegraphics[width=6.3cm,angle=-90]{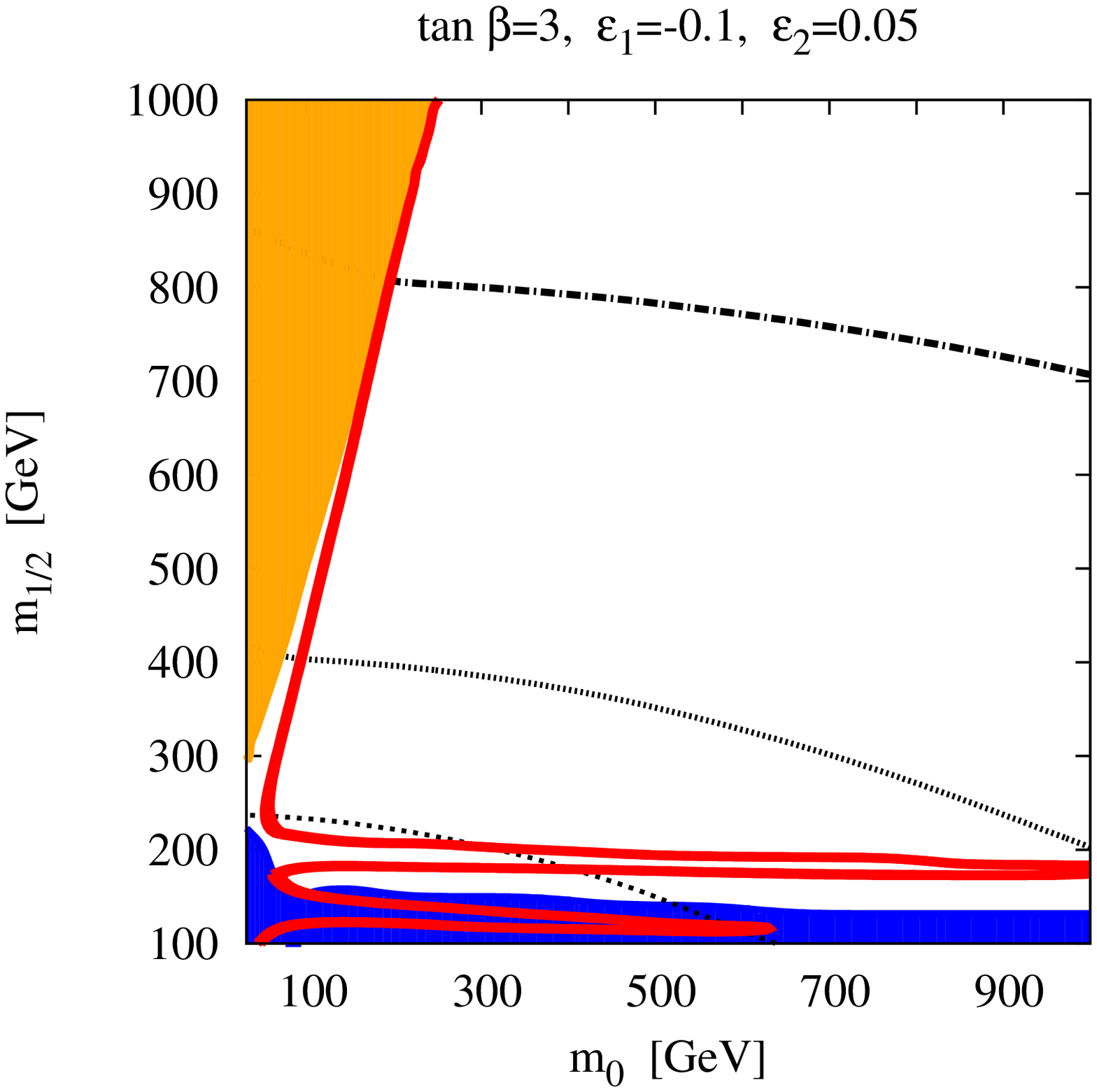}\hspace{-2.3cm}
\includegraphics[width=6.3cm,angle=-90]{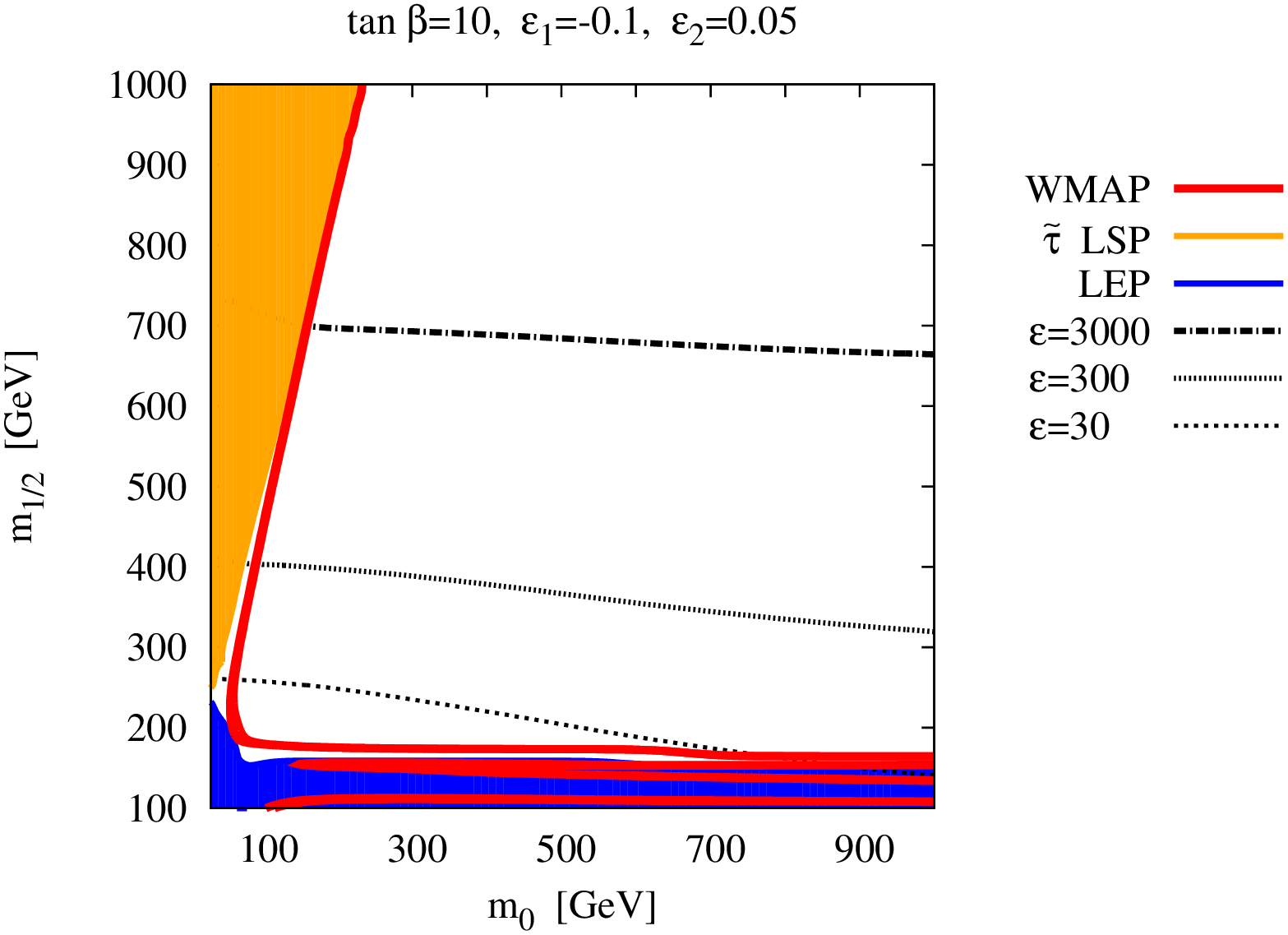}
\end{center}
\vspace{-0.9cm}
\caption{Regions in the $[m_0,\,m_{1/2}]$ plane that can be detected by XENON using exposures
$\varepsilon=30$, $300$ and $3000$ kg$\cdot$year, for our mSUGRA-like scenario. 
The black lines depict the detectability regions: the area below the lines can be probed.
Whenever a line is absent, this means that the whole parameter space can be tested
by the experiment. The blue and orange regions depict the areas that are excluded by
direct LEP chargino searches and the requirement for a 
neutralino LSP respectively. The area above the violet line is excluded by the
metastable vacuum constraint.}
\label{dir1}
\end{figure}

Figure \ref{dir1} shows the exclusion lines (black lines) for 
exposures $\varepsilon=30$, $300$ and $3000$ kg$\cdot$year, 
on the $[m_0,\,m_{1/2}]$ parameter space, for all other parameters as defined in section \ref{ssec:cssm}.
The first-row plots correspond to plain mSUGRA scenarios whereas the second and third to the
`mSUGRA-like' benchmark, with the $\epsilon_1$ and $(\epsilon_1, \epsilon_2)$ parameters turned on 
respectively.
The plots on the left correspond to a choice $\tan\beta = 3$ whereas the right-hand side ones
to $\tan\beta = 10$.
These curves reflect the XENON sensitivity and represent its ability to test and exclude different regions of the mSUGRA and BMSSM relevant model at $95\%$ CL: all points lying below the lines are detectable.
We note that when some line is absent, this means that the whole parameter space can be 
probed for the corresponding exposure.
Furthermore, the regions in orange (light gray) or blue (dark gray) are excluded
due to the fact that the LSP is the stau or because of the null searches for charginos and
sleptons at LEP.
For large $\tan\beta$, an important fraction of the $[m_0,\,m_{1/2}]$ plane,
corresponding to the region above the violet line, generates an unstable vacuum
and is then excluded.
Let us note that the introduction of $\epsilon_2$ alleviates the vacuum stability constraint \cite{Blum:2009na},
and slightly increases the Higgs mass.

As a general rule, the detection prospects are maximised for low values of the $m_0$
and $m_{1/2}$ parameters. For higher $m_0$ values, the masses of the
squarks in the internal propagators increase, penalising the scattering cross-section.
In the same way, the increase of $m_{1/2}$ augments the WIMP mass and
leads to a deterioration of the detection perspectives.
On the other hand, the region of low $m_{1/2}$ is also preferred because in that case the
lightest neutralino is a mixed bino-higgsino state, favouring the $\chi_1^0-\chi_1^0-h$
and $\chi_1^0-\chi_1^0-H$ couplings, and therefore the scattering cross-section. Let us recall
that a pure higgsino or a pure gaugino state does not couple to the Higgs bosons.
On the other hand, the detection prospects are also maximised for low values of $\tan\beta$.
For large values, besides the increase of the lightest Higgs boson mass, the coupling
of the latter to a $\chi_1^0$ pair decreases significantly because it is proportional 
to $\sin 2\beta$, for $|\mu|\gg M_1$.

The introduction of the NR operators gives rise to an important deterioration of the detection prospects.
The main effect enters via the important increase in the lightest CP-even Higgs mass.
This behavior is attenuated for larger values of $\tan\beta$;
the corrections to the Higgs masses being suppressed by $1/\tan\beta$ (see e.g. reference \cite{Dine:2007xi}).
Moreover, as in the bulk region the lightest neutralino is mostly a bino-like state, the impact
on its couplings with Higgs bosons is marginal. 
Nevertheless, let us emphasize again that this deterioration is relative, since we are comparing
with a plain MSSM, which is already excluded because of the light Higgs mass.

Concerning the plots in figure \ref{dir1}, a further remark that can be made is that, 
even for low exposures, a sizable amount of the parameter space can be probed.
The experiment will be particularly sensible to low values of $m_{1/2}$.
However, larger exposures could be able to explore almost the whole parameter space taken
into account. Let us emphasize that, in general terms,  the best detection prospects correspond to low values for
$\tan\beta$ and for the couplings $\epsilon_i$.

\subsubsection{Light stops, heavy sleptons}
\begin{figure}[ht!]
\begin{center}
\vspace{-0.2cm}\hspace{-2.5cm}
\includegraphics[width=6.3cm,angle=-90]{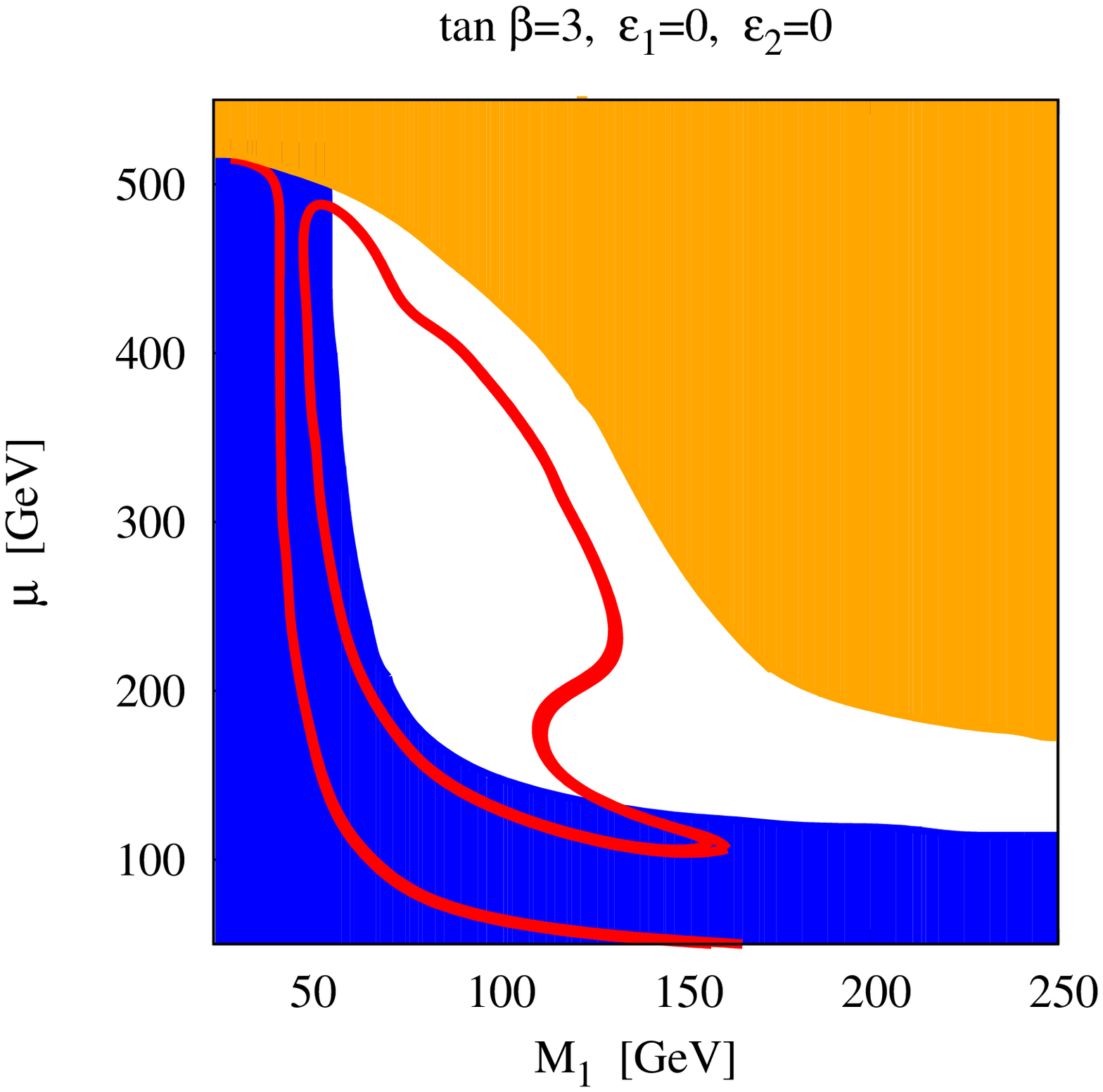}\hspace{-2.3cm}
\includegraphics[width=6.3cm,angle=-90]{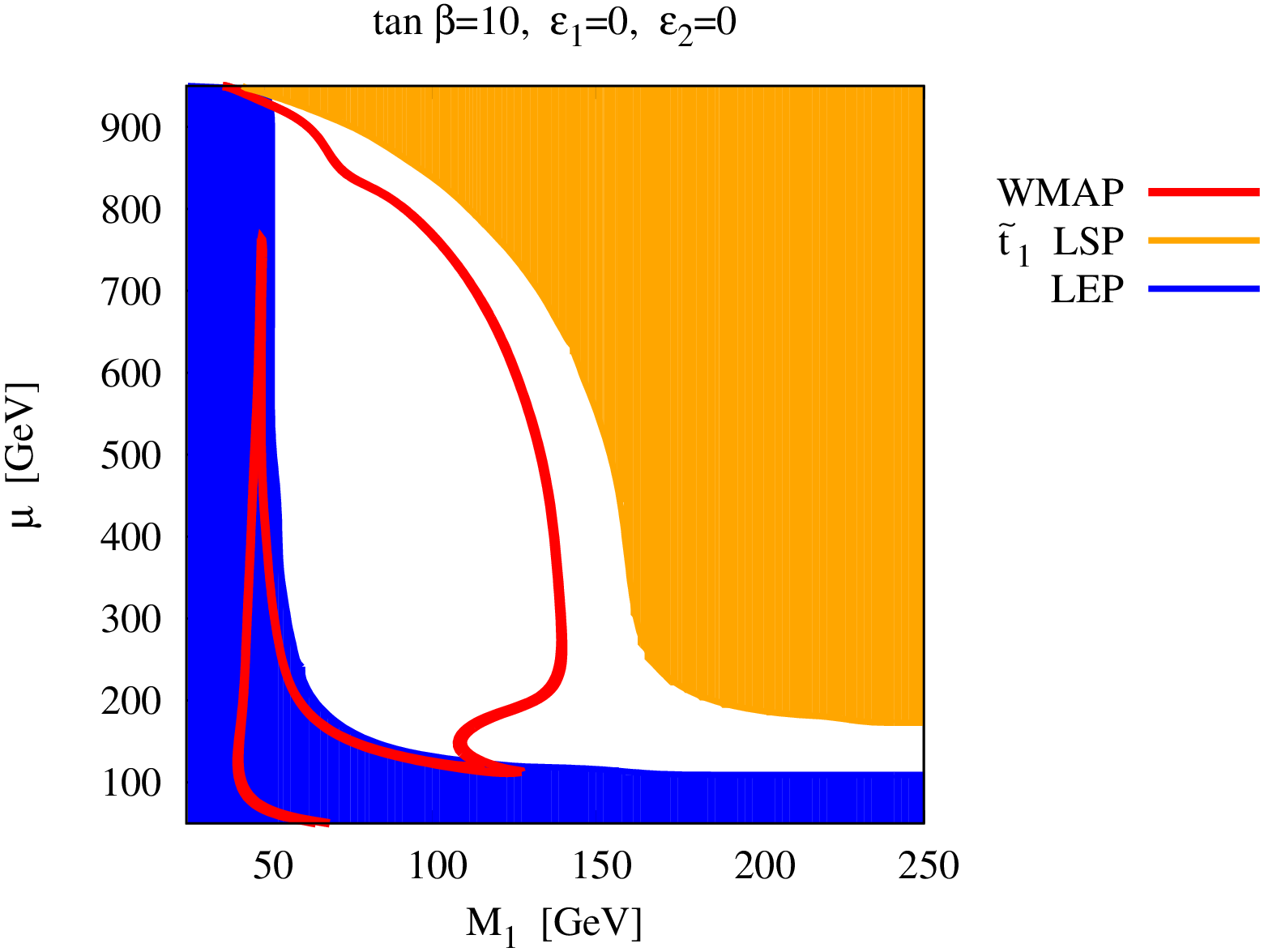}\\
\vspace{-0.4cm}\hspace{-2.5cm}
\includegraphics[width=6.3cm,angle=-90]{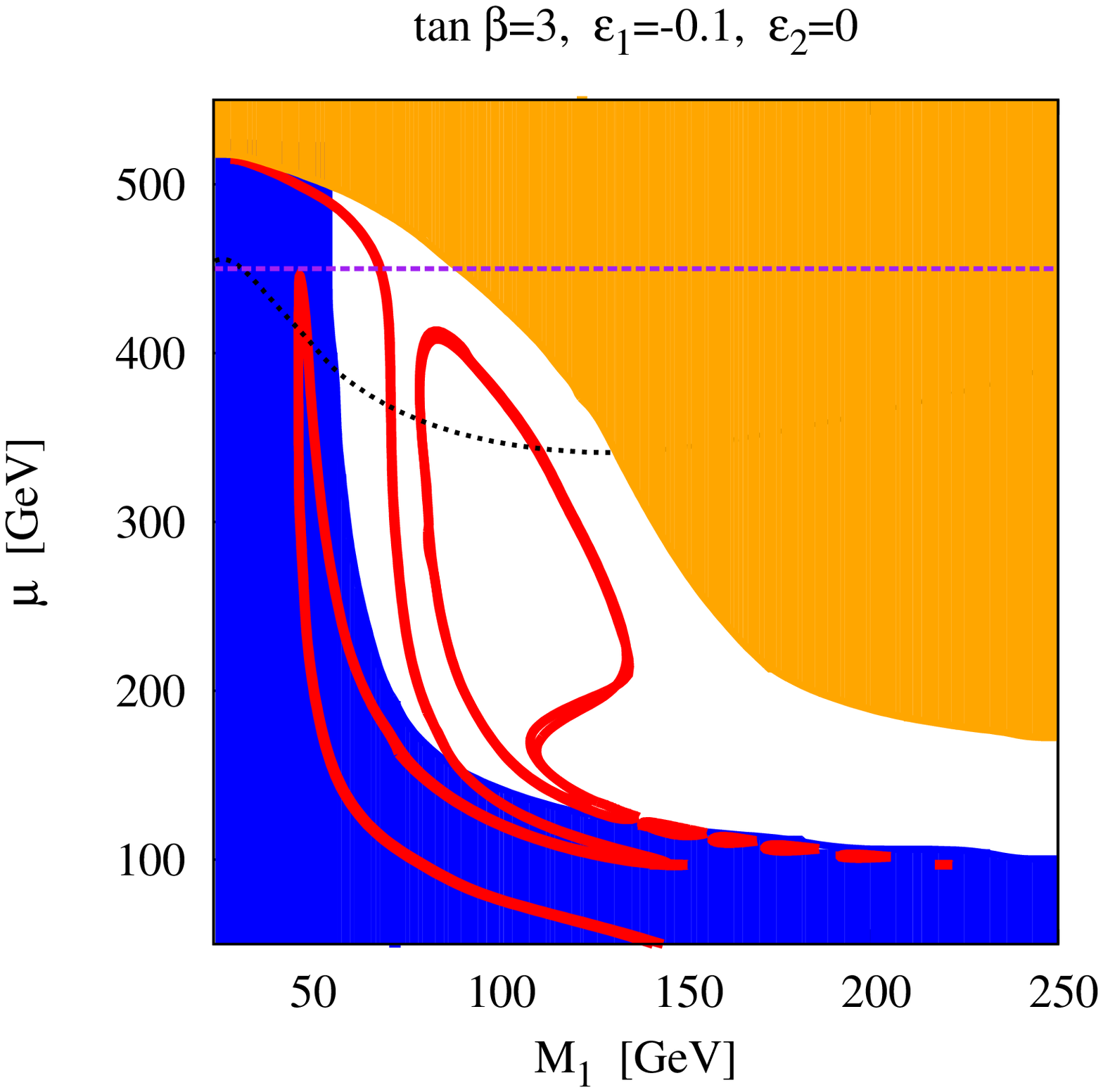}\hspace{-2.3cm}
\includegraphics[width=6.3cm,angle=-90]{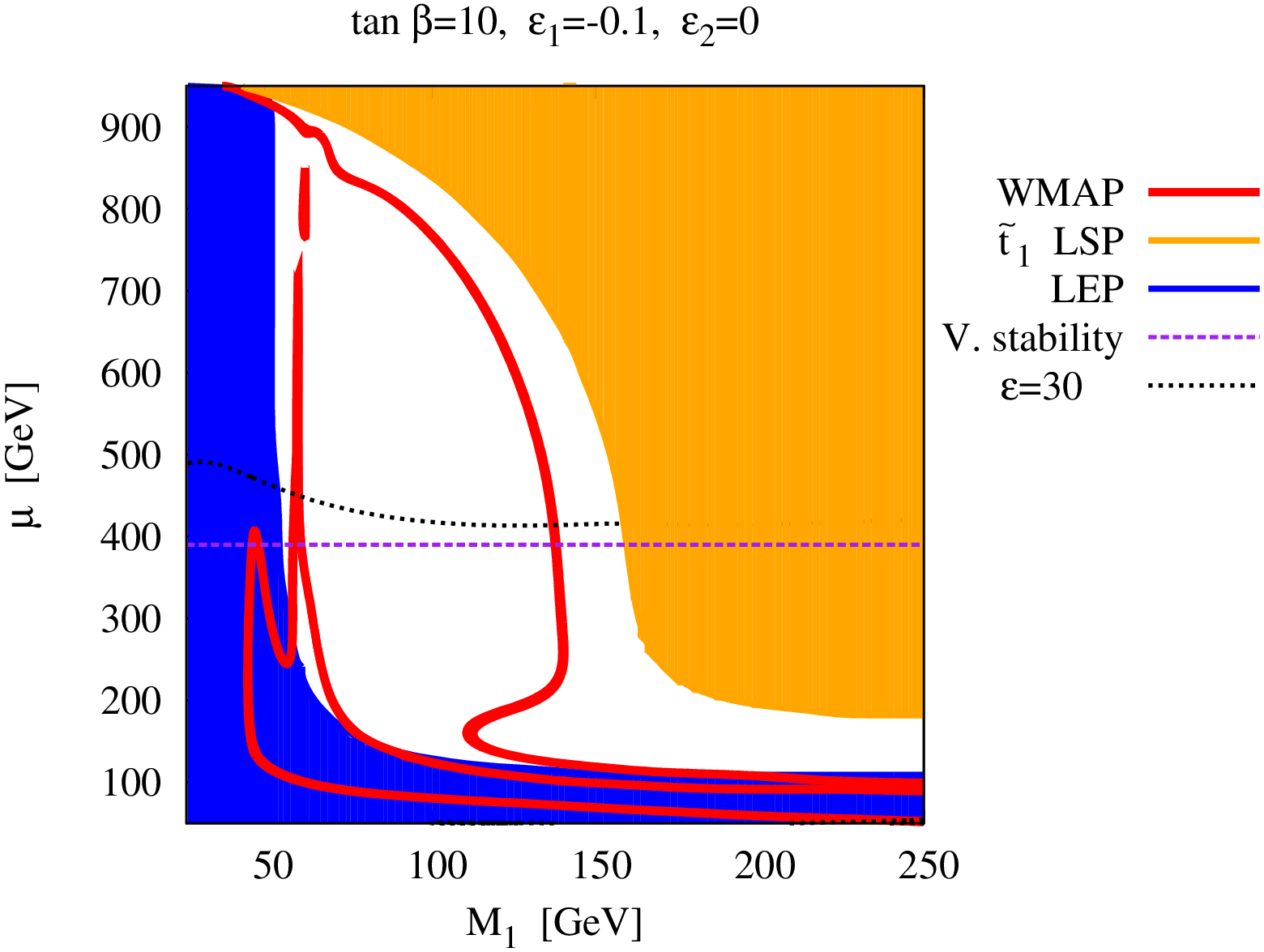}\\
\vspace{-0.4cm}\hspace{-2.5cm}
\includegraphics[width=6.3cm,angle=-90]{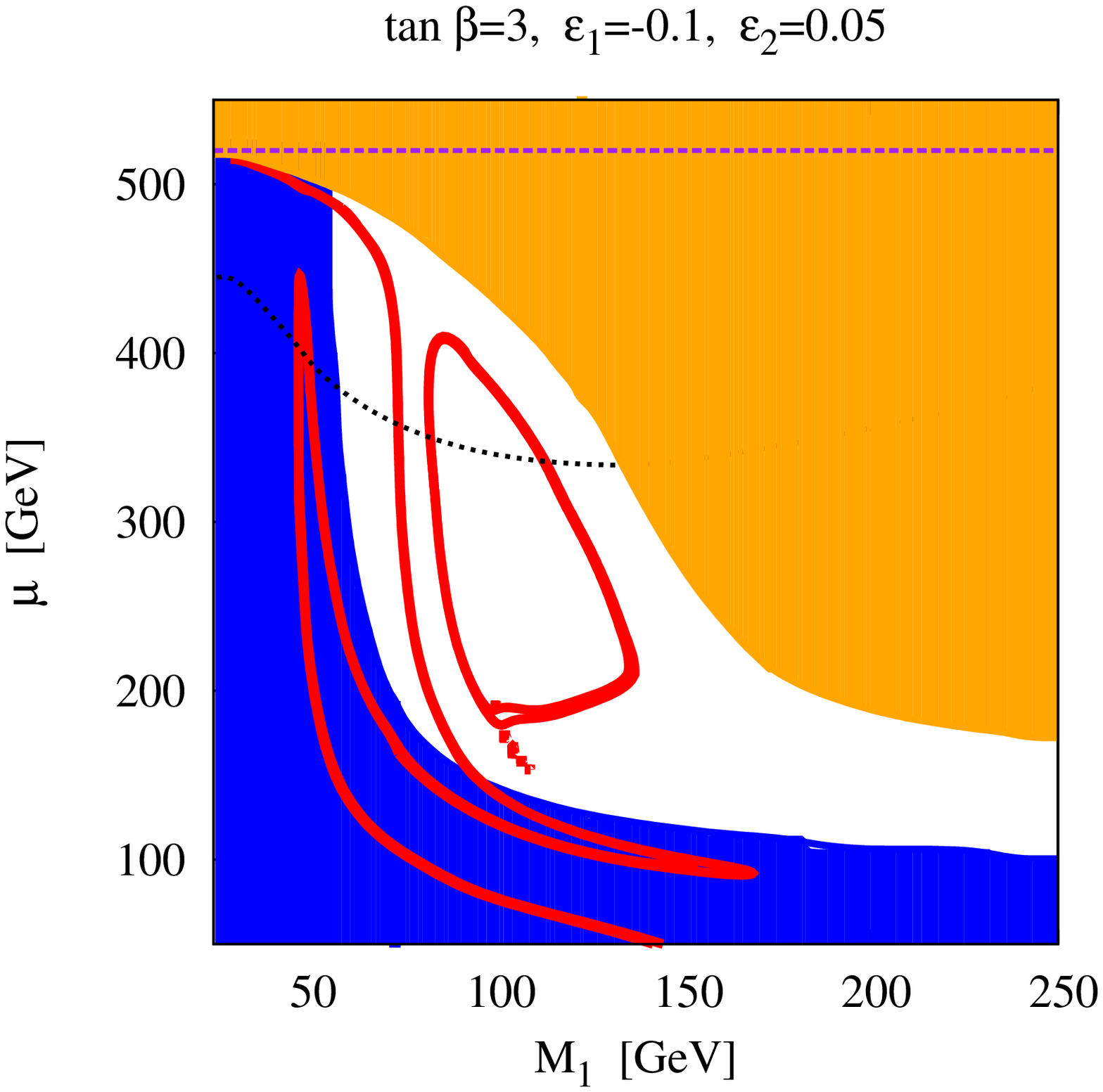}\hspace{-2.3cm}
\includegraphics[width=6.3cm,angle=-90]{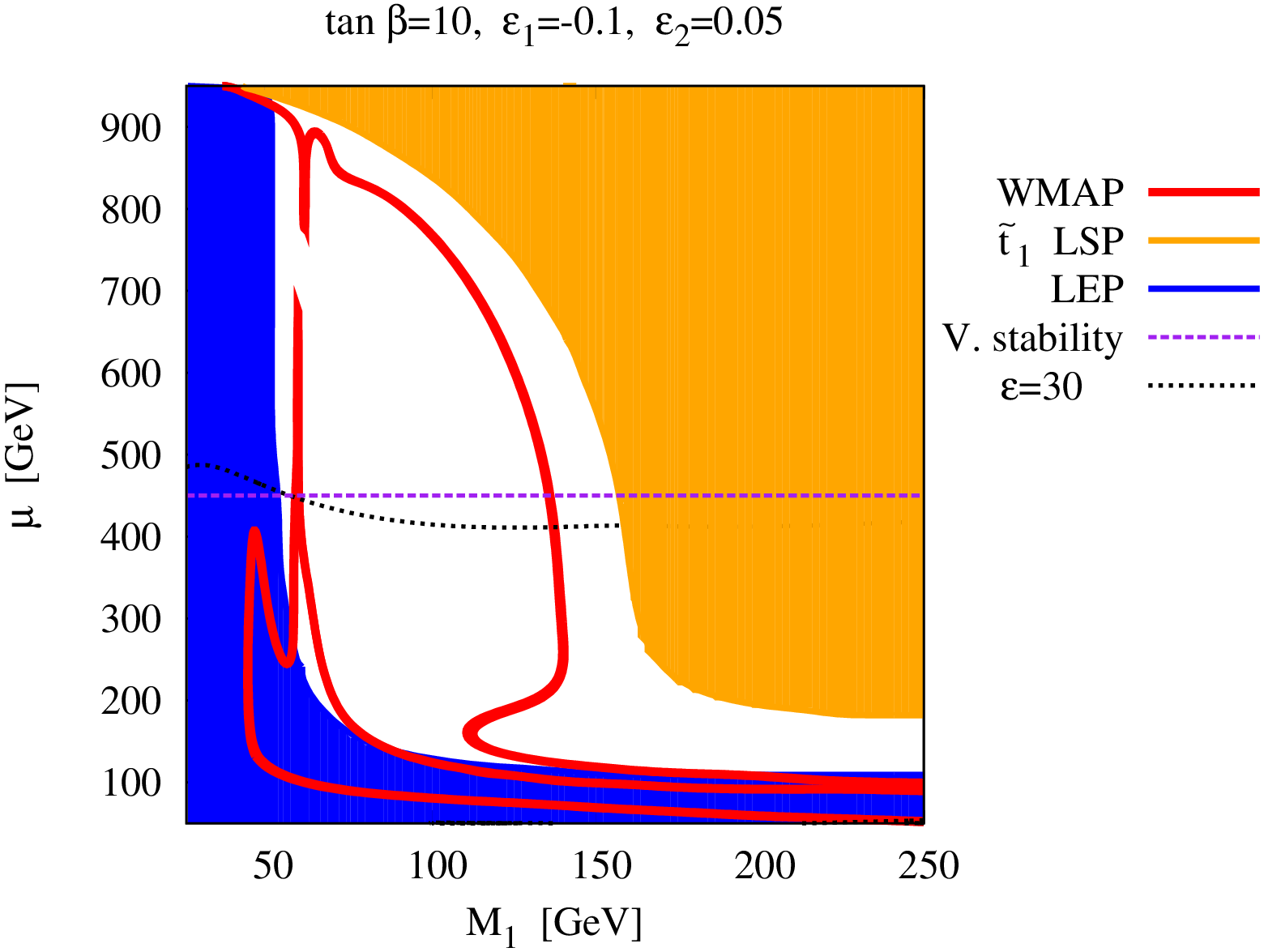}\\
\end{center}
\vspace{-0.9cm}
\caption{Regions in the $[M_1,\,\mu]$ plane that can be detected by
XENON for the scenario with light stops and heavy sleptons. 
The black lines depict the detectability regions for the corresponding XENON
detector with exposures
$\varepsilon=30$, $300$ and $3000$ kg$\cdot$year: the area below the lines can be probed.
Whenever a line is absent, this means that the whole parameter space can be tested
by the experiment. The blue and orange regions depict the areas that are excluded by
direct LEP chargino searches and the requirement for a 
neutralino LSP respectively. The areas above the violet lines are excluded by the
metastable vacuum constraints.
}
\label{dir2}
\end{figure}

Figure \ref{dir2} shows the exclusion lines for XENON with exposures
$\varepsilon=30$, $300$ and $3000$ kg$\cdot$year,
on the $[M_1,\,\mu]$ parameter space, with the other parameters as defined in section \ref{ssec:lshs} for
$\tan\beta=3$ (left panel) and $10$ (right panel).
Here again, the experiment will be sensitive to the regions below the contours.
It can be seen that, in general, the detection prospects are maximised for low values for the $M_1$
and/or the $\mu$ parameters, corresponding to a light $\chi_1^0$.
On the other hand, the scattering cross-section is enhanced near the region
$M_1\sim\mu$, where the lightest neutralino is a mixed bino-higgsino state,
favouring the $\chi_1^0-\chi_1^0-h$ and $\chi_1^0-\chi_1^0-H$ couplings.
Again, the detection prospects are also improved for low values of $\tan\beta$,
mostly because the coupling between the LSP and the Higgs bosons is suppressed by
the factor $\sin 2\beta$, for $|\mu|\gg M_1$.
Let us note that the first line of the figure (corresponding to the case without
the NR operators), besides being excluded by the Higgs mass, is partially ruled out
by XENON10 \cite{Angle:2007uj} and CDMS \cite{Ahmed:2009zw} searches.

When introducing the dimension $5$ operators the detection prospects deteriorate, in a
similar way as in the last subsection, because of the rise of the lightest Higgs mass.
Furthermore, because of a suppression of the $\chi_1^0-\chi_1^0-h$ coupling.
The latter effect is very accentuated in the region where the LSP is higgsino-like.
It is interesting however to notice that, for the case of large $\tan\beta$,
almost the whole area that could not be tested for $\varepsilon=30$ kg$\cdot$year
is already excluded by the vacuum stability constraint 
(i.e. the region above the violet line).
Moreover, in this case the whole parameter space evades the current constraints
from direct detection.

Let us emphasise that this scenario offers moreover exceptionally good detection perspectives.
Even with middle exposures, XENON will be able to detect dark matter in the whole region
for all three benchmarks and in the two models.

\section{Gamma-rays from the Galactic Center}
\subsection{Differential event rate}
The differential flux of gamma--rays generated from dark matter annihilations
and coming from a direction forming an angle $\psi$ with respect to
the galactic center (GC) is
\begin{equation}\label{Eq:flux}
\frac{d\Phi_{\gamma}}{dE_\gamma}(E_{\gamma}, \psi)
=\frac{\langle\sigma v\rangle}{8\,\pi\,m_{\chi}^2}\sum_i
\frac{dN_{\gamma}^i}{dE_{\gamma}}\,Br_i\,\int_\text{line of sight}\rho(r)^2\,dl
\end{equation}
where the discrete sum is over all dark matter annihilation
channels,
$dN_{\gamma}^i/dE_{\gamma}$ is the differential gamma--ray yield of SM particles into photons,
$\langle\sigma v\rangle$ is the total self--annihilation cross-section averaged
over its velocity distribution, $\rho$ is the dark matter density profile, $r$
is the distance from the GC
and $Br_i$ is the branching ratio of annihilation into the $i$-th final state.
The decay of SM particles into gammas has been calculated with
{\tt PYTHIA} \cite{Sjostrand:2006za}. The integration
is performed along the line of sight from the observation point towards the GC.

It is customary to rewrite equation~\eqref{Eq:flux} introducing the
dimensionless quantity $J$, which depends only on the dark
matter distribution (for an explicit calculation see, e.g.\cite{Bergstrom:1997fj,Yuksel:2007ac}):
\begin{equation}
J(\psi)=\frac{1}{R_0\,\rho_0^2}\,\int_{line\ of\ sight}\rho(r(l,\psi))^2\ dl\,,
\label{Jbarr}
\end{equation}
where $R_0\sim 8.5$ kpc is the distance from the Sun to the GC.
After having averaged over a solid angle, $\Delta \Omega$,
the gamma--ray flux can now be expressed as
\begin{eqnarray}\label{Eq:totflux}
\Phi_{\gamma}(E_{\gamma}) &=& 1.55\cdot 10^{-13}\,\mathrm{cm^{-2}\,s^{-1}\,GeV^{-1}\,sr^{-1}}\nonumber\\
&\cdot & \sum_i\frac{dN_{\gamma}^i}{dE_{\gamma}}\left(\frac{Br_i\,\langle\sigma v\rangle}{10^{-29} {\mathrm{cm^3 s^{-1}}}}\right)\left(\frac{100 ~\mathrm{GeV}}{m_{\chi}}\right)^2{\overline{J}}(\Delta \Omega) \Delta \Omega\,.
\end{eqnarray}
The value of $\overline{J}(\Delta \Omega) \Delta \Omega$ depends
crucially on the dark matter distribution.
There are various ways to parametrize the density profile 
\cite{Graham:2005xx,Graham:2006ae,Graham:2006af}. The most usual one is
\begin{equation}\label{profile}
\rho(r)=\frac{\rho_0\,[1+(R_0/a)^{\alpha}]^{(\beta-\gamma)/\alpha}}{(r/R_0)^{\gamma}\,[1+(r/a)^{\alpha}]^{(\beta-\gamma)/\alpha}}\,, 
\end{equation}
where 
$a$ is a characteristic length.

There has been quite some controversy on the values
for the $(\alpha, \beta, \gamma)$ parameters. Some N-body simulations
suggested highly cusped inner regions for the galactic halo \cite{Moore:1999nt}, whereas others
predicted more moderate $\gamma$ values (the basic parameter determining the inner slope
of the profile) \cite{Navarro:1995iw}.
The recent Via Lactea II simulation \cite{Diemand:2008in} seems to partly verify earlier results by
Navarro {\it et al.} and finds their results as being well reproduced by 
$(\alpha,\ \beta,\ \gamma)=(1,\ 3,\ 1)$.
On the other hand, the Aquarius project simulation results \cite{Springel:2008cc} seem to be 
better reproduced by a different parametrization, the so-called Einasto profile \cite{Einasto}:
\begin{equation}
 \rho(r) = \rho_s\ \exp\left[ -\frac{2}{\alpha} \left( \left( \frac{r}{r_s} \right)^\alpha - 1 \right) \right]\,,\qquad \alpha = 0.17\,,
\end{equation}
which does not demonstrate this effect of cuspyness in the inner galactic region. Here $r_s = 20$ kpc is 
a characteristic length, while $\rho_s$ is
a normalization factor, which we fix so as to reproduce the solar dark matter density. This choice 
yields $\rho_s \approx 0.0783$ GeV cm$^{-3}$.

In most numerical simulations, among which the Via Lactea II and Aquarius, it is a common simplification
that the effect of baryons is not taken into account. It has however been pointed out that 
in the presence of baryons there can be adiabatic collapse phenomena taking place near
the galactic center (see, e.g. references \cite{Prada:2004pi,Mambrini:2005vk}), something which could
severely influence the innermost regions of the DM halo, leading to profiles much more cusped
than the ones usually predicted, and hence enhancing the relevant fluxes by important factors.
We thus also consider a profile which, starting from the NFW one, tries to take into account
such effects, leading to an enhanced inner slope; this profile with adiabatic compression is denoted NFW$_c$.
The parameters relevant for the models under discussion
can be seen in table \ref{tab}.

It is worth noticing here that we are neglecting the effect of
clumpyness,
even though other studies showed that, depending
upon assumptions on the clumps' distribution, in principle an enhancement
by a factor $2$ to $10$ is possible \cite{Bergstrom:1998jj,Diemand:2008in}. In this respect, the
following predictions on the gamma-ray flux from the galactic center are
conservative. This shall also be the case in the following analysis regarding 
the detection capacity of the model in the positron and antiproton channels.
\begin{table}
\begin{center}
\begin{tabular}{|c|ccccc|}
\hline
 & $a$ [kpc] & $\alpha$ & $\beta$ & $\gamma$ & $\bar{J}(3\cdot10^{-5}$ sr$)$\\
\hline
Einasto &  -   &   -   &       &   -    & $6.07\cdot10^3$\\
NFW     & $20$ & $1.0$ & $3.0$ & $1.0$  & $8.29\cdot10^3$\\
NFW$_c$ & $20$ & $0.8$ & $2.7$ & $1.45$ & $5.73\cdot10^6$\\
\hline
\end{tabular}
\caption{{\footnotesize Einasto, NFW and NFW$_c$
density profiles with the corresponding parameters,
and values of $\bar{J}(\Delta\Omega)$.
The latter has been computed by means of a VEGAS Monte-Carlo integration
algorithm, imposing a constant density for $r\le 10^{-7}$kpc so as to 
avoid divergences appearing in the NFW-like profiles.}}
\label{tab}
\end{center}
\end{table}

As a final remark, let us repeat the word of caution already mentioned for direct detection.
In order to draw sensitivity lines, we shall be considering a one-particle DM with the
aforementioned density profiles irrespectively of the relic density inferred from the model.
Hence, once again, these lines should be read with respect to the WMAP-compatible regions.

\subsection{Modeling the galactic center background}
HESS \cite{Aharonian:2004wa} has measured the
gamma--ray spectrum of a very bright point-like source very close to the galactic
center in the range of energy $\sim$ [$160$ GeV--$10$ TeV]. The collaboration
claims that the data are fitted by a power--law
\begin{equation}
\phi^{\mathrm{HESS}}_{\mathrm{bkg}}(E) = F_0 ~ E_{\mathrm{TeV}}^{-\alpha},
\end{equation}
with a spectral index
$\alpha=2.21 \pm 0.09$ and
$F_0=(2.50 \pm 0.21) \cdot 10^{-8} ~\mathrm{m^{-2} ~ s^{-1} ~ TeV^{-1}}$.
The data were taken during the second phase of measurements
(July--August, $2003$) with a $\chi^2$ of $0.6$ per degree of
freedom. Because of the constant slope
power--law observed by HESS, it turns out possible but difficult to
conciliate such a spectrum with a signal from dark matter annihilation
\cite{Mambrini:2005vk, Profumo:2005xd}.
Indeed, final particles (quarks, leptons or gauge bosons)
produced through annihilations give rise to a spectrum with a continuously
changing slope. Several astrophysical models have been proposed in order
to match the HESS data \cite{Aharonian:2004jr}.
In the present study we consider the astrophysical background for
gamma--ray detection as the one extrapolated from the HESS data with
a continuous power--law over the energy range of interest
($\approx 1$ -- $300$ GeV).

EGRET \cite{Hunger:1997we} reported the presence of a bright gamma-ray source
at energies below $10$ GeV, which exceeds by far the HESS aforementioned extrapolation. 
However, this source seems not to be confirmed by the 
recent Fermi collaboration data \cite{Porter:2009sg}. We shall hence not take into account this point source.

Finally, we will consider the diffuse background of gamma rays in the region
surrounding the galactic center.
We will describe the spectrum of the
background using the HESS observation from the Galactic Center Ridge \cite{Aharonian:2006au},
which can be described by
\begin{equation}
\phi^{\mathrm{diff}}_{\mathrm{bkg}}(E) = 1.1\cdot 10^{-4}\,E_{\mathrm{GeV}}^{-2.29}\,\mathrm{GeV^{-1} cm^{-2} s^{-1} sr^{-1}}\ .
\end{equation}
In our analysis, we shall consider a solid angle of observation around the
galactic center ($\Delta\Omega=3\cdot 10^{-5}$ sr) and the energy region between
$1$ and $300$ GeV.

\subsection{The Fermi experiment}
The space--based gamma--ray telescope Fermi \cite{Gehrels:1999ri,Peirani:2004wy} was
launched in June $2008$ for a five-year mission. It
performs an all-sky survey covering a large energy range
($\approx 30$ MeV -- $300$ GeV). With an effective area and angular
resolution on the order of $10^4 ~ \mathrm{cm^2}$ and
$0.1^o$ ($\Delta \Omega \sim 3\cdot10^{-5}$ sr) respectively,
Fermi will be able to point and analyze the inner center of the Milky
Way ($\sim 7$ pc).
Concerning the requested condition on the $\chi^2$
for a signal discovery,
we have used an analysis similar to the one considered in the
case of direct detection as defined in
section \ref{detectability}.
Additionally, we take into account a five-year mission run, and an energy
range extending up to $300$ GeV, with $20$ logarithmically evenly spaced
bins.

\subsection{Results}
\subsubsection{Correlated stop-slepton masses}

\begin{figure}[ht!]
\begin{center}
\vspace{-0.2cm}\hspace{-2.5cm}
\includegraphics[width=6.3cm,angle=-90]{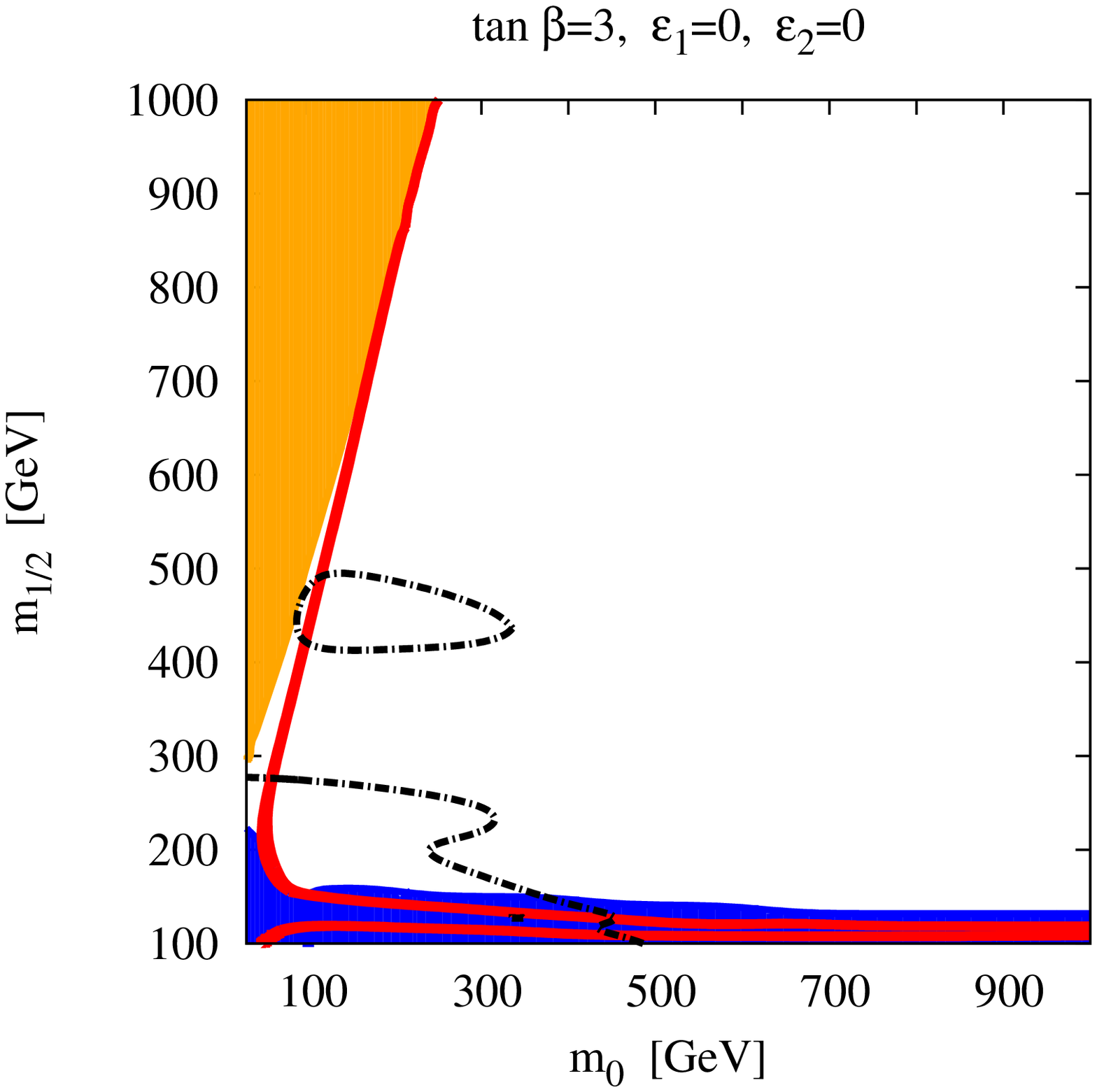}\hspace{-2.3cm}
\includegraphics[width=6.3cm,angle=-90]{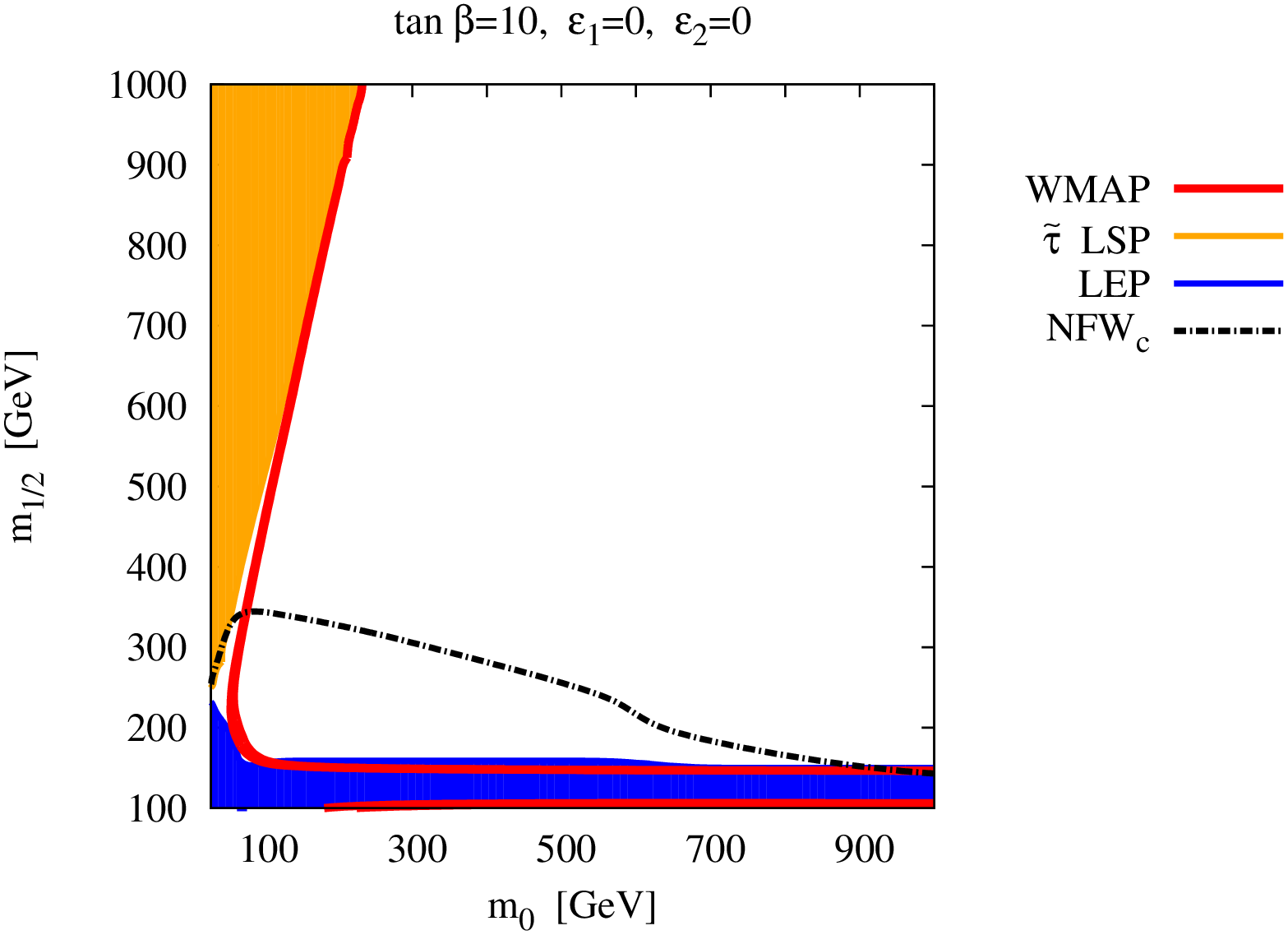}\\
\vspace{-0.4cm}\hspace{-2.5cm}
\includegraphics[width=6.3cm,angle=-90]{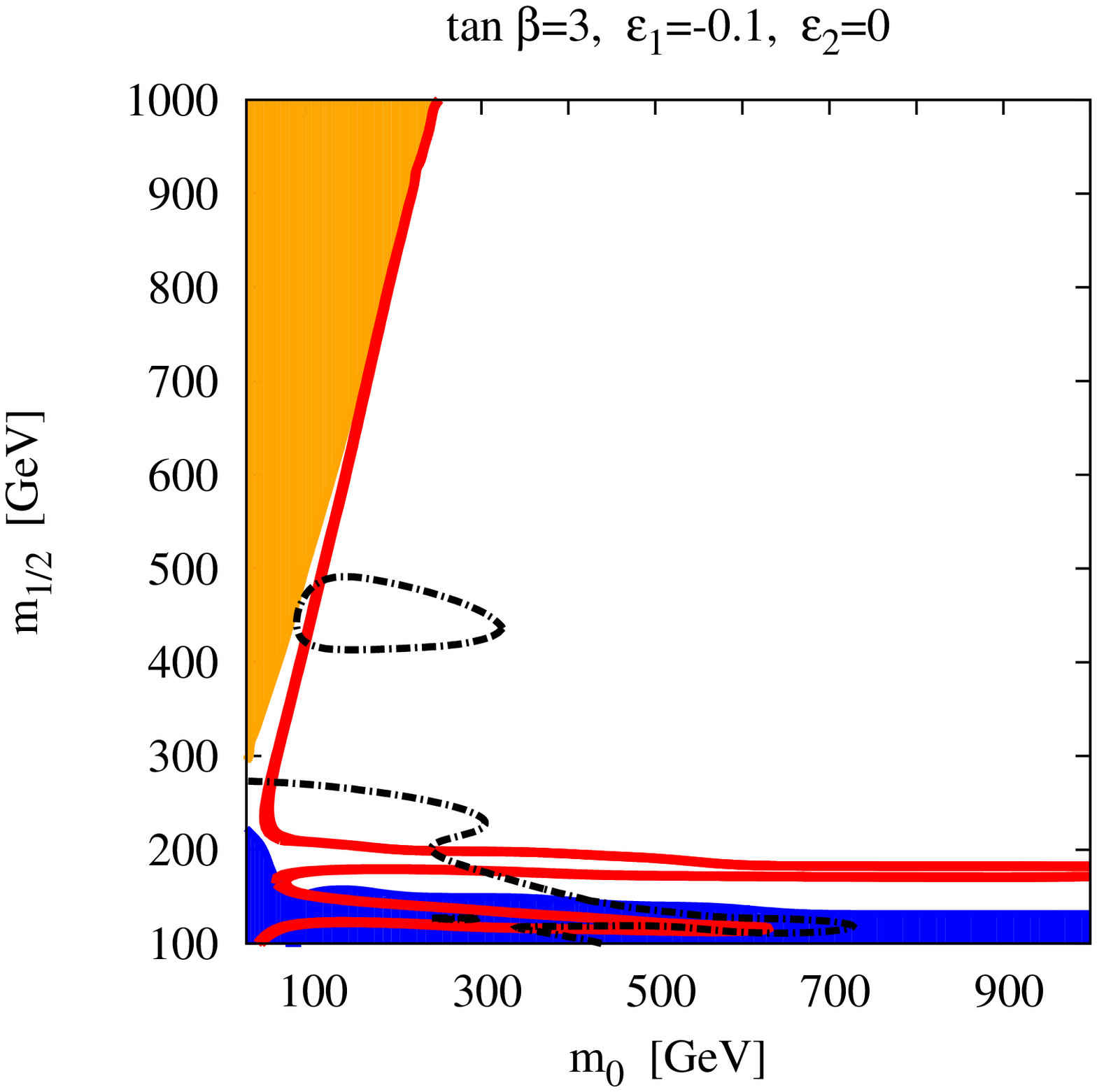}\hspace{-2.3cm}
\includegraphics[width=6.3cm,angle=-90]{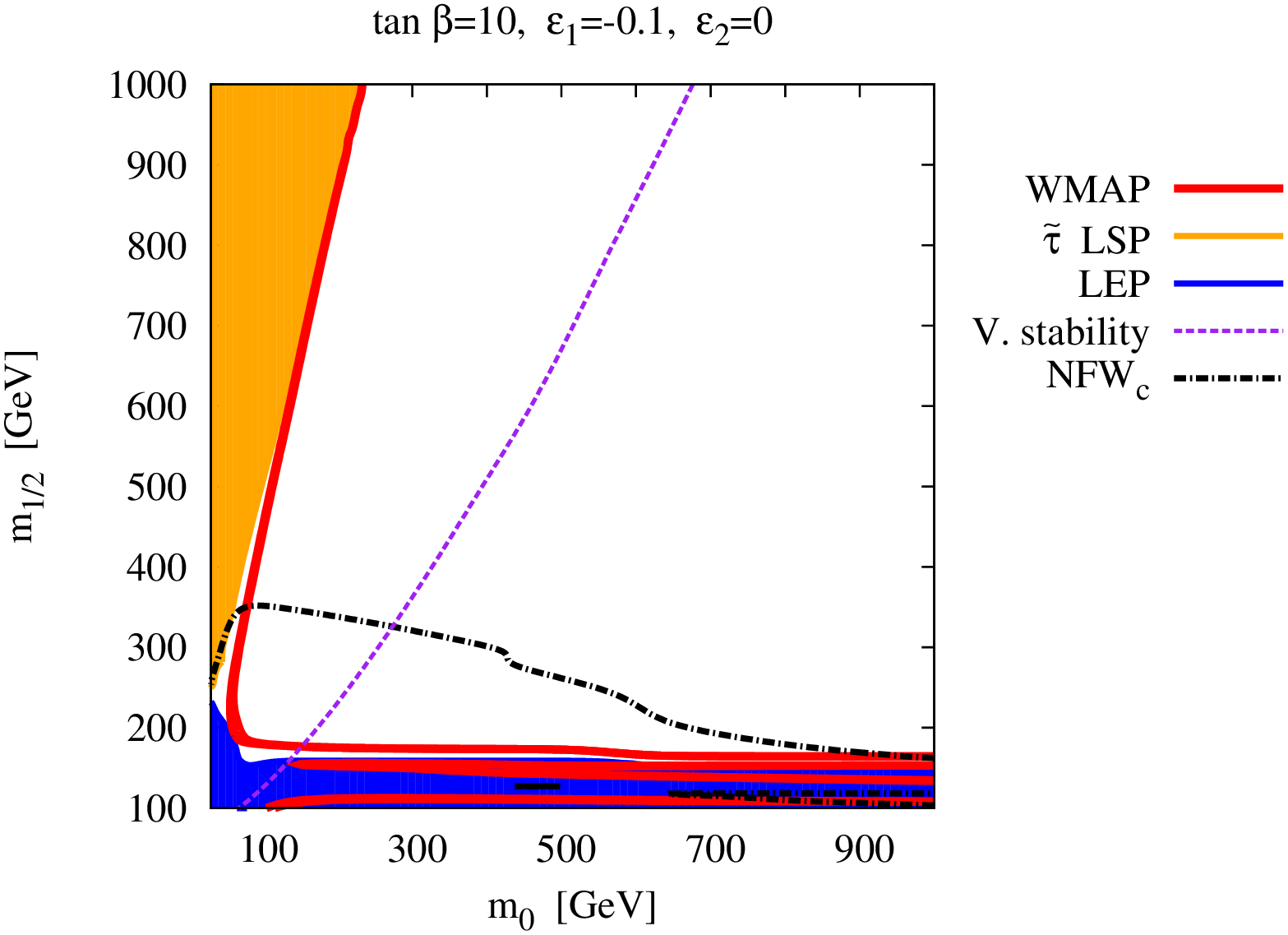}\\
\vspace{-0.4cm}\hspace{-2.5cm}
\includegraphics[width=6.3cm,angle=-90]{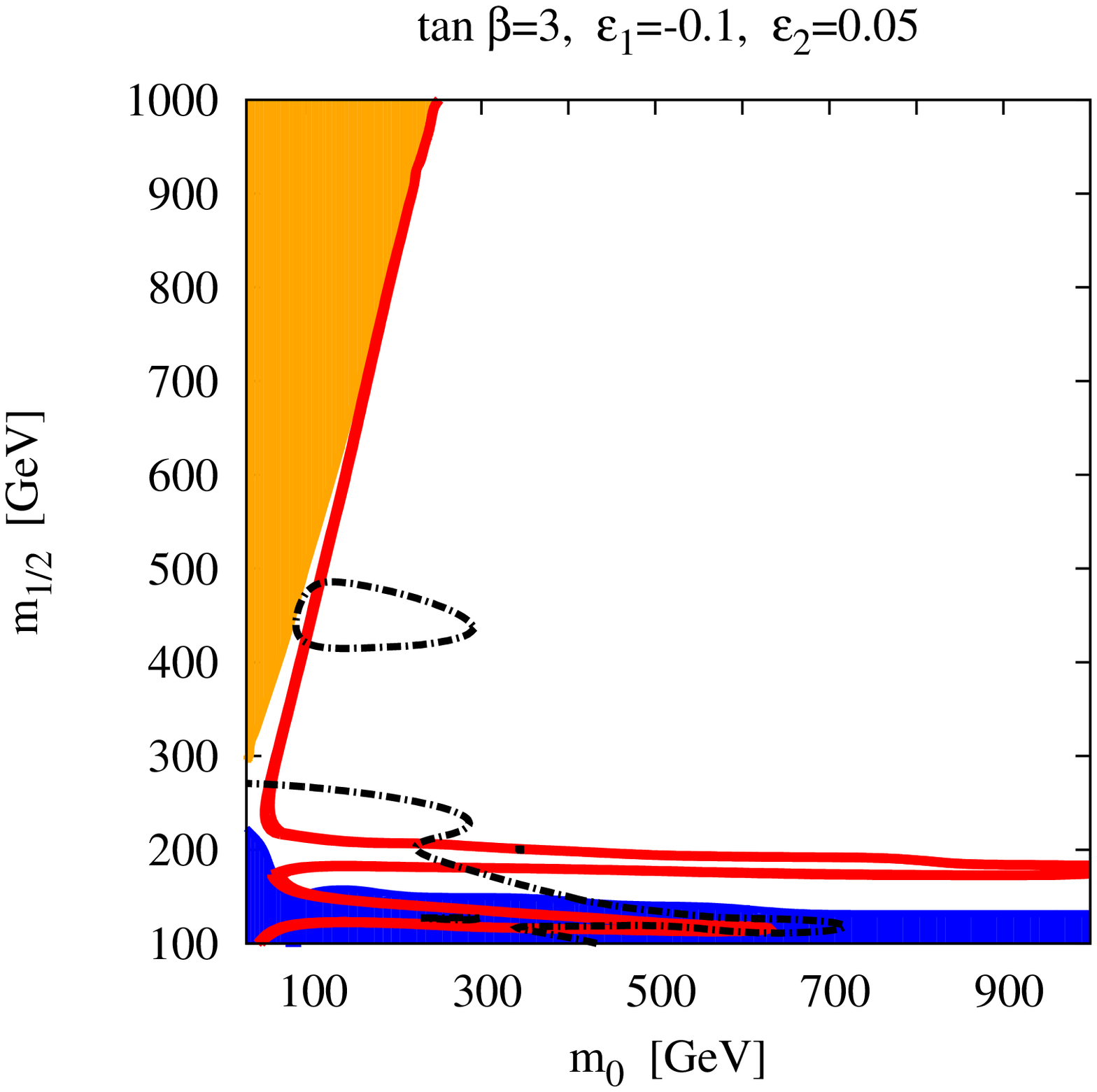}\hspace{-2.3cm}
\includegraphics[width=6.3cm,angle=-90]{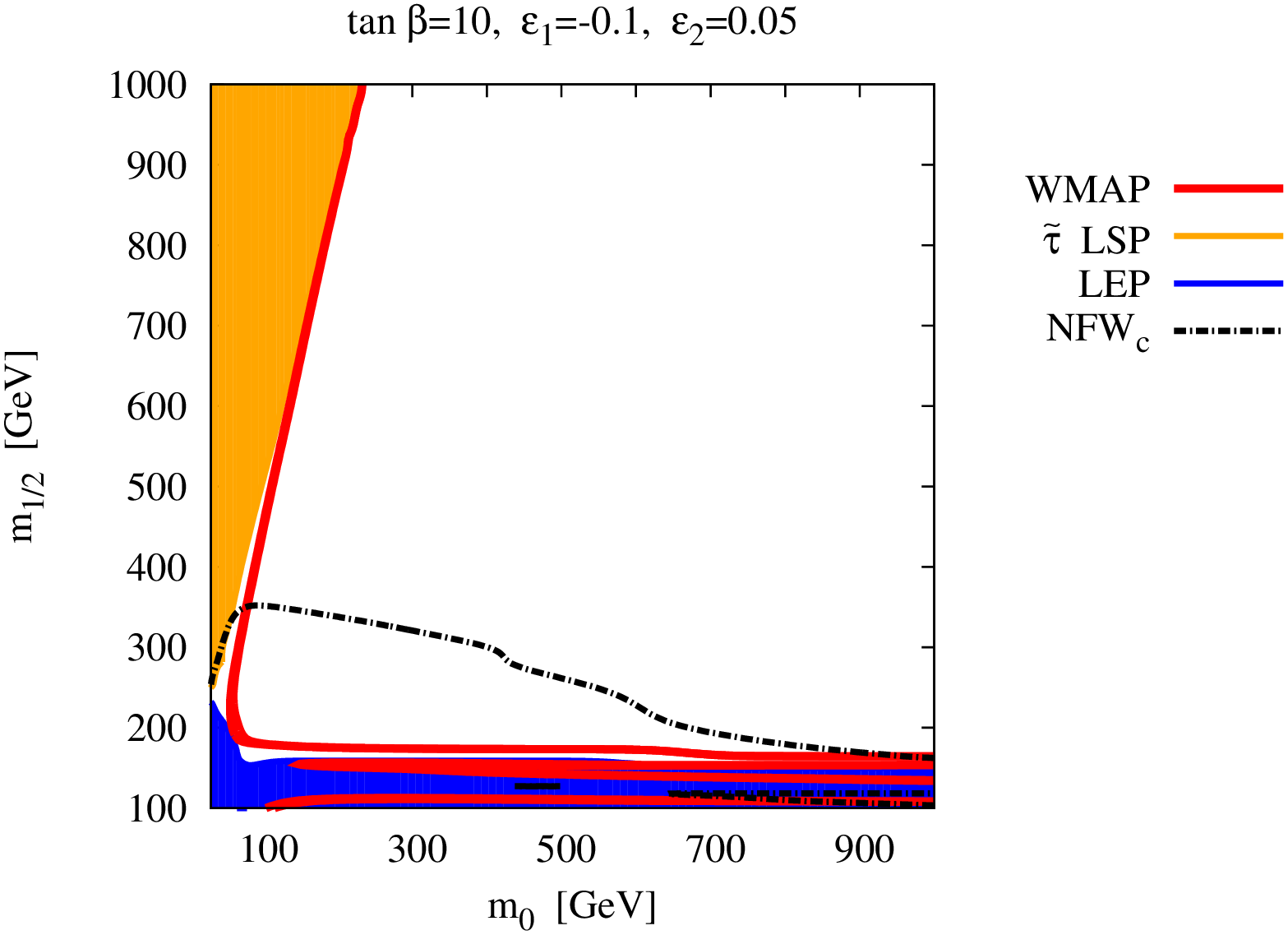}
\end{center}
\vspace{-0.4cm}
\caption{Regions in the $[m_0,\,m_{1/2}]$ plane that can be detected by the
Fermi satellite mission for our mSUGRA-like scenario. 
The black lines depict the detectability regions for the corresponding halo profile assumptions 
and $5$ years of data acquisition: the area below and on the left of the lines can be probed.
The same applies to the top-resonance blob at $m_{1/2}\sim 450$ GeV appearing on the left-hand side
plots. For NFW and Einasto profiles, the model could not be tested.}
\label{gam1}
\end{figure}

In figure \ref{gam1} we present the detectability regions for a $5$-year run of the Fermi
experiment and for $3$ different halo profiles presented in the literature, Einasto, NFW
and NFW$_c$, in the $[m_0,\,m_{1/2}]$ parameter space.
Fermi will be sensitive to the regions below the contours and, for 
$\tan\beta=3$, to the area inside the blob.
The detection prospects are maximised for low values of the $m_0$
and $m_{1/2}$ parameters. For higher $m_0$ values, the masses of the
squarks increase, penalising the annihilation cross-section.
However, the growth of $m_{1/2}$ gives rise to the opening of some relevant production channels,
after passing some thresholds, increasing significantly the $\langle\sigma v\rangle$.
The first one corresponds to a light neutralino, with mass $m_\chi\sim m_Z/2$
($m_{1/2}\sim 130$ GeV). In that case the annihilation is done via the $s$-channel exchange
of a real $Z$ boson, decaying in hadrons ($\sim 70\%$), neutrinos ($\sim 20\%$) and charged
leptons ($\sim 10\%$).
The second threshold appears for $m_\chi\sim m_W$ ($m_{1/2}\sim 220$ GeV). The annihilation cross-section
is enhanced by the opening of the production channel of two real $W^\pm$ bosons in the final state.
This process takes place solely through chargino exchange, since both $Z$ and Higgs bosons
exchange are suppressed by taking the limit $v\to 0$.
The last threshold corresponds to the opening of the channel $\chi_1^0\chi_1^0\to t\bar{t}$
($m_{1/2}\sim 400$ GeV). The diagrams involved in such a process contain contributions from $t$-
and $u$-channel exchange of stops, and from $s$-channel exchange of $Z$'s and pseudoscalar Higgs bosons.
The aforementioned threshold appears as a particular feature on the left-hand side plots:
An isolated detectable region for $m_{1/2}\sim 400$-$500$ GeV and $m_0\lesssim 300$ GeV
corresponding to the annihilation into a pair of real top quarks.

Larger values for the annihilation cross-section can be reached for higher values of $\tan\beta$.
In that case, the production process of a pair of down-type quarks (in particular $b\bar b$ pairs) and
charged leptons, dominates the total cross-section. In fact, the diagrams containing exchanges
of a pseudoscalar Higgs boson or a sfermion are enhanced by factors $\tan\beta$ and $1/\cos\beta$
respectively.
On the other hand, for high values of $\tan\beta$, the channels corresponding to
the annihilations into $W^+W^-$ and $t\bar{t}$ vanish. The first because of the reduction of the
coupling $\chi_1^0-\chi_i^\pm-W^\mp$; the second because of important destructive interference
between diagrams containing the exchange of a $Z$ boson and stops.

For the present scenario, the introduction of the NR operators gives rise to a very mild signature.
Actually, as in almost the whole parameter space the lightest neutralino is bino-like, its couplings
do not vary drastically. Moreover, the increment in the Higgs masses has a small impact on the
$\langle\sigma v\rangle$ factor. For indirect detection prospects, the main effect corresponds to a
slight increase in the LSP mass. Let us emphasize on the fact that, however, the detectable regions
are in the BMSSM case more cosmologically relevant than in the corresponding plain MSSM one.

Concerning figure \ref{gam1}, let us note that the
only astrophysical setup in which some useful information can be extracted
is the NFW$_c$ one. This means that in this scenario, in order to have some positive detection
in the $\gamma$-ray channel, there should exist some important enhancement of the signal by some
astrophysical mechanism (as the adiabatic contraction mechanism invoked in this
case). We note that, and this will be different from the case of antimatter signals, there is
however no important constraint on astrophysical boosts from the Galactic Center.
Gamma-ray detection does not rely, as is the case for positrons that we shall examine in
section \ref{secpos}, that much on local phenomena. 
In this respect, the NFW$_c$ results can be characterized as optimistic (it has been pointed
out that even by changing the gravitational collapse conditions, the results can get even more
pessimistic in the case, e.g., of a binary black hole formation in the GC), but not excluded.

\subsubsection{Light stops, heavy sleptons}
\begin{figure}[ht!]
\begin{center}
\vspace{-0.2cm}\hspace{-2.5cm}
\includegraphics[width=6.3cm,angle=-90]{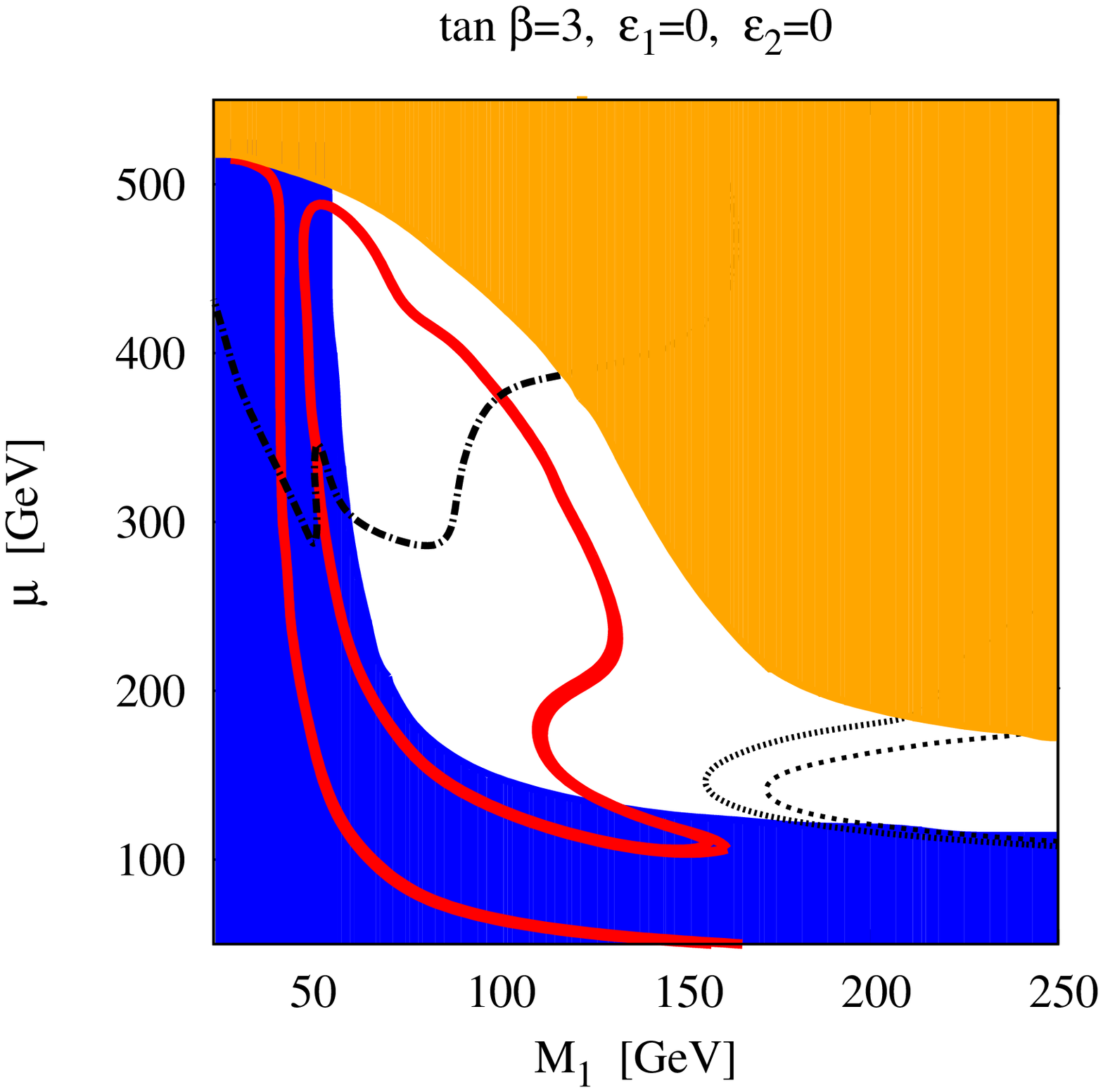}\hspace{-2.3cm}
\includegraphics[width=6.3cm,angle=-90]{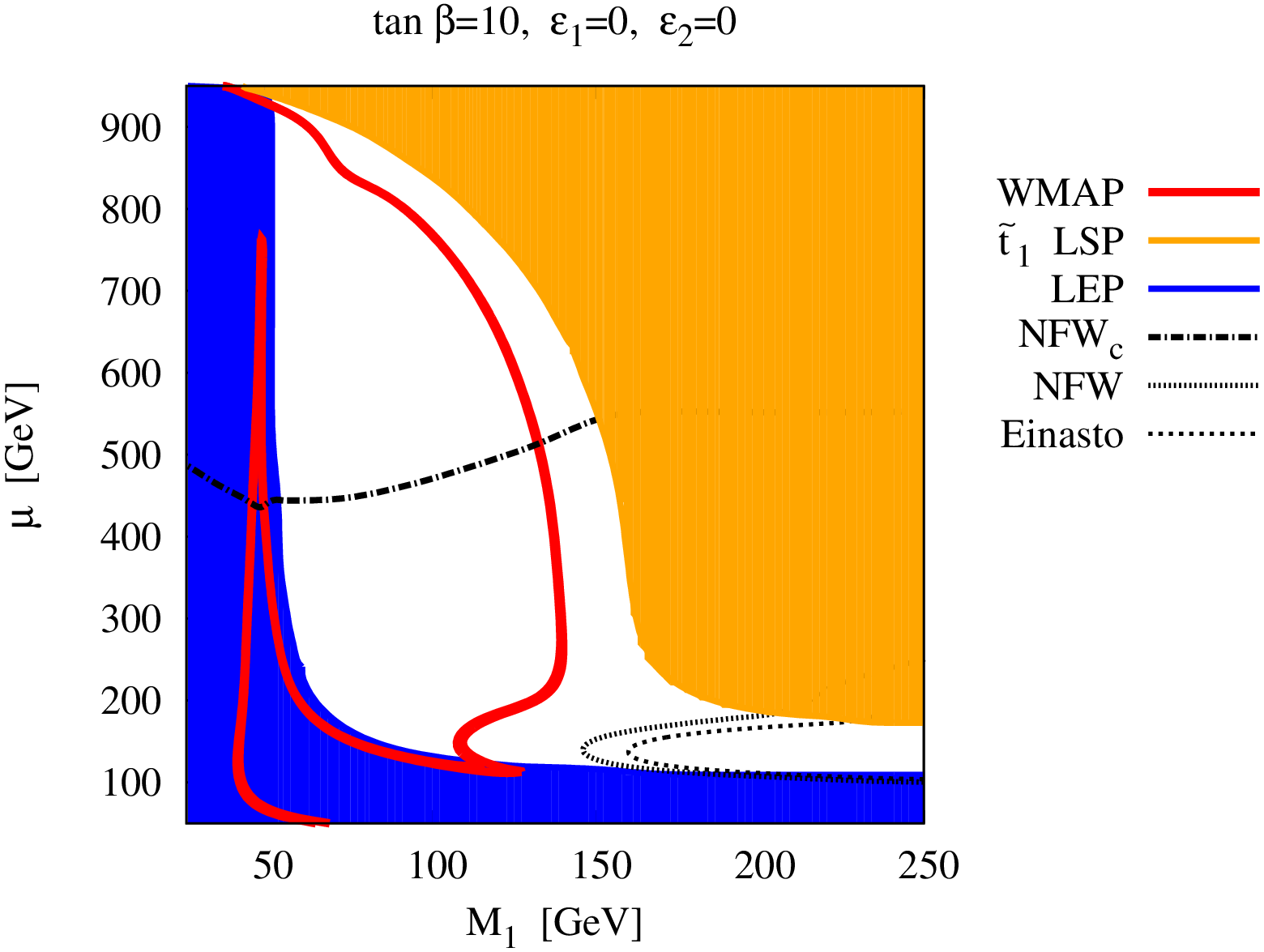}\\
\vspace{-0.4cm}\hspace{-2.5cm}
\includegraphics[width=6.3cm,angle=-90]{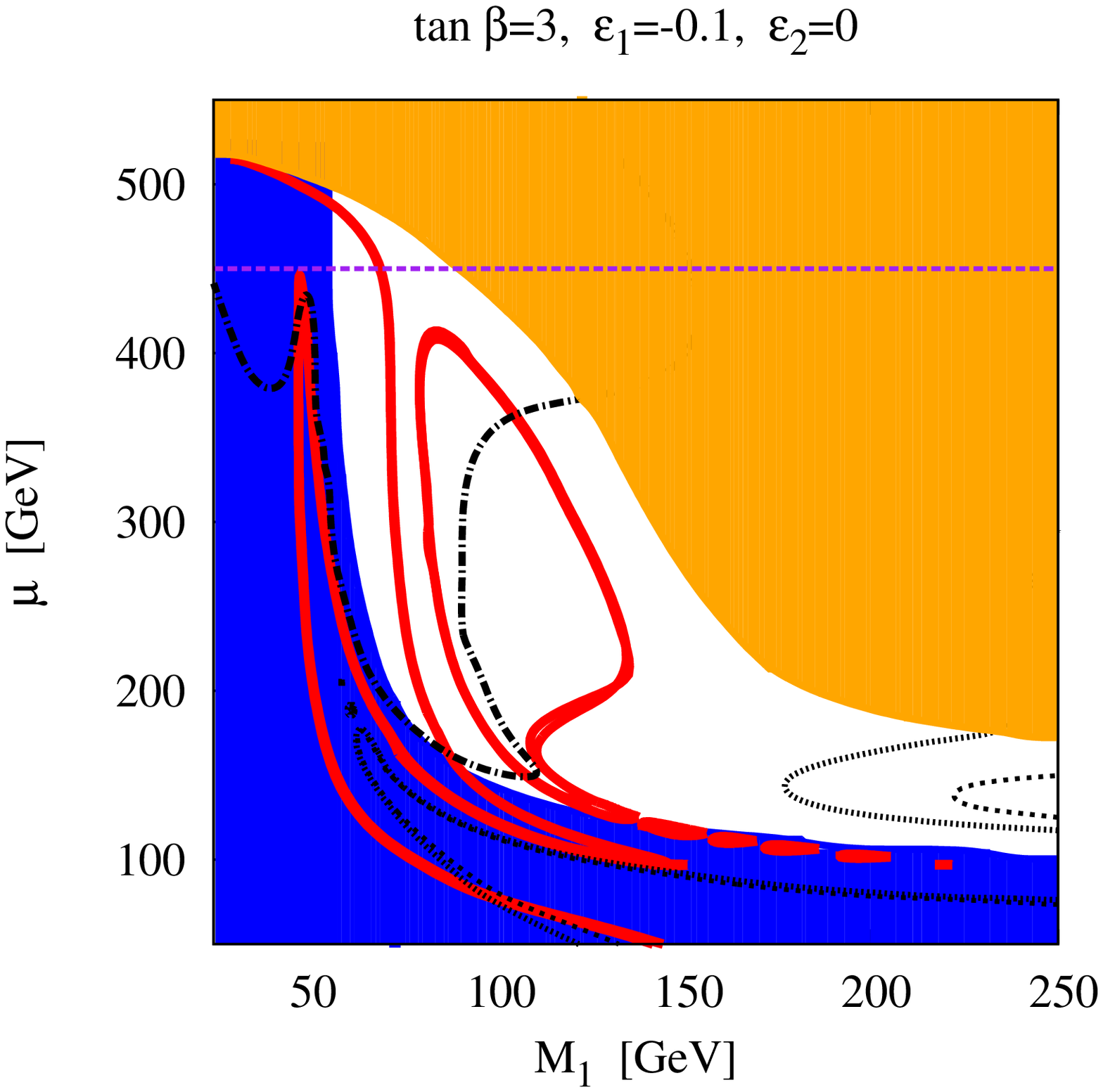}\hspace{-2.3cm}
\includegraphics[width=6.3cm,angle=-90]{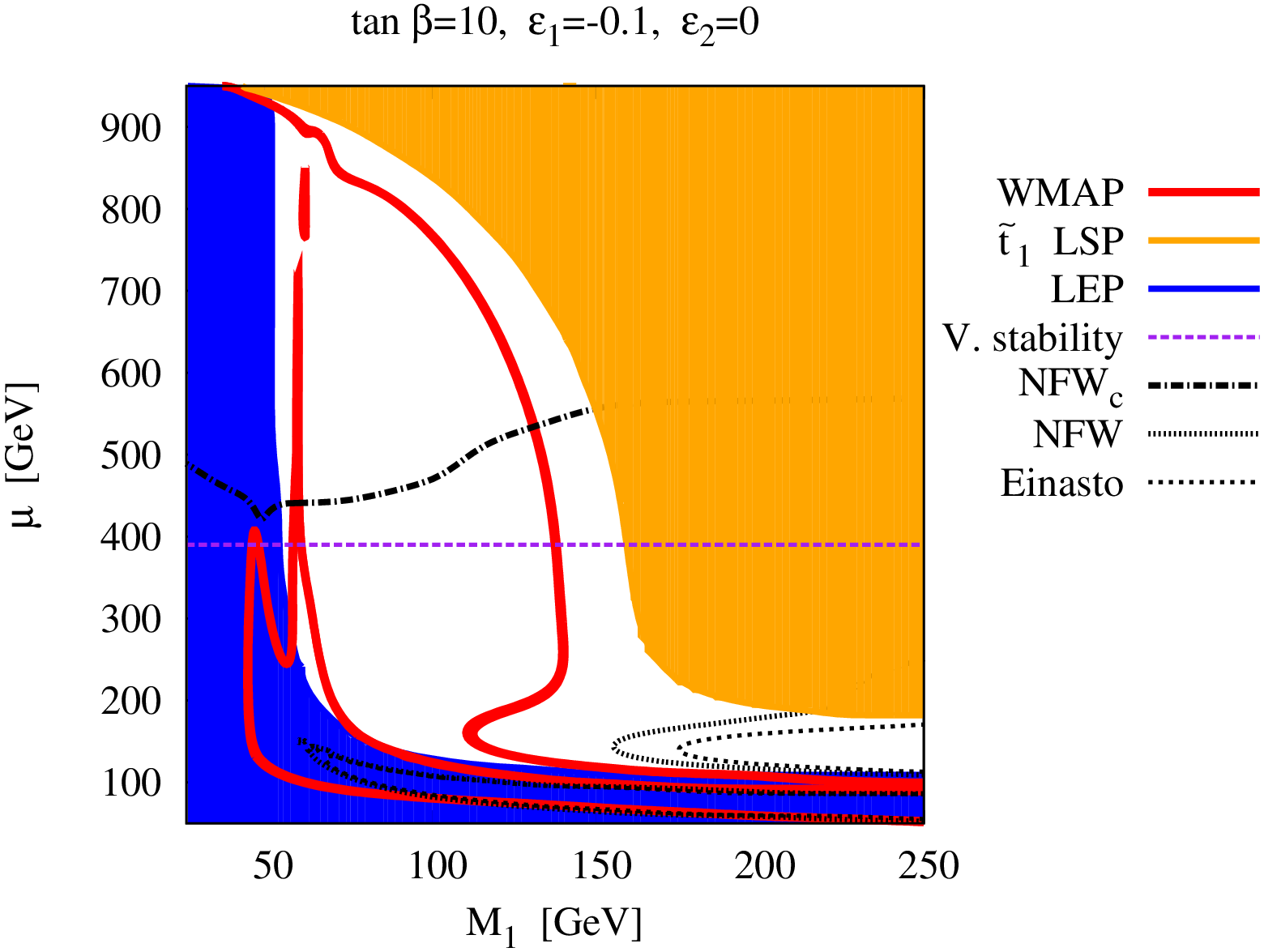}\\
\vspace{-0.4cm}\hspace{-2.5cm}
\includegraphics[width=6.3cm,angle=-90]{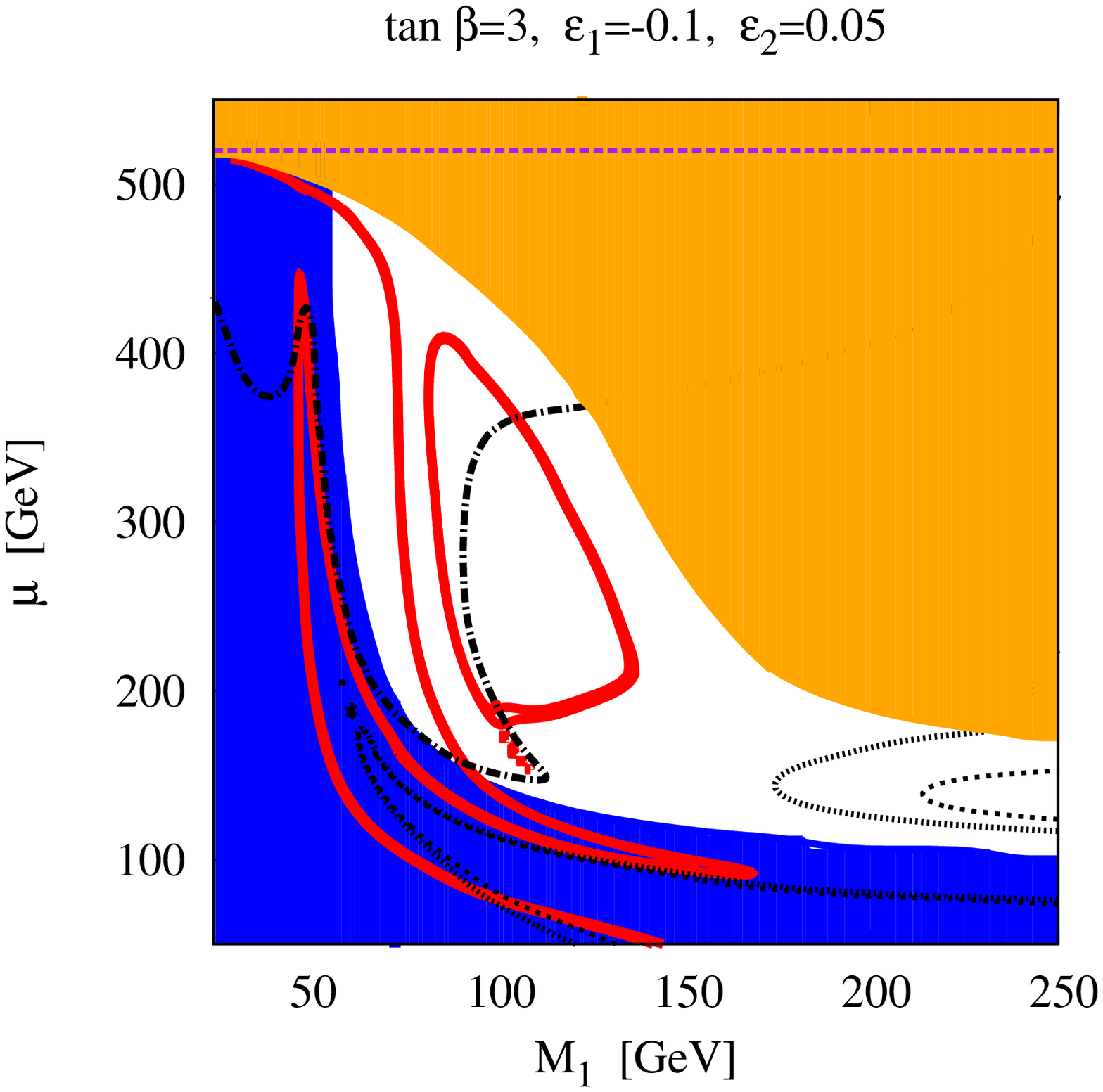}\hspace{-2.3cm}
\includegraphics[width=6.3cm,angle=-90]{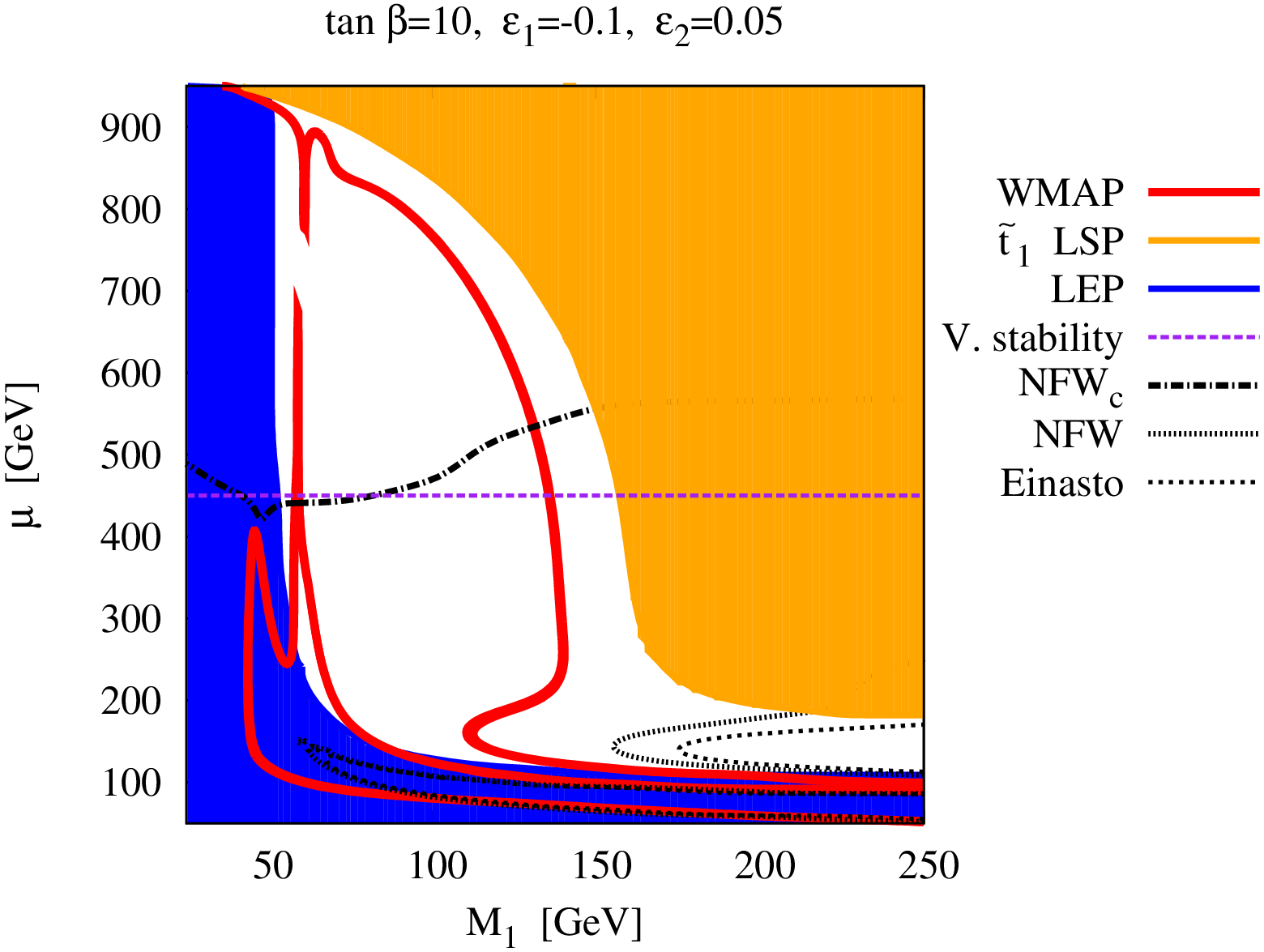}
\end{center}
\vspace{-0.9cm}
\caption{Regions in the $[M_1,\,\mu]$ plane that can be detected by the
Fermi satellite mission for our scenario with light stops and heavy sleptons. 
The black lines depict the detectability regions for the corresponding halo profile assumptions 
and $5$ years of data acquisition: the area below the lines can be probed. The same applies
to lines forming closed regions with respect to the axes, as is the case for the NFW and NFW$_c$
profiles: the parameter space points lying in the interior of these regions yield signals that
are detectable.}
\label{gam2}
\end{figure}

Figure \ref{gam2} presents the results for the second scenario with light stops and heavy sleptons.
The experiment will be sensitive to the regions below/on the right of the contours.
Again, the detection prospects are maximised for low values of the $M_1$
and $\mu$ parameters, corresponding to light WIMPs.
However, the growth of any of the latter parameters gives rise to the opening of some production
channels, enhancing significantly the $\langle\sigma v\rangle$.
The first one appears for $m_\chi\sim m_Z/2$ and corresponds to the $s$-channel exchange
of a real $Z$ boson. The second one concerns the production channel of two real $W$ bosons.
Let us note that in this scenario the neutralino LSP can be as heavy as
$\sim 110$ GeV, implying that the annihilation into a pair of top quarks is never kinematically allowed.
On the other hand, the region where $M_1\gg\mu$ is highly favored for indirect detection due to
the fact that the LSP is higgsino-like, maximising its coupling to the $Z$ boson. Let us recall
that the $Z$ boson does not couple to a pure gaugino-like neutralino.

Large values for the annihilation cross-section can be reached for high values of $\tan\beta$,
mainly because of the enhanced production of $b\bar b$ pairs.
On the other hand, for high values of $\tan\beta$, the threshold corresponding to the opening
of the annihilation into $W^+W^-$ is suppressed or enhanced for $\mu\gg M_1$ or $\mu\ll M_1$ respectively, 
due to the dependence of the $\chi_1^0-\chi_i^\pm-W^\mp$ coupling on the texture of the LSP.

For the present scenario, the introduction of the NR operators gives rise to an important
increase of the $\chi_1^0-\chi_1^0-A$ coupling when $\mu> M_1$, and therefore to a boost
in the annihilation into fermion pairs. On the other hand, as the Higgs boson $h$
becomes heavier, the processes giving rise to the final state $h\,Z$ get kinematically closed.

In the case presented in figure \ref{gam2}, there is a positive detection for all three halo profiles;
however, the regions that can be probed for either the NFW or the Einasto cases are cosmologically
irrelevant.

In fact, they could give rise to a positive detection near the $Z$-funnel and in the region where
the LSP is a higgsino state ($M_1\gtrsim 150$ GeV); nevertheless the first is already excluded by LEP
(at least for minimal scenarios) and the second generates too small a dark matter
relic density, below the WMAP limits.
On the other hand, the profile NFW$_c$ could test a large amount of the parameter space we examine,
particularly for high values of $\tan\beta$. Only the Higgs peak and the regions with a heavy LSP
escape from detection.


\section{Positrons}\label{secpos}
Positrons, as antiprotons that we shall describe in the following, present
the complication of propagating throughout the Intergalactic Medium. Numerous 
treatments, making a different set of assumptions and simplifications have 
been presented in the literature \cite{Baltz:1998xv, Strong:1998su, Lavalle:2006vb}
and applied in the case of the MSSM \cite{Mambrini:2006aq, Baltz:2001ir, deBoer:2002nn, Hooper:2008kv}.
Here we shall use the two-zone diffusion model and its semi-analytical 
solution as described in reference \cite{Lavalle:2006vb}.
In this model, positron and antiproton propagation takes place in a cylindrical region
(Diffusive Zone, DZ) around the galactic center of half thickness $L$; 
the propagating particles being free to 
escape the region, a case in which they are simply lost.
The propagation is described by a diffusion-convection-reacceleration equation:
\begin{equation}
\partial_t\psi+ \partial_z(V_c\,\psi) - \nabla(K\,\nabla\psi) - \partial_E \left[ b(E)\,\psi + K_{EE}(E)\,\partial_E\,\psi\right] = q\,,
\label{masterProp}
\end{equation}
where $\psi = dn/dE$ is the space-energy density of the positrons or antiprotons, 
$b(E)$ is the energy loss rate,
$q$ is the source term,  $V_c \approx (5 - 15)$ km/s is the convective wind velocity wiping
away the positrons or antiprotons from the galactic plane, 
\begin{equation}
K(E) = K_0\,\beta
\left( \frac{E}{E_0} \right)^\alpha
\label{DiffCoeff}
\end{equation}
is the diffusion coefficient, with $\beta$ being the particle's velocity, $K_0$ the diffusion
constant, $\alpha$ a 
constant slope, $E$ the kinetic energy (for positrons in practice the total
one), $E_0$ a reference energy (which we take to be $1$ GeV), and
\begin{equation}
K_{EE} = \frac29\,V_{a}^{2}\,\frac{E^2\,\beta ^4}{K(E)}
\end{equation}
is a coefficient describing reacceleration processes.\\
This is the master equation governing the propagation of cosmic rays
throughout the galactic medium, which we shall be employing in the following
for both positrons and antiprotons, with the details varying, of course, according
to each species ($e^+$ or $\bar p$).

\subsection{Differential event rate}
It has been pointed out in the literature that the convective wind as well
as reacceleration processes can be quite safely neglected for the case of 
positrons \cite{Delahaye:2008ua} (at least to a level sufficient for our purposes).
On the other hand, energy losses should be taken into account, the most important
contributions coming from synchrotron radiation or inverse Compton 
scattering on stellar light and CMB photons. To account for these two processes, we
write the energy loss rate as:
\begin{equation}
b(E) = \frac{E^2}{E_0\,\tau_E}\,,
\end{equation}
where $E$ is the positron energy and $\tau_E=10^{16}$s is the characteristic energy-loss time.
We are then left with the following equation:
\begin{equation}\label{masterposi}
\partial_t \psi - \nabla \left[ K(\vec{x},E)\,\nabla\psi \right]- 
\partial_E \left[b(E)\,\psi\right] = q(\vec{x},E)\,,
\end{equation}
where $K$ is the space diffusion coefficient --steady state is assumed. This coefficient is  taken to be constant in space but depends on the energy as
\begin{equation}
K(E) = K_0\left(  \frac{E}{E_0}\right) ^\alpha.
\end{equation}
Here the diffusion constant, $K_0$, and the spectral index, $\alpha$, are propagation parameters.
Regarding the propagation parameters $L,\,K_0$ and $\alpha$, we take their values from 
the commonly used MIN, MAX and MED models --see table \ref{PropParametersPos}.
The former two models correspond to the minimal and maximal positron 
fluxes that are compatible with the B/C data \cite{Delahaye:2007fr}. 
The MED model, on the other hand, corresponds to  the parameters that best fit the B/C data.\\
The master equation for positron propagation (equation \eqref{masterposi}) gets simplified to its final form
\begin{equation}
K_0\,\epsilon^\alpha \nabla^2 \psi  + 
\frac{\partial}{\partial \epsilon}\left( \frac{\epsilon^2}{\tau_E} \psi \right) + q = 0\,,
\label{masterPos}
\end{equation}
where $\epsilon\equiv E/E_0$. This is the expression we solve to calculate the effects of positron propagation on a signal produced 
at some point in the galaxy.

\begin{table}
\begin{center}
\begin{tabular}{|c|ccc|}
\hline 
&$L$ [kpc]&$K_0$ [kpc$^2$/Myr]&$\alpha$\\
\hline 
MIN & $1$ & $0.00595$ & $0.55$ \\ 
MED & $4$ & $0.0112$ & $0.70$ \\
MAX & $15$ & $0.0765$ & $0.46$ \\
\hline 
\end{tabular}
\caption{{\footnotesize Values of propagation parameters
widely used in the literature and that roughly provide minimal and maximal positron fluxes,
or constitute the best fit to the B/C data.}}
\label{PropParametersPos}
\end{center}
\end{table}
The resulting positron flux  from DM annihilations can be written as
(see reference \cite{Baltz:1998xv,Lavalle:2006vb} for details)
\begin{equation}
\Phi_{e^+} (E)= \frac{\beta_{e^+}}{4\pi} 
 \frac{\left\langle \sigma v \right\rangle}{2}  \left( \frac{\rho(\vec{x}_\odot)}{m_\chi} \right) ^2
\frac{\tau_E}{E^2}
\int_E^{m_\chi} f(E_s)\,\tilde{I}(\lambda_D)\,dE_s\,,
\label{PosFlux}
\end{equation}
where the detection and the production energy are denoted respectively by $E$ and $E_s$, $\vec{x}_\odot$ 
is the solar position with respect to the GC and $\beta_{e^+}$ is 
the positron velocity. $f(E_s)$ is the production spectrum for positrons, $f(E_s) = \sum_{i} dN_{e^+}^i/dE_s$, with $i$ running over all possible annihilation channels. The diffusion length,  $\lambda_D$, is defined by
\begin{equation}
\lambda_D^2 = 4\,K_0\,\tau_E \left(\frac{\epsilon^{\alpha-1} - \epsilon_s^{\alpha-1}}{1-\alpha} \right) .
\end{equation}
The so-called halo function, $\tilde{I}$, contains all the dependence on astrophysical factors. It is given by
\begin{equation}
\tilde{I}(\lambda_D) = \int_{DZ} d^3\vec{x}_s\,\tilde{G}\left(\vec{x}_\odot, E \rightarrow \vec{x}_s, E_s\right)\,
\left(\frac{\rho(\vec{x}_s)}{\rho(\vec{x}_\odot)}\right)^2\,,
\end{equation}
where the integral is performed over the diffusion zone.
The modified Green function $\tilde{G}$ is in its turn defined by 
\begin{equation}
\tilde{G} = \frac{1}{4\pi\,K_0\,\tilde{\tau}} e^{-(r_\odot - r_s)^2/(4 K_0 \tilde{\tau})}\,\tilde{V}\,,
\end{equation}
with $\tilde{V}$ depending on the value of the characteristic parameter $\zeta=\frac{L^2}{4\,K_0 \tilde{\tau}}$ and $\tilde\tau = \tilde{t} - \tilde{t_s} = \tau_E \left[ (\epsilon^{\alpha - 1}/(1 - \alpha)) - (\epsilon_s^{\alpha - 1}/(1 - \alpha)) \right] $.
When $\zeta>1$  --when the diffusion time is small-- boundary conditions can be ignored and the propagation equation can be treated as a $1$D Schrödinger equation.  In that case
\begin{equation}
\tilde{V} = 
\frac{1}{\sqrt{4\pi\,K_0\,\tilde{\tau}}}\,
\exp\left[-\frac{(z_\odot-z_s)^2}{4\,K_0\,\tilde{\tau}}\right]\,. 
\end{equation}
When $\zeta$ is small this approximation no longer holds but  we can express
$\tilde{V}$ as
\begin{equation}
\tilde{V} = \sum_{n = 1}^{\infty} 
\frac{1}{L} \left[ e^{-\lambda_n \tilde{\tau}} \phi_n(z_s)  \phi_n(z_\odot) + 
 e^{-\lambda_n' \tilde{\tau}} \phi_n'(z_s)  \phi_n'(z_\odot)   \right] 
\end{equation}
where
\begin{align}
\phi_n(z)  &= \sin[k_n(L - |z|)]\,,  &k_n =&  \left( n - \frac{1}{2}\right) \frac{\pi}{L}\,,\\ 
\phi_n'(z) &= \sin[k_n'(L - z)]\,,   &k_n' =&  n \frac{\pi}{L} \,,
\end{align}
$\lambda_n=K_0\,k_n^2$ and $\lambda_n'=K_0\,(k'_n)^2$.  

The advantage of this method is that the halo function $\tilde{I}(\lambda_D)$ 
can be calculated (and either tabulated or fitted)
just once as a function of the diffusion length and then be easily used for performing parameter 
space scans which, as in our case, can be rather large.

Let us however note, and this shall also be the case for antiprotons, that this method 
has been proved to have a limited validity. More specifically, at energies below $10$ GeV, 
some of the assumptions and simplifications that we made no longer hold. 
For this reason we shall limit ourselves at energies $\ge 10$ GeV.

\subsection{The background}
In the conventional background model \cite{Strong:2004de}, positrons are produced in 
the interactions between cosmic-ray nuclei and the interstellar medium (ISM). 
This model is however not compatible with the recent data from PAMELA
\cite{Adriani:2008zr} and Fermi-LAT \cite{Abdo:2009zk}.

Even after taking into account several possible uncertainties
due to cosmic ray propagation, the data reveals a clear excess over this background at high energies, $E\gtrsim
10$ GeV. Hence, a new source of high energy positrons is necessary to explain the data.
It also turns out to be exceedingly difficult to reconcile the observed excess with dark matter annihilations.
A large number of models trying to do so have been proposed in the literature, most of which 
turn out to be in conflict with other observational data (see, e.g. the interesting treatments
in references \cite{Cirelli:2008pk,Bertone:2008xr,Bergstrom:2008ag,Gogoladze:2009kv,Cirelli:2009vg,Galli:2009zc,Pato:2009fn,Meade:2009iu,Profumo:2009uf,Huetsi:2009ex,Cirelli:2009bb,Cirelli:2009dv,Papucci:2009gd}).

Although obviously dark matter annihilation could, in principle, account at least for some part
of the observed excess, we feel that the most conservative choice is to consider the whole PAMELA
signal as been due to some -yet unknown- astrophysical process, such as a pulsar 
\cite{Hooper:2008kg,Yuksel:2008rf,Profumo:2008ms} and treat it as a background for the oncoming AMS-02 experiment \cite{Choi:2009qc}.
We shall use as a background the absolute positron flux that can be obtained through a combination
of the PAMELA and Fermi data.

\subsection{AMS-02 and positron detection}
The scheduled AMS-02 mission aims at the detection and measurement of cosmic-ray fluxes
(and also $\gamma$-rays)
coming from various sources. Among these sources could, of course, be dark matter annihilations.

In the case of positrons, AMS-02 will be able to measure the spectrum of
positrons with an average geometrical acceptance of $0.042$ m$^2$ sr in the energy range
above $4$ GeV \cite{Goy:2006pw}.
In our study we consider a $3$-year run, which is actually the collaboration's nominal 
run time, and an energy range extending up to $300$ GeV, with $20$ logarithmically evenly spaced
bins.

\subsection{Results}
\subsubsection{Correlated stop-slepton masses}
The results concerning the detectability perspectives for
the mSUGRA-like scenario in the positron detection channel 
are quite pessimistic.
In fact, since the PAMELA and Fermi measurements mentioned above, 
and according to our conservative treatment, the main issue in the positron channel is an
extreme domination of all measurements by a large background severely obscuring the signal.

Obviously, one could invoke large boost factors of an astrophysical nature
as was the case in the first efforts to explain
the PAMELA excess through dark matter annihilations, a case in which a larger portion of the 
parameter space would be visible. However, it has been pointed out  
that it is highly unlikely to expect large boost factors due, e.g., to substructures
in the halo \cite{Lavalle:1900wn}. In this respect, if we assume a maximal clump-due signal enhancement 
by a factor $\sim 10$,  the only hope for positive detection of a non-LEP excluded area
might come for the bulk region, as it is the only one lying at the limits of detectability.
For the sake of brevity, we omit the relevant plots for the mSUGRA-like benchmark, since
no point of the parameter space can be tested.

\subsubsection{Light stops, heavy sleptons}
\begin{figure}[ht!]
\begin{center}
\vspace{-0.2cm}\hspace{-2.5cm}
\includegraphics[width=6.3cm,angle=-90]{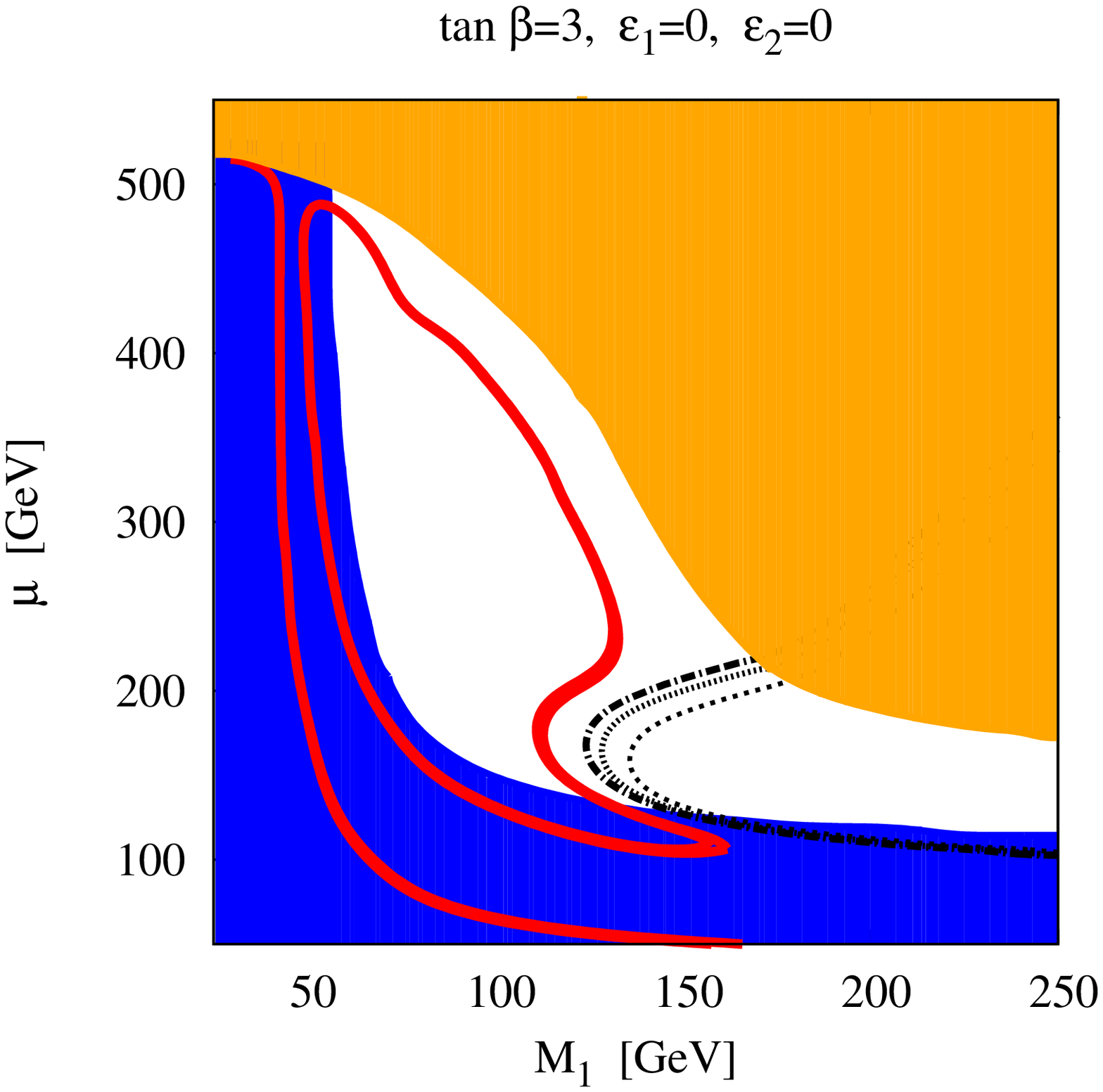}\hspace{-2.3cm}
\includegraphics[width=6.3cm,angle=-90]{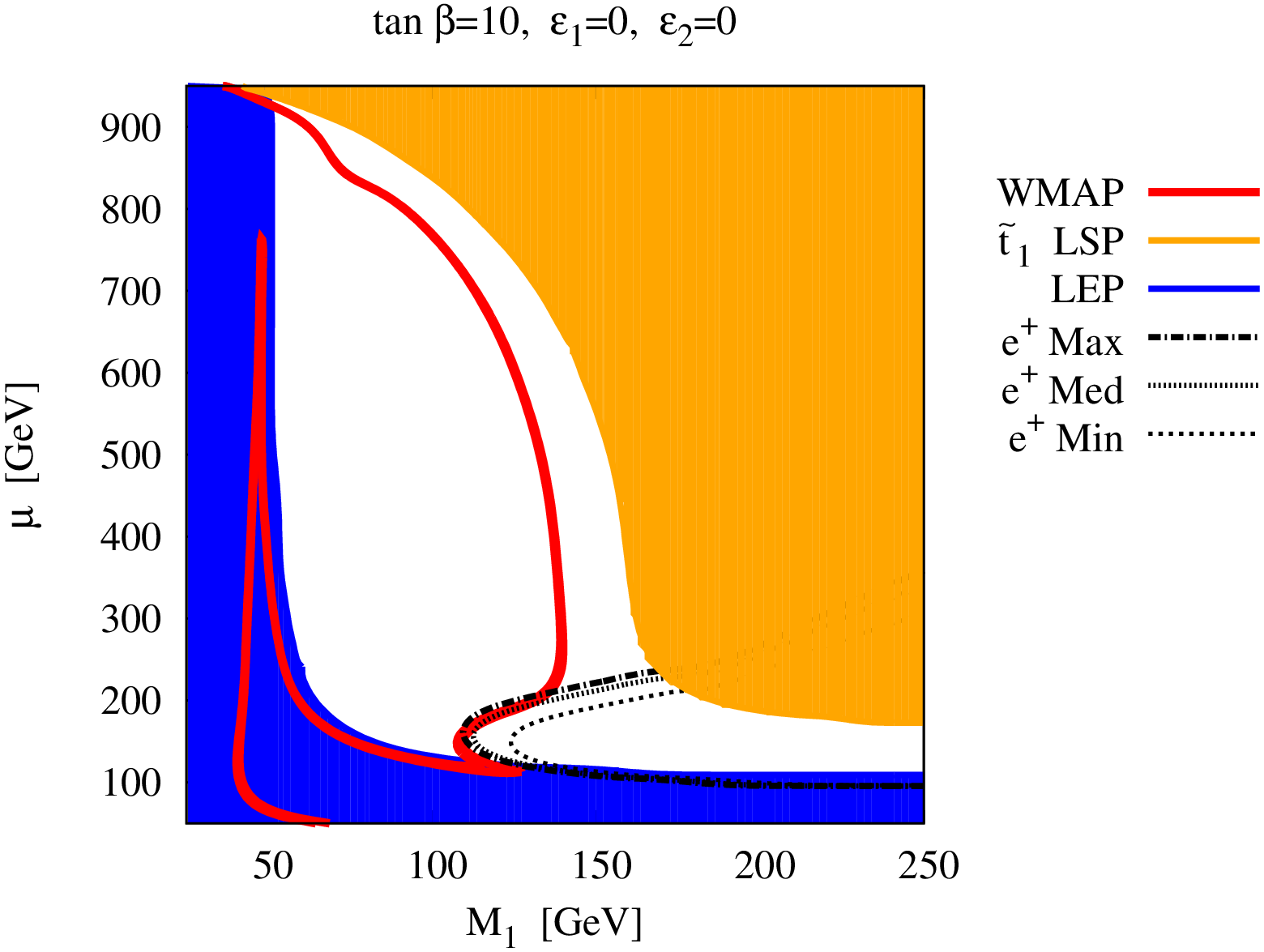}\\
\vspace{-0.4cm}\hspace{-2.5cm}
\includegraphics[width=6.3cm,angle=-90]{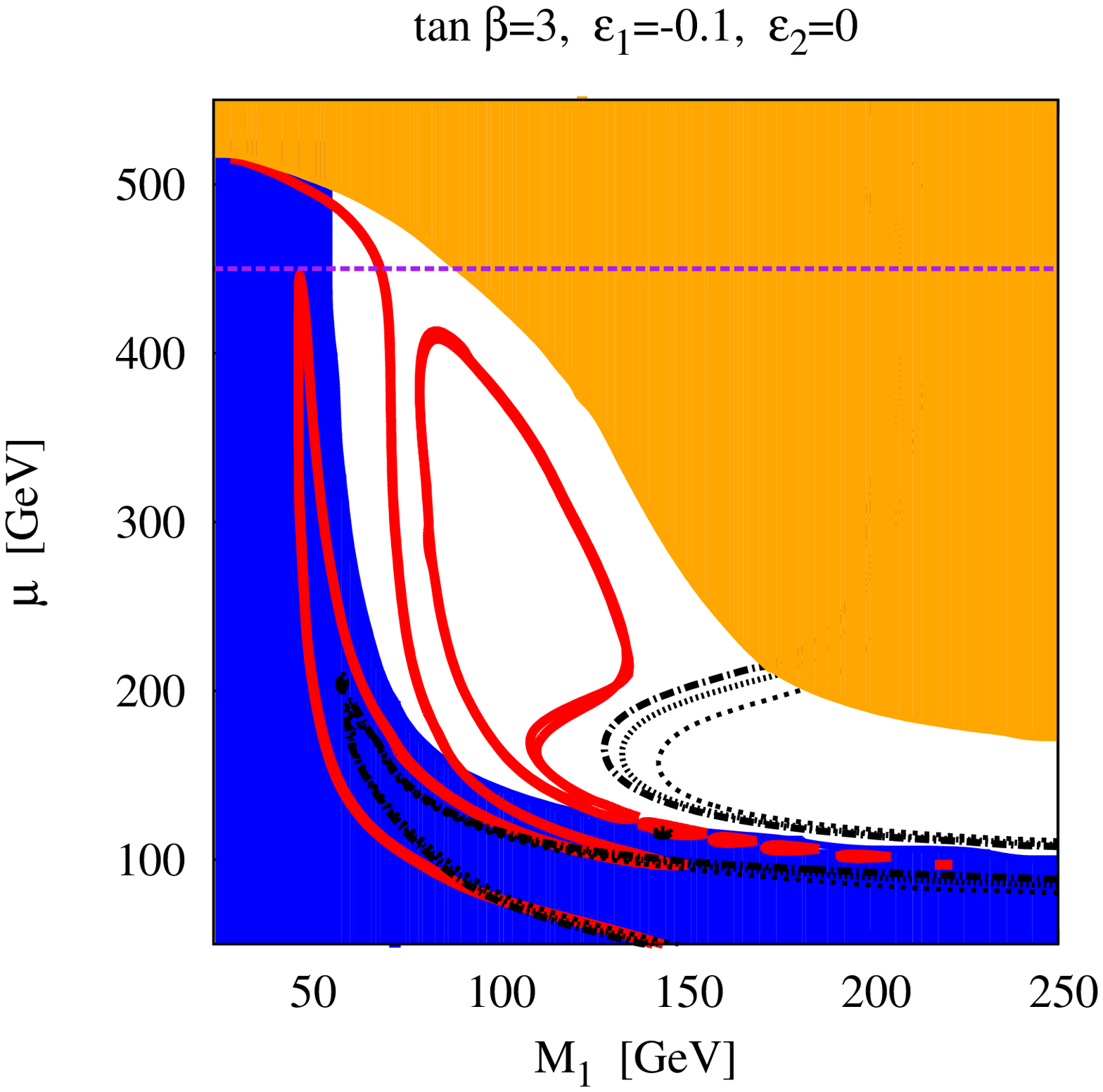}\hspace{-2.3cm}
\includegraphics[width=6.3cm,angle=-90]{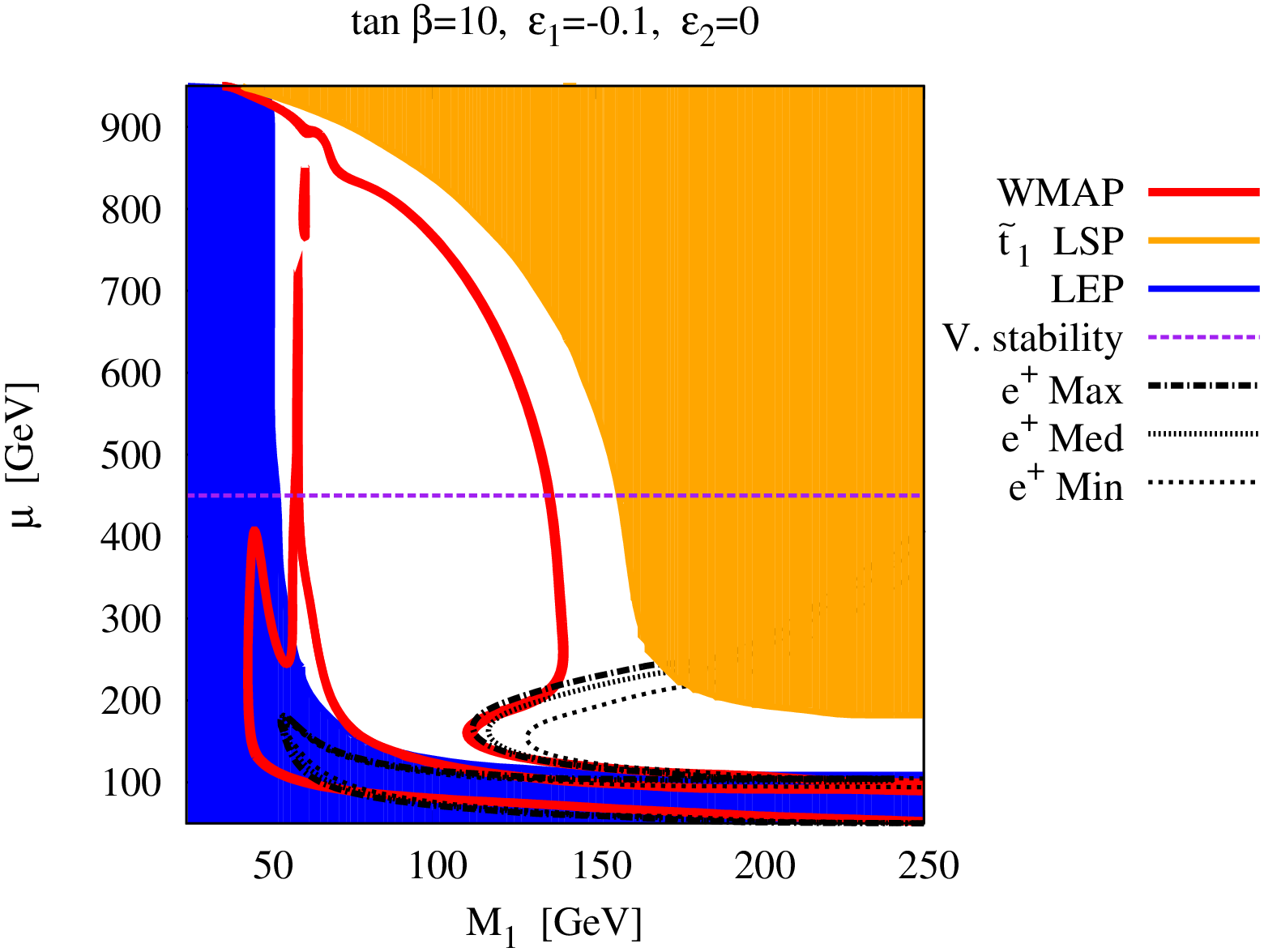}\\
\vspace{-0.4cm}\hspace{-2.5cm}
\includegraphics[width=6.3cm,angle=-90]{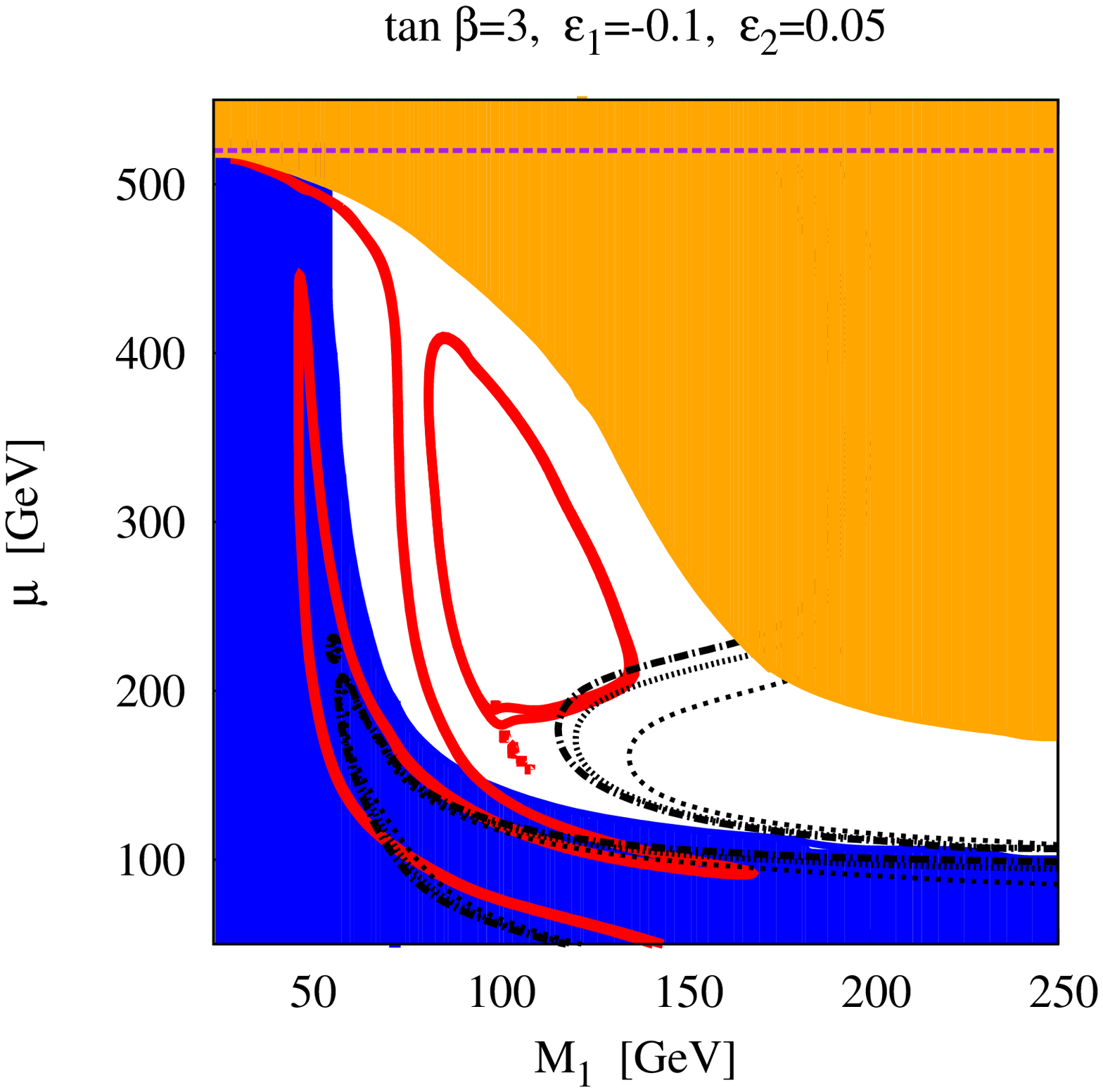}\hspace{-2.3cm}
\includegraphics[width=6.3cm,angle=-90]{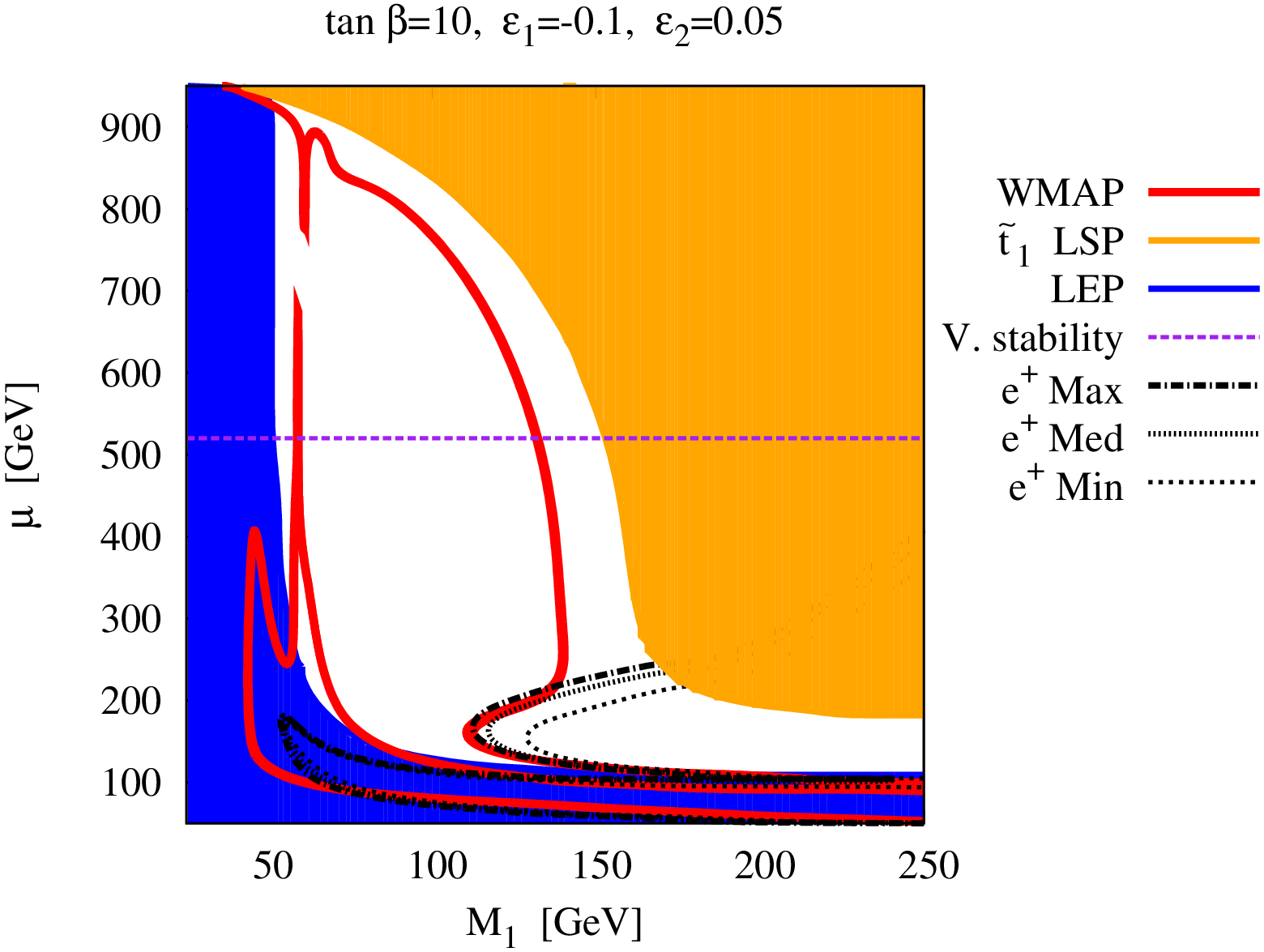}
\end{center}
\vspace{-0.9cm}
\caption{Regions in the $[M_1,\,\mu]$ plane that can be detected by a $3$-year run of the 
AMS-02 satellite mission for the scenario with light stops and heavy sleptons, in the positron channel. 
The black lines depict the detectability regions for the $3$ considered propagation
models, MIN, MED and MAX: 
the parameter space points lying within the regions delimited by the black lines can be
probed, assuming the corresponding propagation models. 
Part of the mixed bino-higgsino region, as well as (marginally) some part of the $Z$ funnel
region can be probed.}
\label{po2}
\end{figure}

In figure  \ref{po2} we present the detection perspectives in the positron channel for our scenario
with light stops and heavy sleptons.
The detectable parameter space regions lie within the zones delimited by the black lines
for the three propagation models: the oval-shaped blobs as well as the banana-shaped ones.
Once again, we notice the general features already present in the $\gamma$-ray channel.
The regions giving rise to a positive detection lie within the zone where the
LSP is a higgsino-like state, with mass $m_\chi>m_W$, in order to have the final state
$W^+W^-$ kinematically available. This region in general does not fulfill the WMAP limit.
However, and this is a novel feature of the BMSSM, with both $\epsilon_1$ and $\epsilon_2$
couplings turned on, a small region of the mixed higgsino-bino regime can be detected for
the MAX (and even the MED) propagation model. As we pointed out before, 
in this regime the total annihilation cross-section can be quite significantly enhanced, leading
to better detection perspectives.


\section{Antiprotons}
\subsection{Differential Event Rate}
Equation \eqref{masterProp} also governs antiproton propagation in the galactic medium.
Nevertheless, the dominant processes vary significantly with respect to the 
positron case.
More precisely, all energy redistributions in the initial 
(injection) spectrum --energy losses, reacceleration, 
as well as `tertiary' contributions (i.e. contributions
from secondary antiprotons produced upon inelastic scattering with the interstellar medium)-- can be ignored. 
Whether these redistributions are important 
or not depends mainly on the antiproton energy. For GeV energies, 
the results may deviate up to $50\%$ from those obtained with the  
(more complete) Bessel function treatment\footnote{In reference \cite{Maurin:2006hy}
a comparison between the two methods can be found (see figure $2$).}.
But for energies around $~10$ GeV, the accuracy
of the method improves dramatically, yielding essentially indistinguishable
results at slightly higher energies. Since the $\bar{p}$ energy region we shall consider
begins at $10$ GeV, we can safely use this simplified approach. 

Let us denote by $\Gamma_{\overline{p}}^{\mbox{\tiny{ann}}} = \sum_{\mbox{\tiny{ISM}}} 
n_{\mbox{\tiny{ISM}}}\,v\,\sigma_{\overline{p} \ \mbox{\tiny{ISM}}}^{\mbox{\tiny{ann}}}$
the destruction rate of antiprotons in the interstellar medium, where $\mbox{ISM} = \mbox{H and He}$, 
$n_{\mbox{\tiny{ISM}}}$ is the average number density of ISM in the galactic disk, $v$ is the
antiproton velocity and $\sigma_{\overline{p} \ \mbox{\tiny{ISM}}}^{\mbox{\tiny{ann}}}$ is the
$\bar{p} - \mbox{ISM}$ annihilation cross-section. 
Implementing the aforementioned simplifications, 
the transport 
equation for a point source (which actually defines the propagator $G$) is:
\begin{equation}
\left[ -K\,\nabla + V_c\,\frac{\partial}{\partial z}
+2\,h\,\Gamma^{\mbox{ann}}_{\bar p}\,\delta(z) \right] G = 
\delta\left(\vec{r} - \vec{r'}\right)\,,
\end{equation}
with $h = 100$ pc being the half-thickness of the galactic disc.
The antiproton propagator at the solar position can then be written (in cylindrical coordinates) as
\begin{equation}
G^{\odot}_{\overline{p}}(r,z) = 
\frac{e^{-k_v\,z}}{2 \pi\,K\,L}\,
\sum_{n=0}^{\infty} c_n^{-1}\,K_0\left(r\sqrt{k_n^2 + k_v^2}\right)
\sin(k_n\,L)\,\sin(k_n\,(L-z))\,,
\label{GreenPbars}
\end{equation}
where
$K_0$ is a modified Bessel function of the second kind and
\begin{eqnarray}
c_n & = & 1 - \frac{\sin(k_n L) \cos(k_n L)}{k_n L}\,,\\
k_v & = & V_c/(2K)\,,\\
k_d & = & 2\,h\,\Gamma_{\overline{p}}^{\mbox{\tiny{ann}}}/K + 2\,k_v\,.
\end{eqnarray}
$k_n$ is obtained as the solution of the equation
\begin{equation}
n\,\pi - k_{n}\,L - \arctan(2\,k_n/k_d) = 0, \ \ n\in\mathbb{N}\,.
\end{equation}
Then, in order to compute the flux expected on  earth, we should
convolute the Green function \eqref{GreenPbars} with the source 
distribution $q(\vec{x}, E)$. For dark matter annihilations in the 
galactic halo, the source term is given by
\begin{equation}
q(\vec{x}, E) = \frac{1}{2}
\left( \frac{\rho(\vec{x})}{m_\chi} \right)^2
\sum_i 
\left( 
\langle\sigma v\rangle \frac{dN_{\bar{p}}^i}{dE_{\bar{p}}}
\right)\,,
\label{eq:q}
\end{equation}
where the index $i$ runs
over all possible annihilation final states. As in the previous cases, the decay of SM 
particles into antiprotons has been calculated with {\tt PYTHIA} \cite{Sjostrand:2006za}.  
Regarding the distribution of dark matter in the Galaxy, $\rho(\vec{x})$, 
we assume a NFW profile. The final expression for 
the antiproton flux on the Earth takes the form
\begin{equation}
\Phi_{\odot}^{\bar{p}} (E_{\mbox{\tiny{kin}}}) = 
\frac{c\,\beta }{4\pi}
\frac{\langle\sigma v\rangle}{2}
\left(   \frac{\rho(\vec{x}_{\odot})}{m_\chi} \right)^2
\frac{dN}{dE}(E_{\mbox{\tiny{kin}}})
\int_{DZ} \left(\frac{\rho(\vec{x_s})}{\rho(\vec{x}_{\odot})} \right)^2
G^{\odot}_{\overline{p}}(\vec{x}_s)\,d^3x\,,
\label{PbarFlux}
\end{equation}
where none of the integrated quantities depends on the antiproton energy. 
Once again, the integral in equation \eqref{PbarFlux}, 
which we compute using a VEGAS Monte-Carlo algorithm, 
needs to be calculated only once for each value of the injection energy, 
which is actually the same as the detection energy. 

Regarding  the propagation parameters $L$, $K_0$, $\alpha$, and  $V_c$, we take their values from 
the well-established MIN, MAX and MED models --see table \ref{PropParameters}. 
The former two models correspond to the minimal and maximal antiprotons 
fluxes that are compatible with the B/C data. 
The MED model, on the other hand, corresponds to  the parameters that best fit the B/C data.
\begin{center}
\begin{table}
\centering
\begin{tabular}{|c|cccc|}
\hline 
&$L$ [kpc]&$K_0$ [kpc$^2$/Myr]&$\alpha$&$V_c$ [km/s]\\
\hline 
MIN & $1$ & $0.0016$ & $0.85$ & $13.5$\\ 
MED & $4$ & $0.0112$ & $0.70$ & $12.0$\\
MAX & $15$ & $0.0765$ & $0.46$ & $5.0$\\
\hline 
\end{tabular}
\caption{{\footnotesize Values of propagation parameters
widely used in the literature and that provide minimal and maximal antiproton fluxes,
or constitute the best fit to the B/C data.}}
\label{PropParameters}
\end{table}
\end{center}

\subsection{The background}
Contrary to the positron case, the most well-known treatment of the 
astrophysical antiproton background by Strong and Moskalenko \cite{Moskalenko:2001ya}, seems to
be compatible with the antiproton data \cite{Adriani:2008zq} from the PAMELA experiment.

In order to parametrize our background, we borrow the simple fit performed by Cirelli {\it et al.}
\cite{Cirelli:2009uv}, which provides a sufficiently good fit for our purposes, to theoretical predictions but also to 
the recent PAMELA data. We pay special attention at reproducing the good 
background normalization at low energies, so as to stay as close as possible to the PAMELA measurements.

\subsection{AMS-02 and antiproton detection}
In the case of antiprotons, AMS-02 will be able to measure \cite{Goy:2006pw} the corresponding
fluxes with an average geometrical acceptance of $330$ cm$^2$ sr above $16$ GeV and up to 
$300$ GeV.
In our study we again consider a $3$-year run and the mentioned energy range divided into 
$20$ logarithmically evenly spaced energy bins.

\subsection{Results}
\subsubsection{Correlated stop-slepton masses}
\begin{figure}[ht!]
\begin{center}
\vspace{-0.2cm}\hspace{-2.5cm}
\includegraphics[width=6.3cm,angle=-90]{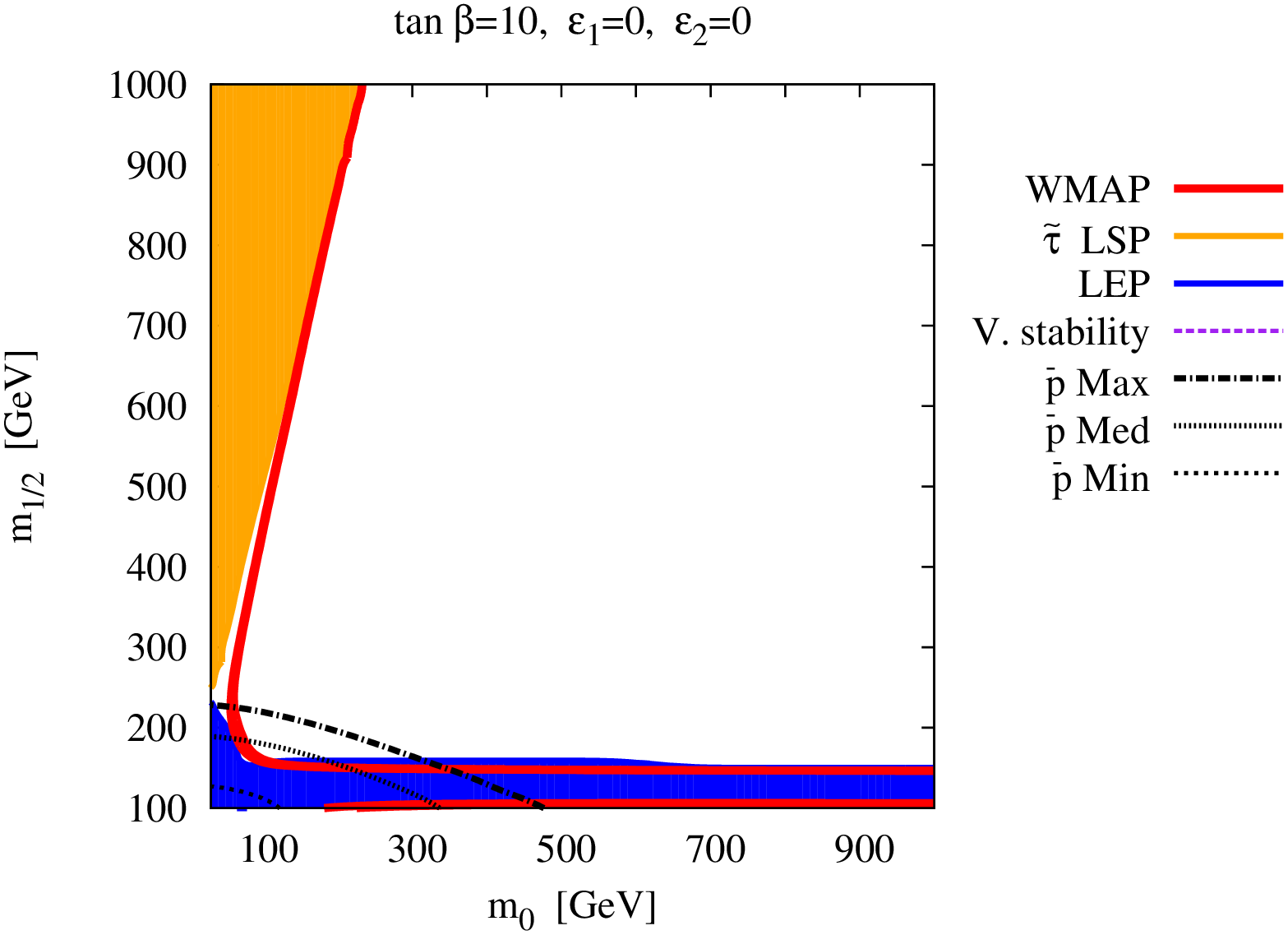}\\
\vspace{-0.4cm}\hspace{-2.5cm}
\includegraphics[width=6.3cm,angle=-90]{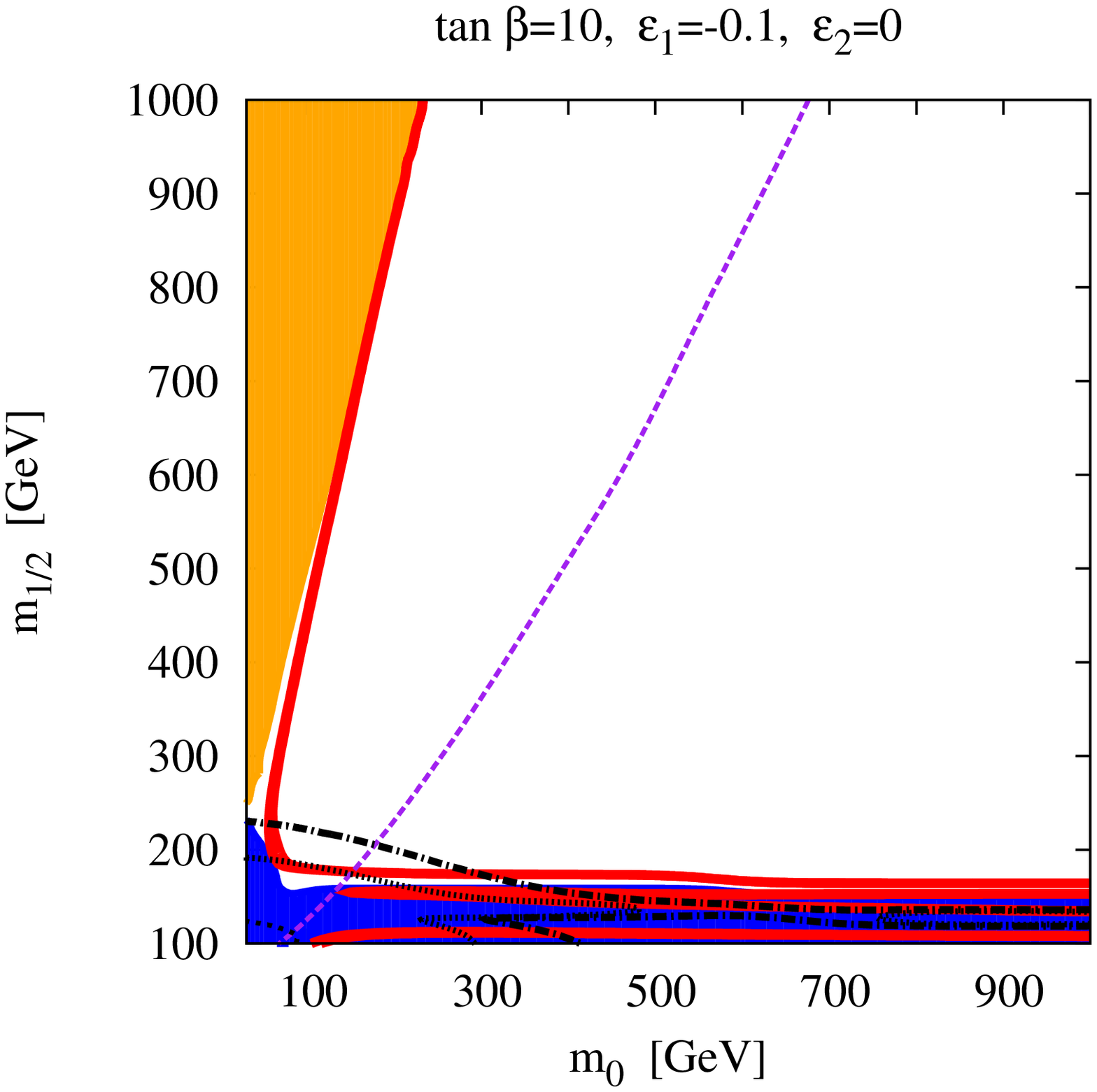}\hspace{-2.3cm}
\includegraphics[width=6.3cm,angle=-90]{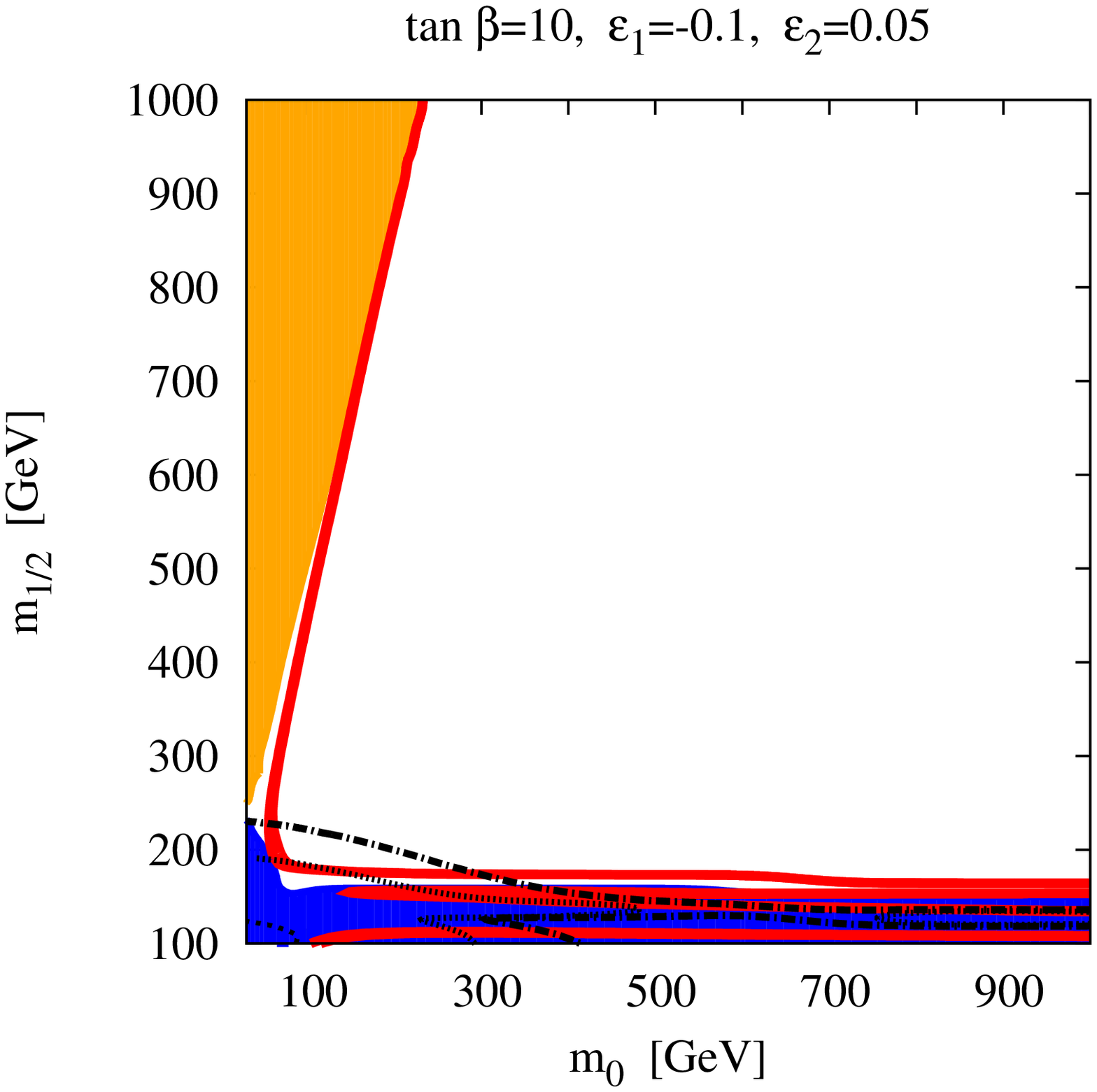}
\end{center}
\vspace{-0.9cm}
\caption{Regions in the $[m_0,\,m_{1/2}]$ plane that can be detected by a $3$-year run of the 
AMS-02 satellite mission for our mSUGRA-like scenario in the antiproton channel. 
The black lines depict the detectability regions for the $3$ considered propagation models: 
the area delimited by the axes and the black lines can be probed for the corresponding propagation
model (i.e. the region towards the lower left corner in each plot).}
\label{an1}
\end{figure}

In figure \ref{an1} we present our results for the detectability  of the BMSSM in comparison to
the MSSM by the AMS-02 experiment for the antiproton channel. The detectable regions lie below the
black lines.
In the case $\tan\beta=3$, the experiment is not sensitive to any point in the parameter space satisfying also the
collider constraints (and, hence, the corresponding results are once again omitted). 
A first remark here should concern the fact that the perspectives for antiproton detection are
significantly ameliorated with respect to the corresponding positron ones, at least for large values of $\tan\beta$.
This could, in some sense, seem quite strange, since the average antiproton yield coming
from a DM annihilation is, in general, smaller than the positron one.
However, we should take into account that these results reflect not only the behaviour of the models
themselves, but also the interplay between the signal and the background. And, as it turns out the $\bar{p}$ channel is a significantly low-background one.

Important areas of the viable parameter space are at the limits of detectability: the bulk region, but also, 
for some cases, the Higgs funnel.
Now, as we stressed out before, the possible enhancements due for example to substructures are quite constrained.
Given however that some regions are marginally out of reach, it would not be impossible to state that 
even small boosts could render important (in a qualitative sense, due to their cosmological relevance) 
regions of the parameter space detectable by AMS-02.

\subsubsection{Light stops, heavy sleptons}
\begin{figure}[ht!]
\begin{center}
\vspace{-0.2cm}\hspace{-2.5cm}
\includegraphics[width=6.3cm,angle=-90]{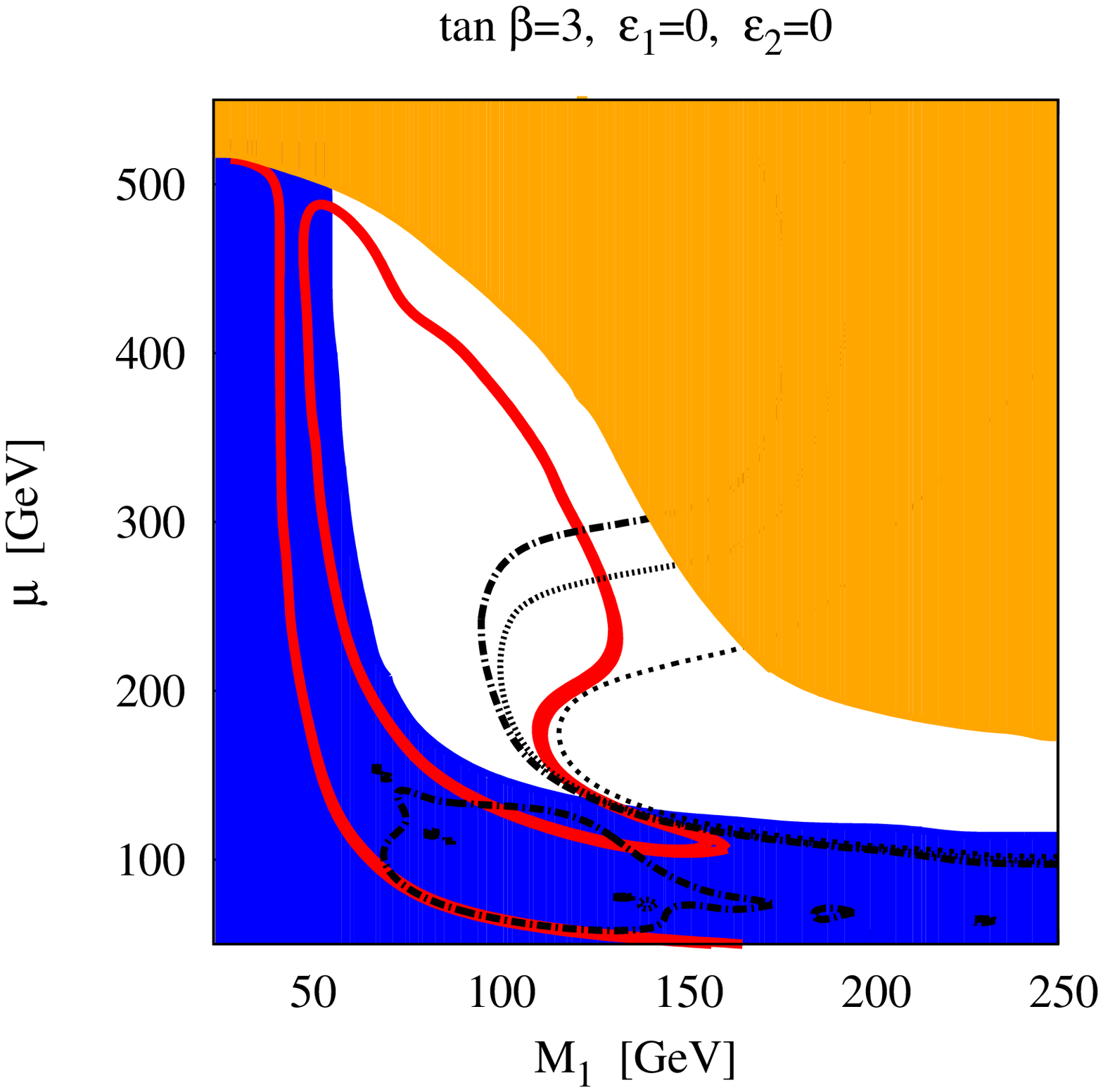}\hspace{-2.3cm}
\includegraphics[width=6.3cm,angle=-90]{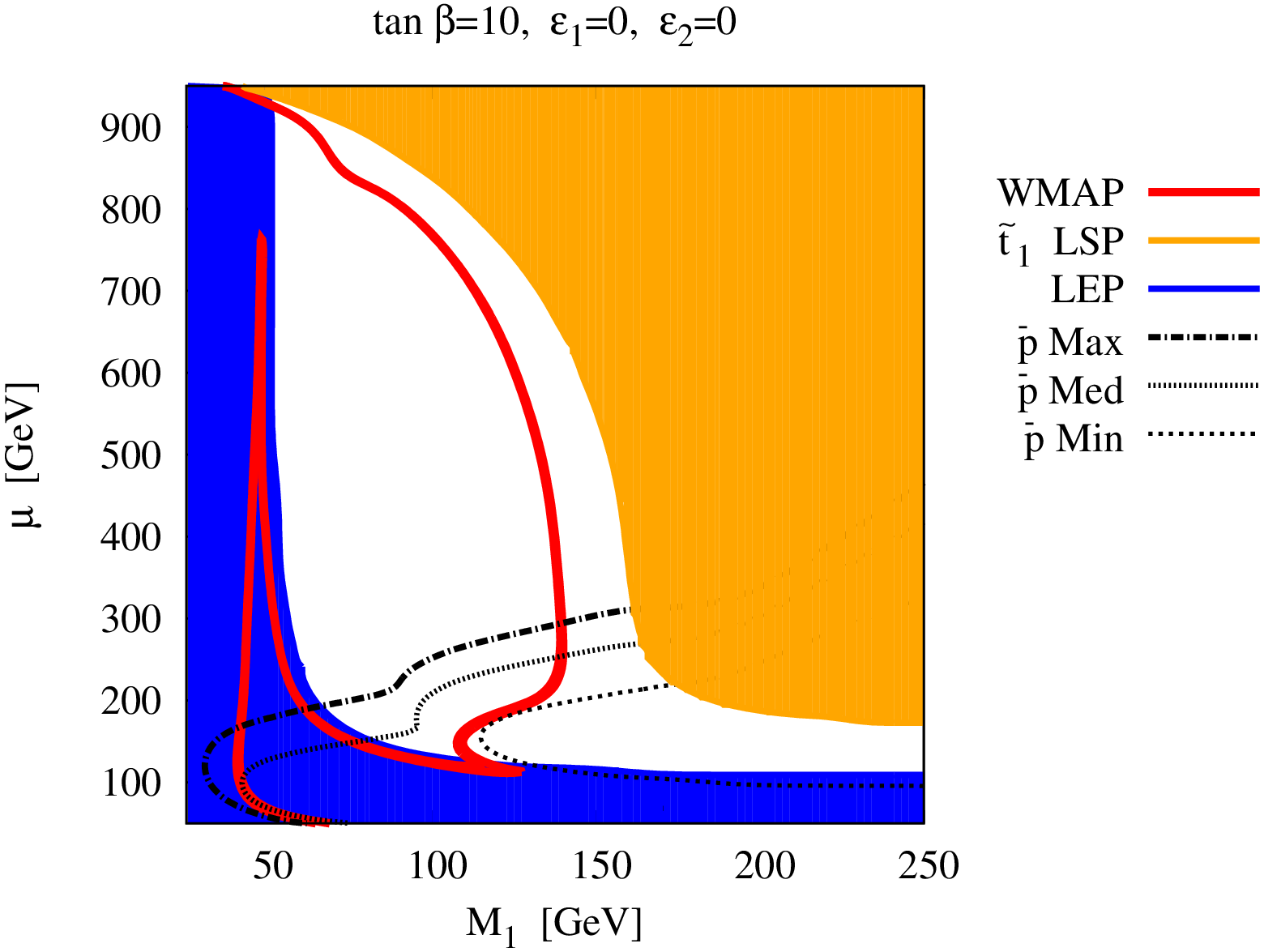}\\
\vspace{-0.4cm}\hspace{-2.5cm}
\includegraphics[width=6.3cm,angle=-90]{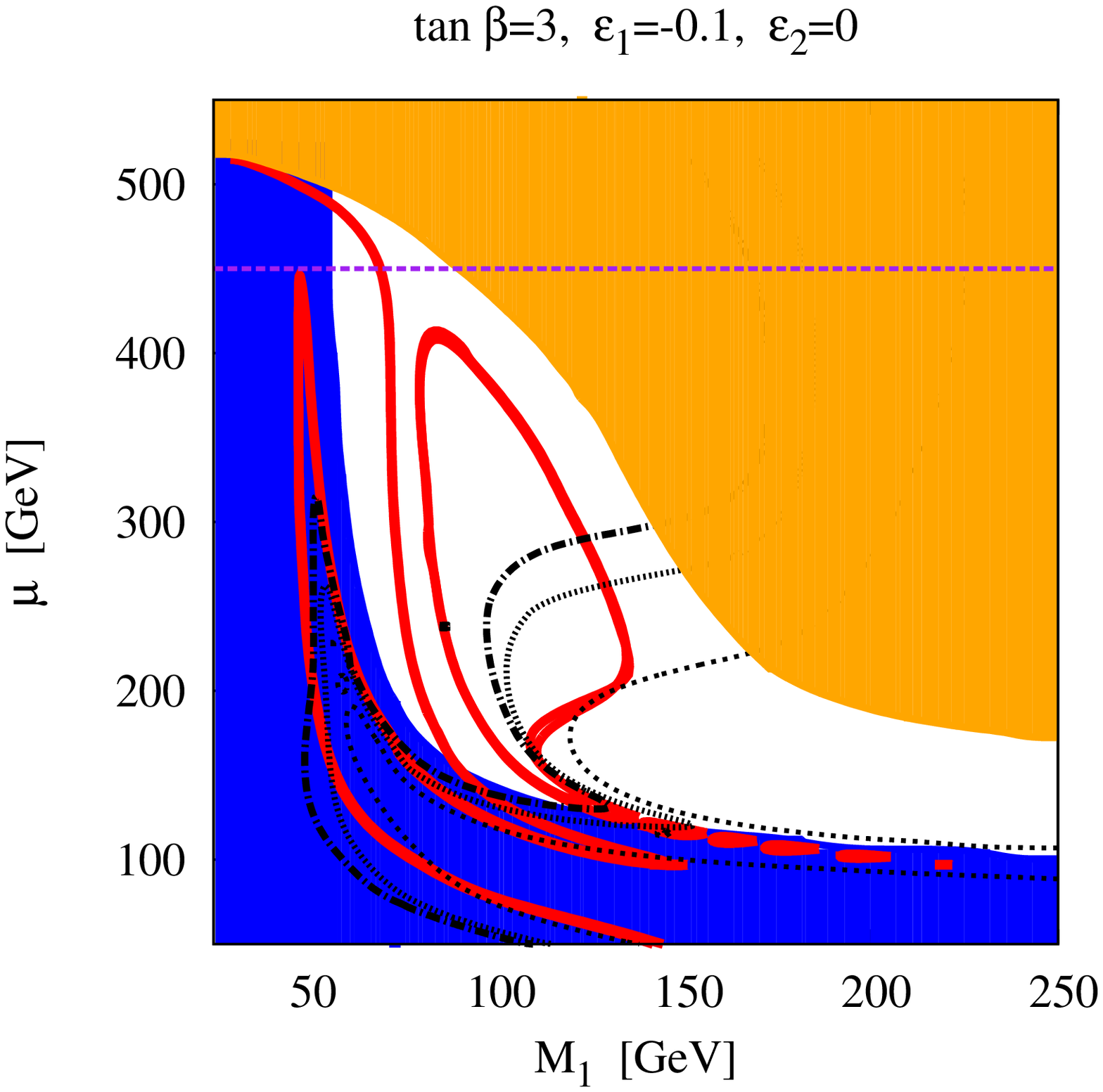}\hspace{-2.3cm}
\includegraphics[width=6.3cm,angle=-90]{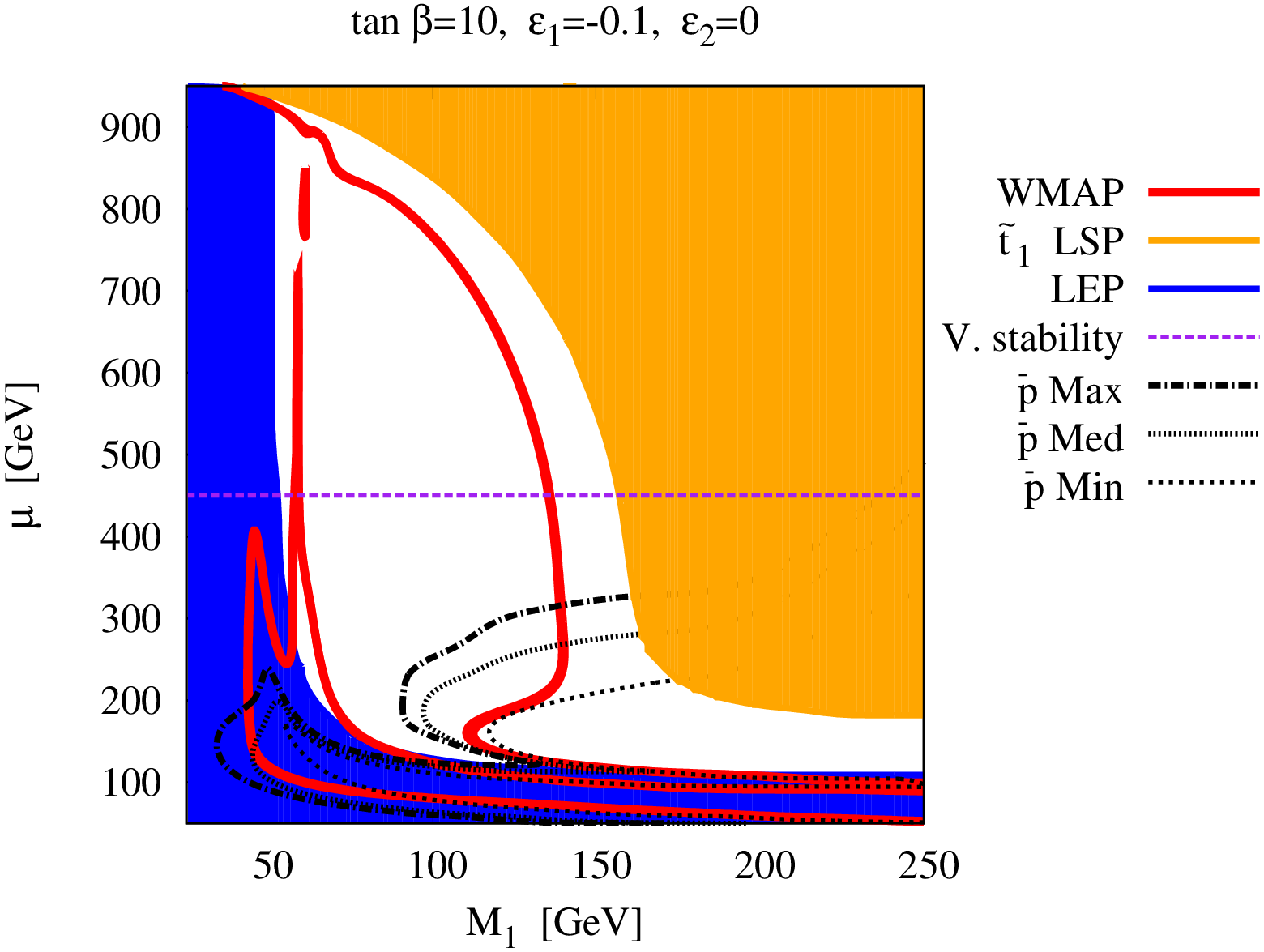}\\
\vspace{-0.4cm}\hspace{-2.5cm}
\includegraphics[width=6.3cm,angle=-90]{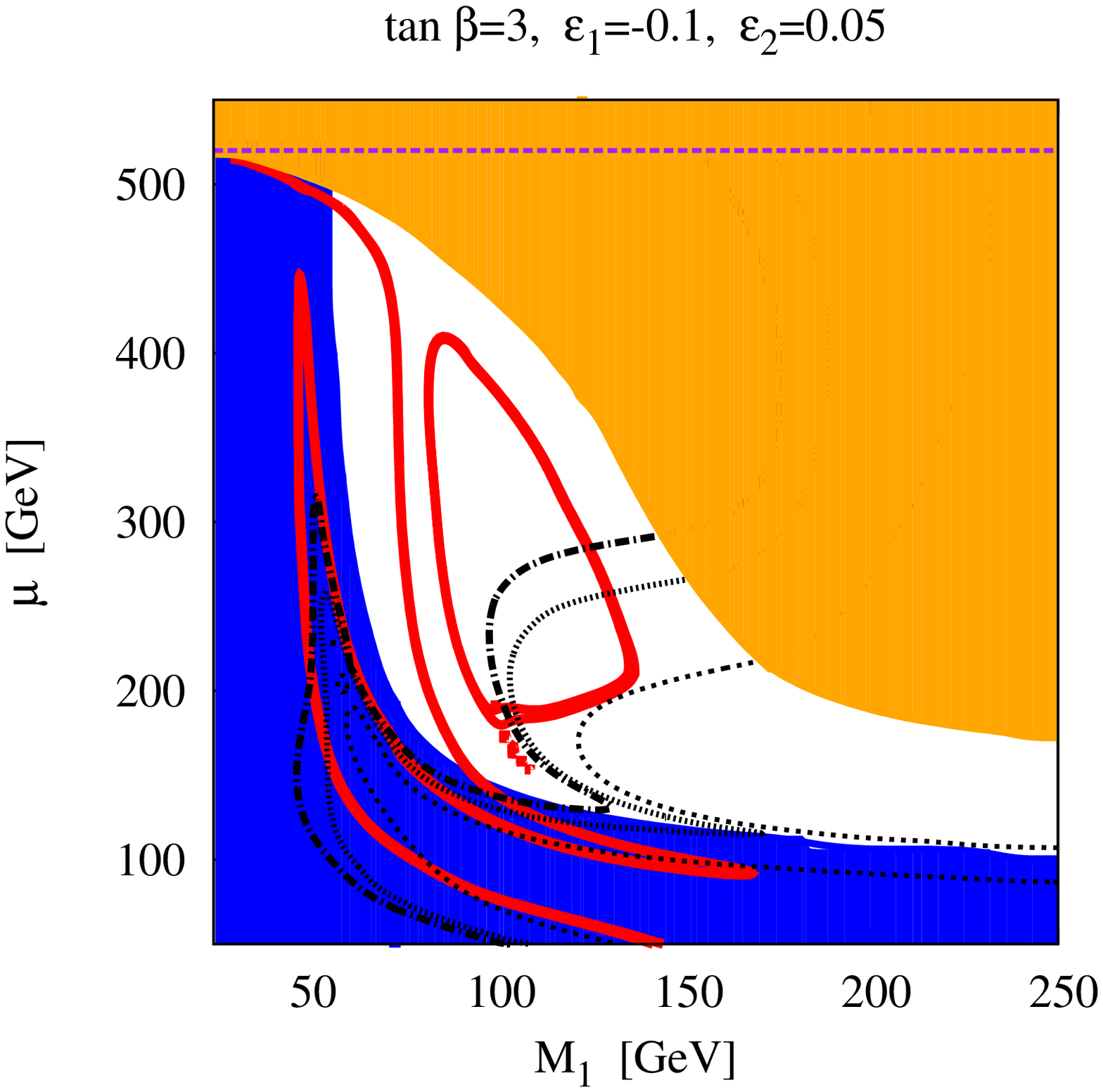}\hspace{-2.3cm}
\includegraphics[width=6.3cm,angle=-90]{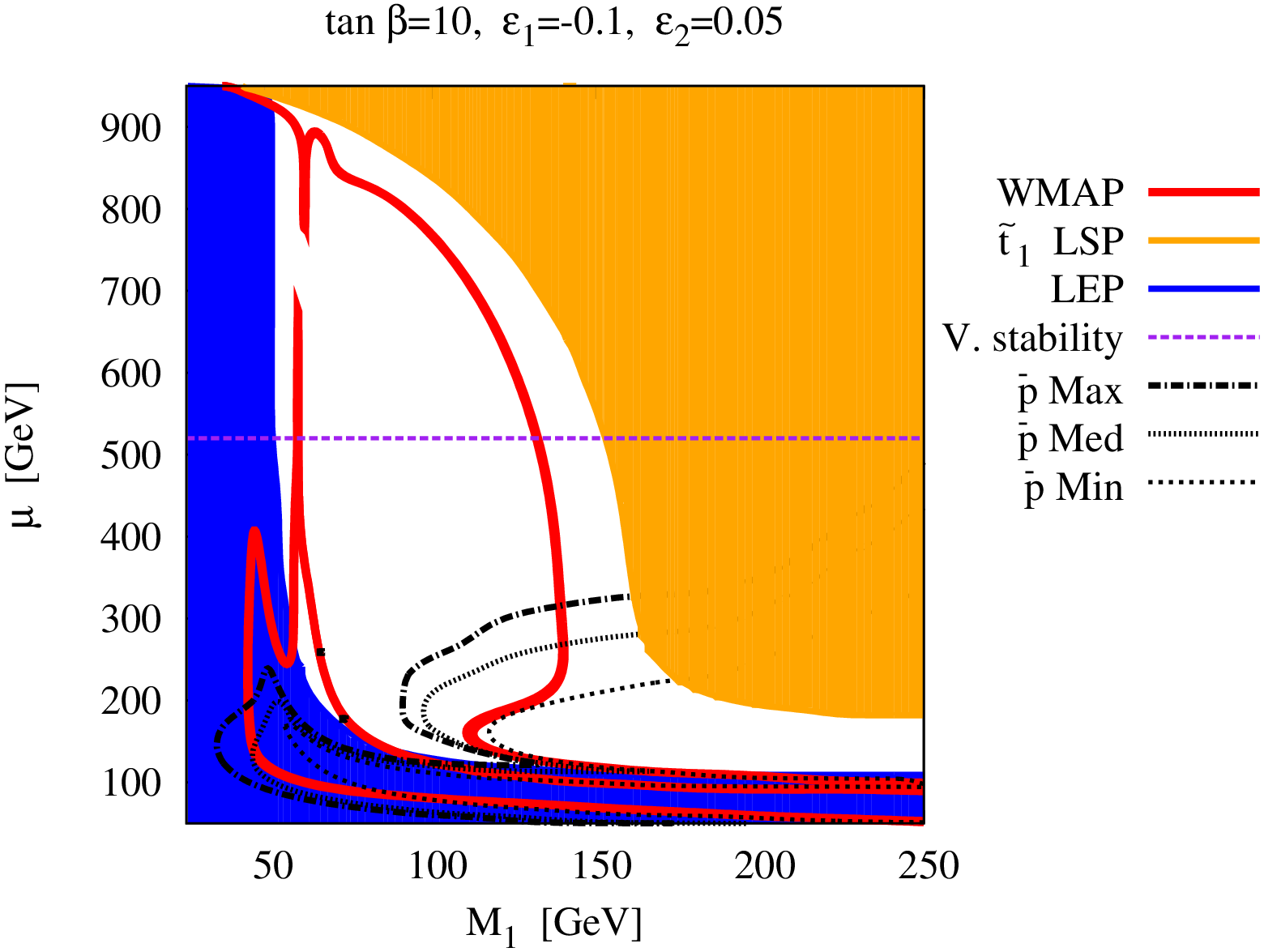}
\end{center}
\vspace{-0.9cm}
\caption{Regions in the $[M_1,\,\mu]$ plane that can be detected by a $3$-year run of the 
AMS-02 satellite mission for the scenario with light stops and heavy sleptons, in the antiproton channel. 
The black lines depict the detectability regions for the $3$ considered propagation models: 
the areas delimited by the axes and the black lines can be probed for the corresponding propagation
model (i.e. the regions towards the lower right side in each plot), as well as the areas delimited by closed
lines.}
\label{an2}
\end{figure}

Figure \ref{an2} presents the results for antiprotons and for the second scenario under consideration.
AMS-02 will be able to probe the regions lying within the oval-like blobs and the banana-shaped regions
delimited by the black contours and the axes.
Once again, the BMSSM turns out to be more favorable for DM detection than the ordinary case
of light stops and heavy sleptons without NR operators.
Detectable regions fall either into the case of the higgsino-like neutralino regime, or in the
low-mass $Z$ funnel region. We point out that an important part of the area where the dark matter
relic density is fulfilled via coannihilation with the lightest stop could also be tested.

\section{Conclusions}
\label{sec:con}

In this paper, we presented detection perspectives for neutralino dark matter
in the framework of the so-called BMSSM, where 
the superpotential is enriched by the addition of dimension-$5$ 
non-renormalizable operators in the Higgs sector. 
The main motivation to add non-renormalizable operators to the MSSM Higgs sector is
to reduce the fine-tuning that is required by the lower bound on the Higgs mass.
These new terms re-open the bulk region, already excluded because it gives
rise to a too light Higgs boson. Focusing on the new available regions,
we have studied four of the most popular detection modes: direct detection, as
well as three different channels of indirect detection, $\gamma$-rays from the galactic center, positrons
and antiprotons. We placed ourselves in the framework of the experiments XENON, Fermi and AMS-02.

According to our results, the most favourable detection mode seems to be, by far, the direct one.
XENON, even for low exposures, can provide a sizable parameter space coverage and, in case of a
negative detection, can exclude a significant part of the model. Moreover,
regions of the supersymmetric parameter space that were favored by the
WMAP constraint, but were excluded within the MSSM, become viable within the BMSSM and could
be potentially detected.

From the indirect detection point of view, we have shown that gamma-rays are the most favoured probe
in the benchmarks of the BMSSM we examined. We stress nevertheless that this is the case assuming
quite optimistic astrophysical conditions, namely an important collapse of dark matter in the galactic
center due to the presence of baryons, yielding a significantly spiked inner region profile.
We should also note again that we considered a $5$-year Fermi mission, whereas a period of $3$ years
was considered for the AMS-02 experiment.
On the other hand, antiprotons can provide significant information without having to assume
extreme conditions for the galactic medium: we only used the standard propagation models
used in numerous other analyses in the literature. However, the detection prospects could be
enhanced taking into account  DM substructures, which could provide a -small- amelioration
of the signals by at most one order of magnitude.
For positrons, our treatment
for the background seems to lead us to conclude that the PAMELA data, providing a large supplementary
background with respect to previous estimates, seem to render a positive detection
in this channel more difficult.

As a final remark, it is interesting to note again that for the low $\tan\beta$ values we considered
here, as we wanted to stick to the original motivation for the model (uplifting the Higgs mass without
over-constraining the stop sector), direct detection seems dominant. 
It is quite well-known that increasing $\tan\beta$ usually tends to 
ameliorate the indirect detection prospects, contrary to direct detection ones. This is an 
element which can give us an idea of the complementarity of the various detection 
modes. As underlined in \cite{Bernal:2008zk,Pato:2009fn}, different detection techniques offer the possibility
for a more complete parameter space coverage. Moreover, they can serve as a means for independent
confirmation of discoveries and/or comparison with other constraints. 
Given the controversy that has been generated during the past few years
on the nature of the various excesses that have occasionally been observed, we feel that combining information
from various sources is an essential element in our searches.

\section*{Acknowledgements}
The authors would like to thank G. Bélanger for useful discussions about the implementation of the model in {\tt micrOMEGAs}. 
AG would like to
acknowledge enlightening discussions with F. Bonnet, M. Fornasa, M. Gustafsson, M. Pato and M. Passera. 
We would like to thank Y. Mambrini  and S. Palomares-Ruiz for careful reading of the 
manuscript and valuable remarks. The work of NB and AG is 
supported in part by the E.C. Research Training Networks under contract MRTN-CT-2006-035505. 
AG is also supported by the french ANR TAPDMS.

\bibliographystyle{utphys}
\bibliography{biblio}

\end{document}